\documentclass[a4paper,12pt]{article}
\usepackage[utf8]{inputenc}
\usepackage[T1]{fontenc}
\usepackage{lmodern}  
\usepackage{geometry}
\usepackage{physics}
\usepackage{amsmath, amssymb}
\usepackage{graphicx}
\usepackage{hyperref}
\usepackage{authblk}
\usepackage[numbers,compress,sort]{natbib}
\usepackage{tikz}
\usepackage{pgfplots}
\pgfplotsset{compat=1.17}

\usetikzlibrary{lindenmayersystems}

\pgfdeclarelindenmayersystem{Koch}{\rule{F -> F-F++F-F}}

\usepackage{url}

\title{\textbf{Condensates: A Bridge to Quantum Gravity}}

\author{Piero Nicolini}
\affil{%
Dipartimento di Fisica, Universit\`a degli Studi di Trieste,\\
Strada Costiera 11, 34151 Trieste, Italy\\
\and
Istituto Nazionale di Fisica Nucleare (INFN), Sezione di Trieste,\\
Via Alfonso Valerio 2, 34127 Trieste, Italy\\
\and
Institut f\"ur Theoretische Physik, Johann Wolfgang Goethe-Universit\"at,\\
Max-von-Laue-Str. 1, 60438 Frankfurt am Main, Germany\\
\texttt{piero.nicolini@units.it}
}
\date{September 23, 2026}

\begin{document}

\maketitle

\begin{abstract}
Particle accelerators have carried fundamental physics to remarkable
precision, but the Planck scale remains fifteen orders of magnitude
beyond any foreseeable collider. We argue that this impasse is not a
temporary shortfall but an insurmountable one. The way out runs
through condensed matter physics: laboratory tests of generalized
uncertainty relations, analog systems, and --- most provocatively ---
frameworks in which spacetime itself is a condensate. We review the
string-loop picture, in which quantization turns the world sheet into
a three-dimensional fractal, the corpuscular description of black
holes as self-sustained graviton condensates, and the phase-transition
structure connecting the Riemannian phase to a pre-geometric one. The
conclusion is deliberately provocative: the question of quantum
gravity is no longer a question of energy, but of crossing the bridge.
\end{abstract}

\begingroup\small

\vspace{0.5em}
\noindent This article is an invited contribution under review for the Special Issue ``The Universe Observed With Particle Detectors: Celebrating the Scientific Legacy of Prof. Guido Barbiellini Amidei'' of the journal \textit{Condensed Matter}.
\endgroup

\vspace{1em}

\newpage

\section{Introduction}

Every bridge begins on one bank and ends where the terrain is still unmeasured.
This essay argues that quantum gravity, long expected to arrive from the far
side of the energy frontier, is already reachable from the bank we stand on:
condensed matter.

What would happen if time stopped in 1974? The sentence sounds like a
provocation, but it states an undeniable truth, at least for anyone assessing
the current status of research in physics. Despite gigantic investments and
improved technological capabilities, physics as a fundamental science suffers
from a long list of unsolved problems: the nature of dark matter and dark
energy, the origin of the matter--antimatter asymmetry, the statistical origin
of black hole entropy, and the cosmological constant problem. Rather than
enumerating further puzzles, we should focus on the crux of the matter.

Historically, the answer has been the search for fundamental objects
governing the microscopic world, whose properties emerge at larger scales in
phenomena of everyday experience -- a question as old as ancient Greek atomists.
This is what makes physics unique among the sciences: its predictive power rests
on the ability to connect observed phenomena to microscopic fundamental
structures. No other field has an equivalent of this reductionist program.
Cell dynamics in biology, volcanoes and earthquakes in earth science, stock
market dynamics in economics: all remain at the level of phenomenological
description, without a comparably deep theoretical understanding. If this is
the main success of physics, the actual question is to understand what counts
as fundamental.

This explains why, from Rutherford's scattering experiments to LHC
collisions, particle physicists like Guido Barbiellini have probed
increasingly shorter length scales.
In 1974, the discovery of the $J/\Psi$ \cite{Aubert:1974js,Augustin:1974xv} provided crucial evidence for the
Standard Model and virtually marked the end of what we might call `old
physics'. This is more than the birth of a new particle: the $J/\Psi$
resolved into a bound state of charm quarks, objects which by confinement
cannot be isolated, so that the very concept of what counts as an
elementary particle changed from that moment on. Following Hawking's
announcement at Caltech---before Feynman---about black holes emitting
thermal radiation~\cite{Hawking1975,Nicolini2025}, the year
1974 also marks the
beginning of `new physics', a regime where gravity plays a significant
role at microscopic scales. 
Unfortunately, no significant signals of the new physics,
nor of particle sectors beyond the Standard Model, have been observed
since. The fundamental scale is the Planck energy $E_\mathrm{P}\sim 10^{19}$~GeV, 
fifteen orders of magnitude beyond the LHC and unreachable by any foreseeable technology. 
Quantum gravity and its connected open problems are thus stranded without a
laboratory. One has to look for alternatives to conventional particle
accelerators.

Guido Barbiellini was already aware of the limits of accelerator physics,
and of how distant the Planck scale is. For this reason, he devoted
significant efforts to signal detection in astrophysics by means of
satellites, such as the Italian Space Agency's AGILE project, of which he
became a co-principal investigator.
Astrophysical observations push the energy frontier, but they inherit a
severe limitation: unlike a laboratory experiment, they cannot be
repeated, tuned, or designed. A gamma-ray burst is observed once, at a
random energy, with no control on the setup. What is needed is not a
bigger telescope or a bigger accelerator, but a different kind of
laboratory altogether. In the following, we argue that such a laboratory
already exists: condensed matter systems, where quantum gravity
phenomenology can be tested, simulated, and possibly realized.

\section{The High-Energy Impasse}
\label{sec:heimpasse}
If we put it in simple terms, quantum gravity is not only plagued by theoretical
hurdles, but also by the absence of observational data. Such a difficulty in
detecting gravity in the quantum realm is connected to a simple character of
gravity: its extreme weakness. Gravitational effects only show up in the
presence of large masses (like planets, stars), whereas they are negligible in
the microscopic world. As a piece of evidence we can think of the simplest
quantum mechanical system, the hydrogen atom, and consider the energy levels of
the electron by including a gravitational coupling between the mass of the
electron and that of the proton:
\begin{equation}
V(r)=\frac{e^2}{r} \left( 1 - \frac{G m_e m_\mathrm{p} }{e^2 }\right)
\label{eq:potential}
\end{equation}
Observed data rule out effects of this kind with extreme
accuracy~\cite{Kramida2010}. This is no surprise, but simply the direct
consequence of the gravitational coupling constant. When expressed in natural
units it has the following dimensions
\begin{equation}
[G]= \mathrm{energy}^{-2}.
\label{eq:G}
\end{equation}
The value of $G$ defines the above energy scale, which by definition is the
aforementioned Planck energy. Since $m_e\sim 0.5$ MeV, $m_\mathrm{p}\sim 1$ GeV
and $e^2\sim 1/137$, the correcting term in \eqref{eq:potential} is of the order
of $10^{-40}$. This is a problem that is more severe than the generic lack of
observation of signals beyond the Standard Model. In practice, such a problem,
called hierarchy problem, not only posits the question of the completeness of
the Standard Model, but depicts in clear terms how insurmountable the hurdle is
to bring gravity into the quantum world. About 30 years ago
Antoniadis~\cite{Antoniadis1990}, Arkani-Hamed, Dimopoulos and
Dvali~\cite{ADD1998,ADD1999} and again Antoniadis et al.\ (AADD)~\cite{AADD1998}
with the large extra dimensions, Randall and Sundrum (RS1 and
RS2)~\cite{Randall1999a,Randall1999b} with brane-world models on one side and
Appelquist, Cheng and Dobrescu with the universal extra dimensions~\cite{Appelquist2001} on the other
side tackled the hierarchy problem by proposing string inspired mechanisms to
lower the Planck scale to a scale potentially accessible to the Large Hadron
Collider or near future accelerators. Despite the technical differences, the
basic idea of such proposals is connected to the great uncertainty in measuring
gravity below a millimeter. Such an experimental inability, however, became a
theoretical advantage, because it allowed us to conjecture the existence of two
regimes for gravity, a conventional weak gravity regime and a new, unexpected
strong gravity regime which shows up only below a certain scale $R$ of the order
of 1 mm or smaller. In such a new regime, additional spatial dimensions would
open up and a new coupling constant $G_\mathrm{F}$ depending on some powers of a
new fundamental scale $M_\mathrm{F}\lesssim R^{-1}\ll E_\mathrm{P}$ would govern
gravity in place of $G$. The great expectations back at the time were that
$M_\mathrm{F}$ could be ``just beyond the corner,'' namely in a range 1--10 TeV
offering the access to the new physics. To guarantee the consistency with the
old physics tested so far, $M_\mathrm{F}$ must exceed the highest energy scale
ever reached in particle detectors, that at the time was about 1 TeV, namely the
peak energy of the Tevatron at the Fermi National Accelerator Laboratory. This
value implied a bound for the length scale $R$, namely
$R\lesssim 10^{-19}$ m. The brilliant idea behind both the large extra dimension
and brane world scenarios was the possibility of evading such a constraint on
$R$, by assuming that, while gravity can propagate in the full higher dimensional
spacetime, the other fundamental interactions are confined to a 3 dimensional
brane, where Standard Model predictions hold. In this way $R$ can extend up to a
millimeter, allowing us to lower $M_\mathrm{F}$ to the terascale with only two
extra spatial dimensions. The most striking prediction of having a strong
gravity regime was the possibility of observing matter collapsing in a particle
detector. Despite the details of the collapse being, and still remaining, poorly known,
a microscopic black hole will inevitably form, having a mass equivalent to that
of the colliding particles~\cite{Dimopoulos2001,Giddings2002,Banks1999}.
Given the cross section
\begin{equation}
\sigma\left(\mathrm{p}\mathrm{p}\rightarrow \mathrm{BH}\right)=
\pi r_\mathrm{g}^2 (M)\sim 5\ \mathrm{nb}
\end{equation}
for a black hole mass $M\sim 1$ TeV~\cite{Dimopoulos2001,Giddings2002}, and the
LHC design luminosity
$L\sim 1\times 10^{10}$ b$^{-1}$ s$^{-1}$~\cite{LHC2025},
black holes would form at an extremely generous rate
\begin{equation}
\frac{dN}{dt}=\sigma L\sim 50\ \mathrm{s}^{-1},
\end{equation}
turning the LHC into a factory able to make more than a billion microscopic
black holes over a year! This conclusion has generated a terrific amount of
follow up work and has become a topic in newspapers, movies, science fiction,
safety concerns and doomsday scenarios.
The life of black holes in particle detectors was expected to be equally
astonishing. Due to their size, such black holes would decay due to Hawking
radiation on a time scale of roughly $1/\mathrm{TeV}\sim 10^{-27}$ s, having
extreme temperatures of the order of $1~\mathrm{TeV}\sim 10^{16}$ K. The black
hole thermal emission, mostly in terms of lower mass, lower spin particles,
would trigger a cascade of particle production events in terms of
\textit{Bremsstrahlung} and pair production mechanisms. The net result of the
event would be the formation of two atmospheres, a quark-gluon plasma and an
electron-positron-photon plasma~\cite{Casanova2005,Bleicher2010}. Since QCD
particle scattering processes become relevant at temperature
$T_\mathrm{QCD}\sim 175$ MeV, smaller than the corresponding QED critical
temperature $T_\mathrm{QED}\sim 50$ GeV~\cite{Heckler1997}, the quark-gluon
plasma would dominate the emission spectrum and finally hadronize on a length
scale of 1 fermi or equivalently in $\sim 10^{-24}$ s. In the process the black
hole would cool down to an effective temperature much smaller than the initial
Hawking temperature.
The related signatures are very distinctive. Due to the explosive nature of the
event, the geometry of the emission is expected to be spherical, corresponding
to a suppression of di-jet events, hadronic to leptonic activity of roughly 5:1.
More importantly, the black hole will radiate both on the brane and in the full
higher dimensional spacetime. This means that particle detectors should observe a
missing energy with respect to that of the colliding particles. At the end stage
of evaporation, the black hole would emit a few hard visible quanta, or
alternatively would collapse into a remnant, a black hole object expected to
escape detection unless charged.
The net result of experimental observations at the LHC is that no black hole has
been observed so far, making the full paradigm about a possible solution of the
hierarchy problem under tremendous stress~\cite{CMS2026}.
Possible explanations of such very disappointing results are of three kinds,
with decreasing degree of pessimism: 1) we have produced black holes, but we
are not able to detect them because we were looking for the wrong signatures \cite{Nicolini:2005vd,Nicolini:2008aj,Rizzo:2006zb,Nicolini:2011nz,Gingrich:2010ed};
2) Extra dimensions do exist, but the LHC working energy was just below
$M_\mathrm{F}$ -- so we fell short before getting the first ever quantum
gravity signal; 3) The full idea of solving the hierarchy problem with extra
dimensions is wrong, there is no hope to observe gravity effects in particle
accelerators. In this third scenario, gravity itself acts as a barrier: below
the Planck scale the gravitational radius of any colliding particles is
smaller than their Compton wavelength, so that no black hole can form at
all~\cite{Mureika2012}.

In the most pessimistic of the above scenarios, one has to face a variety of
problems when considering the possibility of building a Planck energy
accelerator, including the size~\cite{Roser2023,Loeb2015,Casher1995,LoebMedium}.
Being the relativistic Larmor radius
\begin{equation}
r_{\rm c}={\frac {p_{\perp }}{|q|B}}={\frac {\gamma m v_{\perp }}{|q|B}}
\end{equation}
proportional to the linear momentum, the actual size of a Planckian accelerator
would scale as
\begin{equation}
r_{\rm P}=\left(\frac{E_{\rm P}}{E_{\rm LHC}}\right)\left(
\frac{B_{\rm LHC}}{B_{\rm P}} \right)r_{\rm LHC}
\sim 10^{19}\ \mathrm{m}\sim 10^3\ \mathrm{l.y.}
\end{equation}
for equivalent technology, i.e., $B_{\rm P}\sim B_{\rm
LHC}$~\cite{Siegel2024,Danchev2026}. 
Particle physics could still have something relevant to say, but with
facilities alternative to particle accelerators -- a route whose viability
Barbiellini anticipated decades ago, as discussed above.
Cosmic rays hitting
the Earth atmosphere offer also a formidable testbed, if we think that the 1991
Oh-My-God particle, the highest energetic event ever observed, had an energy of
$10^{11}$ GeV~\cite{Bird1995}. Cosmic rays, however, posit another kind of
technological problem: due to their intensity scaling with respect to energy
\begin{equation}
I(E)\propto E^{-3}
\end{equation}
any Earth based detector must be large in order to catch a decent number of
events per unit of time in the high energy region of the spectrum: above
$E\sim 10^{10}$~GeV, one expects on average less than one event per km$^2$ per
year~\cite{Swordy2001,DeAngelis2018}. For instance the detection area at the
Pierre Auger Observatory is 3,000 km$^2$, about 12 times the area of the city of
Frankfurt~\cite{Abraham2004}.

Despite many efforts and ongoing attempts, we realistically have to consider
the hypothesis that we will never get to the Planck scale with such kind of
experiments.

\section{Condensed Matter as a Laboratory}

Nothing is infinite, including financial resources. High energy physics is no
exception. Even if accelerator-based experiments can provide significant or
promising results for the solution of the quantum gravity puzzle, they may run
out of money. In other words, even if it works, high energy physics is not
cheap, and its costs increase with the energy. In analogy with the Malthusian
growth of populations against finite resources~\cite{Malthus1798}, we can
conjecture that costs grow exponentially with energy,
\begin{equation}
\$=\$(E)=\$_0\, e^{E/E_0},
\label{eq:costlaw}
\end{equation}
where $E_0$ is a reference energy scale. We stress that \eqref{eq:costlaw} is
a conjecture, not an established scaling law: construction, maintenance,
electricity and personnel costs all grow with the collision energy, plausibly
each with a different power of $E$. This explains the existence of
cancellation risks, namely the possibility that experiments are terminated and
facilities abandoned, as happened with the Superconducting Super Collider in
1993~\cite{SSC1993}.

Fortunately, there is an alternative, offered by condensed matter physics. By
this, we mean more generally phenomena not belonging to so-called high energy
physics. This is of course a pedagogical simplification for the sake of
clarity only, because physics as a discipline cannot be divided into
compartments. Physicists, however, tend to think in terms of sub-fields, and
organize themselves, socially and academically, into communities, to the
detriment of the quality of research and of full scientific understanding.

Historically, there are nevertheless many points of contact between the two sub-fields -- e.g.\ Goldstone
bosons~\cite{Goldstone1961}, string theory born as a theory of
hadrons~\cite{Veneziano1968}, the Schwinger effect for vortex nucleation
dynamics~\cite{Desrochers2025} -- contacts that often led to
groundbreaking results in physics.
Now, the crux of the question is that the role of condensed
matter physics is twofold: it can be both an ``experimental laboratory'' for
quantum gravity, and a ``theoretical laboratory'' for quantum gravity. By
experimental, we mean the possibility that current or near future high
precision low energy systems can actually provide the first quantum gravity
data ever. The terminology ``theoretical laboratory'' seems
self-contradictory, but here it simply indicates the testing of theoretical
conjectures in a phenomenological framework or energy regime where physics is
better understood.

The list of current high precision experiments is rich. Although a genuine
quantum gravity signal has never been detected, there is an array of
experiments aiming to set compelling bounds on parameters descending from a
quantum gravity formulation. This is the case for deformed quantum
commutators. According to theoretical investigations about the collision of
strings at Planckian energy~\cite{Amati1989,AmatiVeneziano1991,Kempf1995},
conventional quantum commutators are just the low energy limit of actual
commutators at the Planck scale:
\begin{equation}
[\textbf{x},\textbf{p}]=i\hbar\left(1+\alpha \textbf{x}^{2}+\beta \textbf{p}^{2}\right)
\label{eq:gup}
\end{equation}
Here we can consider $\alpha\approx 0$, corresponding to no uncertainty in
momentum, or equivalently no maximal length scale. From \eqref{eq:gup}, one
finds that the parameter $\beta$ has the dimension of an inverse momentum
squared, i.e.\ a length squared in natural units, while it is dimensionless in
Planck units $l_\mathrm{P}=1/M_\mathrm{P}=1$.

Experiments with optomechanical resonators have set a bound $\beta\sim
10^{12}$, corresponding to the existence of a minimal resolution length
$\sim 10^{6} l_\mathrm{P}$ over the spacetime~\cite{Pikovski2012}. To achieve
a similar bound with particle accelerators, one would have to reach
$10^{13}$ GeV, something like 100 times more energetic than the Oh-My-God
particle. Further experiments with micro- and nano-oscillators have improved
the above result, setting a bound $\beta\sim 3\times 10^{7}$, which implies a
minimal length $\sim 10^{4} l_\mathrm{P}$~\cite{Bawaj2015}. Quartz resonators
have, however, set the benchmark in the field, with $\beta\sim 4\times
10^{4}$ and a minimal length $\sim 10^{2} l_\mathrm{P}$~\cite{Bushev2019}.

There also exists an array of ideas for forthcoming experiments, including
cryogenic technology~\cite{Bekenstein2012,Bekenstein2014}, laser
interferometry~\cite{Hogan2012}, scanning tunnelling
microscopy~\cite{DasVagenas2008} and cold atom gases in the form of
Bose-Einstein condensates~\cite{Howl2023,Aziz2025}.

It is evident that the above tests have by far surpassed the results obtained
with accelerator physics so far. It is already possible to draw some
conclusions: extra-dimensional scenarios and the solution of the hierarchy
problem are in big trouble. Still we miss any quantum gravity data or direct evidence to rule them
fully out. This is the only hope to save such proposals, for instance by
conjecturing that on the brane quantum gravity effects are actually
Planckian for some unknown mechanism.

The other reason for strong interest in condensed matter systems is the
possibility of testing theories by studying phenomena which behave similarly
to what we expect to happen at the Planck scale. For instance, the analogy
between gravitational systems and acoustics can offer a reliable theoretical
testbed for theoretical formulations that would otherwise be difficult to
understand. In addition, when an analogy is fully established, it can open
the way to experimental verification by means of a cheap low energy
apparatus. Unruh has probably set the first milestone in this program by
recognizing similarities between an event horizon and any critical
surface~\cite{Unruh1981}. By critical surface we mean a surface where a
certain phenomenon no longer occurs, or where a transition to a new regime
takes place. As an example, we can consider a fluid moving along the $z$-axis
in a tube passing through a de Laval nozzle, namely a narrower region where
the fluid can increase its speed. 

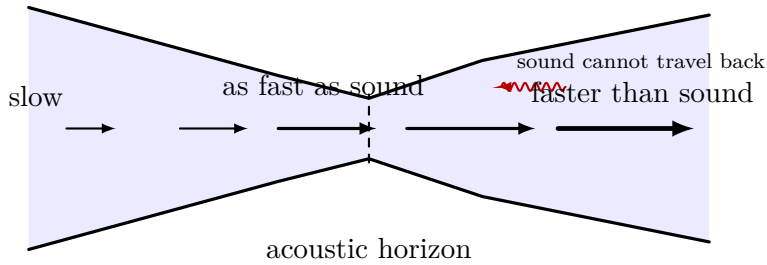
\begin{figure}[htbp]
\centering
\begin{tikzpicture}[scale=1.0,line cap=round]
\fill[blue!8]  (-4.5,1.6) -- (-1.2,0.7) -- (0,0.4) -- (1.5,0.9) -- (4.5,1.5)
               -- (4.5,-1.5) -- (1.5,-0.9) -- (0,-0.4) -- (-1.2,-0.7)
               -- (-4.5,-1.6) -- cycle;
\draw[very thick] (-4.5,1.6) -- (-1.2,0.7) -- (0,0.4) -- (1.5,0.9) -- (4.5,1.5);
\draw[very thick] (-4.5,-1.6) -- (-1.2,-0.7) -- (0,-0.4) -- (1.5,-0.9) -- (4.5,-1.5);
\draw[dashed,thick] (0,-0.45) -- (0,0.45);
\draw[-latex,thick]       (-4.0,0) -- (-3.35,0);
\draw[-latex,thick]       (-2.5,0) -- (-1.6,0);
\draw[-latex,very thick]  (-1.2,0) -- (0.1,0);
\draw[-latex,very thick]  (0.5,0)  -- (2.2,0);
\draw[-latex,line width=1.8pt] (2.5,0) -- (4.3,0);
\draw[decorate,decoration={snake,amplitude=1.5pt,segment length=4pt},
      -latex,thick,red!70!black] (2.6,0.55) -- (1.7,0.55);
\node[above left]  at (-3.9,0.15) {\small slow};
\node[above]       at (-0.6,0.3)  {\small as fast as sound};
\node[above right] at (2.0,0.2)   {\small faster than sound};
\node at (3.6,0.9) {\scriptsize sound cannot travel back};
\node[below] at (0,-1.3) {\small acoustic horizon};
\end{tikzpicture}
\caption{Sound trapped in a de Laval nozzle. The fluid flows from left to
right and accelerates as the channel narrows (arrow length: flow velocity).
At the narrowest point the flow speed equals the local speed of sound.
Beyond that line, sound waves can no longer travel back: the red wave is
dragged along by the flow. The dashed line is an acoustic horizon for
phonons, the analogue of a black hole horizon for photons; sound
propagation is governed by the acoustic metric \eqref{eq:acousticmetric}.}
\label{fig:laval}
\end{figure}

If the fluid speed at the nozzle equals the
speed of sound, an acoustic horizon forms, separating subsonic from supersonic
regions of the fluid~(cf.\ Fig.~\ref{fig:laval}). The analogy works because it is not only geometric but
also physically grounded. While an event horizon works as a trapping surface
for photons, an acoustic horizon acts as a trapping surface for phonons, the
elastic perturbations of the fluid. One can see this by expanding density,
pressure and displacement of a zero viscosity, locally irrotational fluid in
terms of average bulk variables $\left(\rho_0, p_0, \psi_0\right)$ and
perturbations $\left(\rho_1, p_1, \psi_1\right)$. The resulting equation of
motion for the perturbation $\psi_1$ is that of a scalar field minimally
coupled to a background metric $g_{\mu\nu}$,
\begin{equation}
\Box \psi_1=\frac{1}{\sqrt{-g}}\partial_\mu \left( \sqrt{-g} g^{\mu\nu}\partial_\nu \psi_1 \right)=0,
\label{eq:boxpsi}
\end{equation}
where
\begin{equation}
ds^2 =  g_{\mu\nu} dx^\mu dx^\nu =\frac{\rho_0}{c_\mathrm{s}^2}\left[-c_\mathrm{s}^2 dt^2 +\left(dz-v_0(z)dt\right)^2\right]
\label{eq:acousticmetric}
\end{equation}
is the (1+1)-dimensional acoustic line element. Here $c_\mathrm{s}$ is the
local speed of sound, defined as
$c_\mathrm{s}^{-2}\equiv\left.\frac{d\rho}{dp}\right|_{p=p_0}$, and $v_0(z)$
is the average bulk speed. In the more general (3+1)-dimensional case,
\eqref{eq:boxpsi} and \eqref{eq:acousticmetric} provide an exact mapping
between a fluid undergoing supersonic motion and a scalar field propagating
in a curved background spacetime. Similarly, other phenomena, such as the
refracting critical angle and surface gravity waves, offer such a mapping to
a spacetime having an event horizon~\cite{Barcelo2011}.

The most intriguing application of such a formalism lies in the possibility
of studying the evaporation of black holes by means of acoustic analogue
systems. According to Hawking, black holes can emit thermal radiation like a
black body with a temperature given by their surface gravity~\cite{Hawking1975}.
One of the
major theoretical problems of Hawking's prediction is that a black hole, by
emitting particles and radiation, loses energy, shrinks and becomes hotter,
as its temperature is inversely proportional to its mass. The process is
runaway and non-linear: when the black hole temperature and the black hole
mass are of the same order of magnitude, $T\sim M$, back-reaction effects
become relevant, since the energy of the emitted particles couples via
gravity to the mass of the black hole itself. The other crucial problem of
Hawking's prediction is related to the possibility of detecting such thermal
radiation. In the most optimistic case known in astrophysics, namely the case
of solar mass black holes--$M\sim M_\odot$ where $M_\odot \sim 10^{30}$
kg--the black hole temperature is $T\sim 10^{-7}$ K, which is far below the
temperature of the cosmic microwave background, $T_\mathrm{CMB}\simeq 2.73$
K. This confirms the intuition that quantum mechanical effects are
negligible for sun-sized macroscopic objects. In practice, one would need
microscopic black holes to observe relevant emission, but this case would
correspond to going back to the high energy physics realm, probably a
``no-go'' case for what has been discussed in Section~\ref{sec:heimpasse}.

For the above reasons, in the recent past research activity has aimed to
clarify both analytically and numerically the amount of back reaction effects
for an acoustic black hole during its evaporation life
cycle~\cite{Balbinot2005,Carusotto2008,Balbinot2008}. These theoretical
studies have paved the way for the realization of an acoustic black hole in a
laboratory, in terms of very low temperature atomic Bose-Einstein condensates,
to meet the low viscosity condition behind \eqref{eq:boxpsi} and
\eqref{eq:acousticmetric}~\cite{Lahav2010}. This has led to the first ever
observation of Hawking radiation in an analogue gravity
system~\cite{Steinhauer2014}. More importantly, such an acoustic black hole
displayed Hawking radiation matching the above theoretical predictions, a
fact that represents as of today the most striking evidence in support of
Hawking's work as well as the first and only evidence of a quantum gravity
phenomenon~\cite{Steinhauer2016,Carusotto2016}.

Since then, further experiments have confirmed the above
results~\cite{MunozDeNova2019,Kolobov2021}. Other platforms, based on ion
rings~\cite{Horstmann2011}, helium-3 A films~\cite{Volovik1999,Volovik2001,Volovik2021}
and ultrashort laser technology, have been proposed to detect Hawking
radiation and have triggered an intense discussion on the
topic~\cite{Belgiorno2010,Schutzhold2011,Belgiorno2011}. Interestingly, the
discovery of the Hawking emission has been accompanied by the detection of
black hole superradiance~\cite{Torres2017}, a phenomenon predicted by
Zel'dovich, Starobinsky, Misner, Unruh and other
authors~\cite{Misner1972,Starobinsky1973,Unruh1974,PressTeukolsky1972,Zeldovich1971,Zeldovich1972}.
All such work is important evidence that there has to be a fundamental
relation between physical phenomena at different energy scales, in particular
between particle physics, cosmology and condensed matter
physics~\cite{Volovik2006}.

One of the criticisms one can raise is that such analogies are purely
kinematic. Although the phenomenology of both gravitational systems and their
analogue systems can be described with the same mathematics, interactions are
inherently different. Gravity in particular, resulting from the exchange of a
spin-2 field, is rather peculiar and drastically differs from any force
governing condensed matter physics. There is, however, another possibility
that can supersede the concept of analogy and lift it to a more powerful
level of description. Maxwell's equations offer a simple illustration. The
only candidate magnetic monopole event ever recorded, on Valentine's Day
1982 in a superconducting loop detector~\cite{Cabrera1982}, was never
confirmed; if monopoles exist, they must be hard to isolate, because of a
very strong, although finite, force. The intensity of the magnetic
interaction is reflected by the value of the coupling constant
\begin{equation}
\alpha_{\mathrm m}\gg \alpha_{\mathrm e},
\end{equation}
which is typically $\alpha_{\mathrm m}\approx 34$~\cite{Dirac1931}. Therefore, if monopoles exist, the net
result is the appearance of a feeble although non-vanishing magnetic charge
distribution $\rho_\mathrm{m}$ and magnetic current $j_\mathrm{m}$ in the
equation for the divergence of the magnetic field, and in Faraday's equation
\begin{align}
\nabla \cdot \mathbf{B} &= 4\pi \rho_{\mathrm{m}}, \nonumber\\
\nabla \times \mathbf{E} + \frac{1}{c}\frac{\partial \mathbf{B}}{\partial t}
&= -\frac{4\pi}{c}\, \mathbf{j}_{\mathrm{m}}.
\label{eq:maxeq}
\end{align}
In the Maxwell limit the magnetic force is infinite
($\alpha_{\mathrm m}\to \infty$) and both the distribution and current vanish,
$\rho_\mathrm{m}\approx 0$, $j_\mathrm{m}\approx 0$. 

The novelty of \eqref{eq:maxeq} is a new symmetry of Maxwell's equations
under the exchange $\mathbf{E}\to\mathbf{B}$, $\mathbf{B}\to-\mathbf{E}$,
together with the corresponding interchange of electric and magnetic
charges and currents~\cite{Carroll2019}. The tensorial form of this
symmetry makes use of the Hodge dual of the field strength, defined as
${}^\star F_{\mu\nu}\equiv \frac{1}{2}\,\epsilon_{\mu\nu\alpha\beta}\,
F^{\alpha\beta}$, where $\epsilon_{\mu\nu\alpha\beta}$ is the totally
antisymmetric Levi-Civita tensor. In four dimensions, the map
$F\to{}^\star F$ is an involution acting on fields and sources alike,
\begin{align}
F &\longrightarrow {}^\star F, &
j_{\mathrm{e}}^\mu &\longrightarrow j_{\mathrm{m}}^\mu, &
j_{\mathrm{m}}^\mu &\longrightarrow -\,j_{\mathrm{e}}^\mu ,
\end{align}
and is called duality. It is this map that relates the two sectors: a weak
interaction ($\alpha_{\mathrm e}\ll 1$) in the physical space is mapped
into a strong interaction ($\alpha_{\mathrm m}\gg 1$) in the dual space,
and the other way around. We have also learned that dualities can be
stronger than analogies, which, by contrast, do not capture any properties
of interactions.

In analogy to the Maxwell equation case, in which a simple field
redefinition reveals a hidden symmetry, 
about thirty years ago Maldacena conjectured the existence of a duality
relation between weakly curved gravity in Anti-de Sitter space (AdS) and a
conformal field theory living on a space of one dimension fewer, namely the
boundary of the Anti-de Sitter
space~\cite{Maldacena1998,Witten1998,Gubser1998}. This idea, labeled as
AdS/CFT correspondence, has had a tremendous follow-up, in particular in the
framework of the high energy nuclear physics program, which includes the
understanding of the deconfinement onset as well as the exploration of the
nuclear matter phase diagram. While heavy-ion collisions have been able to
produce quark-gluon plasmas in the laboratory, conventional perturbative QCD
calculations are unable to describe their dynamics, due to the strong
correlations of their constituents. The AdS/CFT correspondence, on the other
hand, can evade such limitations with calculations in the space where the
coupling is not large. One of the major results in this field has been the
determination of the ratio between shear viscosity $\eta$ and entropy density
$s$
\begin{equation}
\frac{\eta}{s}\approx \frac{1}{4\pi k_\mathrm{B}}
\label{eq:etaovers}
\end{equation}
for a quark-gluon plasma, which represents the lower bound for a larger class
of field theories~\cite{Kovtun2005}. Experiments have partially corroborated
the above result~\cite{Luzum2008}. Despite the potential of the paradigm and
the applications in condensed matter physics under the label AdS/CMT, there
exist important weak points. First, CFT is not QCD~\cite{McLerran2007}, nor
does it have anything to do with condensed matter theory~\cite{Anderson2013}.
This means that, apart from results that turn out to be universal as
\eqref{eq:etaovers}, further predictions are in general not valid. Second, the
program of gauge-gravity duality aims to extend the case of AdS/CFT to more
physically meaningful field theories, but the correspondence itself, in its
most illuminating example, is only conjectured rather than proven. 
Third, the topic has been developed to a large extent by nuclear
theorists, who invested four decades of expertise in nuclear matter into
the new correspondence. Their contribution should not be
underestimated: they turned a formal conjecture into a working tool,
capable of matching experimental data in a regime where perturbative
methods fail. What has been slowed down, rather, is the original ambition
in the opposite direction: exploiting a quantum field theory to gain
insight into the quantum behavior of gravity itself. In that direction
very little has been done, and we are far from getting anything useful
for quantum gravity.

\section{The Spacetime Condensate}

Up to now we have provided enough evidence to convince the reader of the
necessity of a global view, where condensed matter physics plays a crucial
role in understanding quantum gravity, both theoretically and
experimentally. Nevertheless, the issue is even more surprising. We propose
here an even more radical view: spacetime is actually a condensate! From
this perspective something very powerful starts, because there is no longer
any need for an analogy or a duality.

We recall that by condensate we mean a material phase where particles
occupy the same lowest-energy quantum state in such a way that the quantum
character shows up at larger scales. Apart from superconductors and
superfluids, Bose-Einstein condensates, a collection of highly packed bosons
at low temperature, display all the expected properties of a condensate:
atomic wavefunctions overlap and amplify quantum effects, while weakly
repulsive interactions allow for a macroscopic growth of the material. For
their unique properties such condensates are called the fifth state of
matter. Metaphorically, we can visualize the properties of a condensate by
comparing a group of people randomly walking in a corridor with a squad of
soldiers marching at the same beat. For the group of people it is still
possible to define density, speed, directions, but only on average. For the
squad, the collective motion is just the amplified version of each
soldier's footsteps. For the spacetime one encounters the same thing: many
individual degrees of freedom become one collective quantum state. The
problem is that for the spacetime we do not really know the exact nature of
such degrees of freedom; we may call them atoms, in a loose sense, to
exploit the analogy with the condensate. We just know that there exists a
phase in which such atoms are in a disordered state. A phase transition,
however, can occur as such atoms lump together in a sort of coarse-graining.
Spacetime atoms have the property of carrying bits of information, i.e.,
letters of an alphabet. A single letter has no specific meaning unless it
combines into a word or a sentence. Similarly, atoms have to reorganize
themselves on a larger scale to deliver information. This means that
before the phase transition one cannot speak of spacetime geometry, but of
an entity which is its precursor: the pregeometry. As in the case of a text,
letters follow specific rules in order to combine into meaningful words.
Similarly, the pregeometry has some basic notions of space, but cannot have
any metric properties, like distance or direction. On the other hand, when
atoms clump together and letters form words, some structure emerges,
corresponding to the ``birth'' of the manifold. We call this process
emergence of the spacetime: it is very similar to putting marmalade in a
fridge. The pregeometry is like marmalade at room temperature, namely a
gluey liquid mess which is more or less the same at every point. When it
cools down, each chunk of marmalade freezes in a certain position and can
no longer move. As a result, a certain rigid structure develops. 

The fridge does not create
the ingredients of the marmalade, it just lets them condense into a metric
space. One can invent further analogies, for instance in terms of water,
volcanic lava or ferromagnets. We chose the marmalade for the simple reason
that it is not transparent. This conveys the idea that the atoms of the
spacetime, basically the letters of the alphabet, correspond to the minimal
area below which we can resolve the position of an object. Inside such a
minimal area, we cannot see through. This is identical to single letters,
which cannot be broken into smaller pieces and in general do not carry a
full meaning when not tied to other letters. Below a single letter there is
no meaningful sign in writing.

The quantum spacetime behaves as a structure-less, pregeometric, Planckian
entity that cools down, in the large distance limit, to the ordinary
spacetime geometry. To understand the spacetime condensate it is, however,
imperative to identify three elements: (a) the microscopic degrees of
freedom; (b) the role of the temperature; (c) the properties the spacetime
acquires and loses by switching phases. Concerning (b) and (c) we can
assume, on the grounds of dimensional analysis, that some energy should
play the role of the temperature. In practice, the cooling phase would
represent a low energy, long distance, weak curvature limit. The spacetime
property (c), which is basically the order parameter of the transition,
can be nothing else than the spacetime metric. To address (a), rather than
giving a precise definition, it is useful to consider an example of how
such atoms actually work.

Suppose you have two masses $m$ and $M$ at a certain distance $r$. We
assume some uncertainty in the position of the masses. For instance, we can
center the origin of the axis in $M$ and say that the position of $m$ is
known up to $\lambda=\hbar/mc$, the reduced Compton wavelength. Suppose now
we sit on $m$ and we want to have access to the information related to
$M$. The reasoning appears vague. If we had a star with mass $M$ and radius
$r$, we could interpret the word ``information'' by calculating the
entropy of the star. Here, however, we do not have such a thing, we have
empty space. Still, we can think that some information related to $M$ is
stored in the space around it. Even if we do not exactly know what such
information is, we can at least say that the information increases as we
get closer to $M$, and the other way around as we get further.

In terms of entropy, namely ignorance, we temporarily assume it changes as
follows
\begin{equation}
\Delta S= \eta k_\mathrm{B}\left(\frac{\Delta r}{\lambda}\right)
\label{eq:entropy}
\end{equation}
namely linear dependence with respect to the position. Here $\eta$ is a
dimensionless number and $k_\mathrm{B}$ is the Boltzmann constant, for
dimensional consistency. To compute the entropy we would like to have a
statistical description in terms of microscopic states. If we had a star,
we would be able to address the issue fully. In this case, we have really
no clue. We just guess that the number of degrees of freedom is
proportional to the area around $M$, and we elevate our assumption to the
rank of principle, namely the ``Holographic Principle''. In practice, we
say that the number of atoms of our system is
\begin{equation}
N=\frac{A}{\ell^2}
\label{eq:degfreed}
\end{equation}
where $A=4\pi r^2$ and $\ell$ is a length scale we do not know much about.
At this point, if we want to keep playing this game of the thermodynamics
of space, we have to introduce some temperature. We keep thinking of the
space around $M$ as a gas, and we introduce its internal energy
\begin{equation}
U=\frac{f}{2} N  k_\mathrm{B} T.
\label{eq:internal}
\end{equation}
Here $f$ is the number of degrees of freedom, and since we move in one
dimension we can think of it as equal to unity. Nevertheless, we keep it
generic because we do not really know much about this construction. Since
we do not have any actual gas but just a mass $M$ without kinetic energy,
we could say that the internal energy is just the rest mass, namely
$U=Mc^2$. At this point, given that there is some entropy change, there
could be a force emerging from the relation
\begin{equation}
F\Delta r =T \Delta S.
\label{eq:force}
\end{equation}
From \eqref{eq:entropy} and \eqref{eq:force}, and again from
\eqref{eq:internal} and \eqref{eq:degfreed}, one gets
\begin{equation}
F=\eta k_\mathrm{B} T \frac{mc}{\hbar}
=\left(\frac{2\eta}{f}\right) \frac{c^3}{\hbar}\frac{Mm}{N}
=  \left(\frac{\eta}{2\pi f}\right)\left(\frac{\ell^2 c^3}{\hbar}\right)
\frac{Mm}{r^2}.
\label{eq:entropicforce}
\end{equation}
We obtain a force that goes with the inverse of the distance squared. This
results from our choice of the dependence of $N$ on the surface rather than
on the volume of the sphere centered in $M$. The force can only be
gravitational, because there are no other charges involved. Furthermore,
the force has a microscopic origin: it emerges as a long range effect of
those $N$ atoms. We also see that the coupling constant must equal the
Newton's constant, to match what we observe macroscopically. This means
that the unknown length $\ell$ is actually what we call the Planck length,
namely
\begin{equation}
\ell=l_\mathrm{P}\equiv\sqrt{\frac{\hbar G}{c^3}}
\simeq 1.616255(18)\times 10^{-35}\,\mathrm{m}.
\end{equation}
The final step is to fix the coefficient $\eta$, which must read
$\eta=2\pi f$. Accordingly, we can write \eqref{eq:entropy} as
\begin{equation}
\Delta S= \Delta S_0\equiv \left( f k_\mathrm{B}\right)
\left.\frac{\Delta A}{4\ell^2}\right|_{r=\ell}\left( \frac{\ell}{\lambda}\right)
\label{eq:arealaw}
\end{equation}
where $\Delta A$ denotes the variation of the area $A$ and $\Delta S_0$
denotes the minimal amount of entropy. The above equation resembles the
Bekenstein-Hawking formula for the entropy of black holes. In our case,
however, we do not have any general relativity description, nor do we have
a black hole. The conclusion is therefore universal: gravity emerges from
some microscopic theory; it is the low energy condensate of a collection
of Planckian atoms. Interestingly, the above description offers the
possibility of predicting measurable deviations from Newton's law,
emerging from a microscopic theory alternative to the one considered here.
For instance, one can modify \eqref{eq:degfreed} as follows:
\begin{equation}
N=N_0\longrightarrow N_0\left[1+\frac{n(A)}{ A}\right]
\end{equation}
where $N_0=A/\ell^2$ and $n(A)$ is a generic function of $A$ accounting
for terms beyond the linear one. This has an impact on the definition of
the coefficient $\eta$ and on the expression of the force itself
\eqref{eq:entropicforce}. Alternatively, one can keep $N=N_0$ as in
\eqref{eq:degfreed} and consider corrections to the Bekenstein-Hawking
formula \eqref{eq:arealaw},
\begin{equation}
\Delta S= \Delta S_0 \longrightarrow \Delta S_0
\left[1+\frac{\partial s(A)}{\partial A}\right],
\end{equation}
where $s(A)\equiv S(A)- A$ models the sought effects. In this way one can
implement small scale, large scale, or both regimes of
corrections~\cite{Nicolini2010,Verlinde2011}.

These considerations lead naturally to the next question: if gravity is an
entropic, emergent force, then the atoms of the condensate must set a
fundamental resolution. How small can resolution get?

From the example above we have seen the existence of a fundamental area in
which space is pixelized like tiles of a mosaic or letters of an alphabet.
However, there is a more profound question to address: to what extent can
we consider such a length $\ell$ minimal? Well, in quantum field theory we
already have the concept of minimal length. The (reduced) Compton
wavelength $\lambda$ sets the maximal resolution to observe a particle of
mass $m$ in a given volume. Any attempt to compress the volume would cost
energy $E$. When such energy is equivalent to the rest mass of the
particle, $E\sim 1/\lambda \sim m$, any further increase of energy would
correspond to the production of further particles in a given volume,
making any attempt to observe the initial particle meaningless. On the
other hand, any attempt to compress the volume containing a mass is
limited by the fact that, at a given moment, gravitational forces become
relevant as distances decrease. Eventually, the initial mass can collapse
into a black hole and cannot be made smaller than its gravitational radius
$r_\mathrm{g}= 2mG/c^2$. As a result, we have two physically meaningful
minima for any attempt to compress matter, as customarily done in particle
detectors. We notice, however, that such expressions are respectively
inversely proportional and proportional to the mass parameter. This means
it is possible to determine the overall minimum, which occurs at the
confluence of the two curves. Not surprisingly
\begin{equation}
\frac{\hbar}{mc}\sim \frac{Gm}{c^2}
\Longrightarrow m^2\sim \frac{\hbar c}{G}=m_\mathrm{P}^2,\quad
m_\mathrm{P} \simeq 2.176434(24)\times 10^{-8}\,\mathrm{kg}.
\end{equation}
At this energy $E\sim E_\mathrm{P}$, one realizes a particle-black hole
configuration, being $\lambda\sim r_\mathrm{g}\sim l_\mathrm{P}=1/m_\mathrm{P}$
~\cite{Adler2010,Nicolini2010,Padmanabhan1997,Amati1989,Maggiore1993}.

Even if length scales smaller than $l_\mathrm{P}$ may exist geometrically,
there is no way in the world to probe distances smaller than that. In
practice we will never measure any distance smaller than this, not even in
principle, namely with the best possible devices we can think of. Since
what is physically meaningful is what can be measured, in this sense
$l_\mathrm{P}$ is the minimal length in the universe.
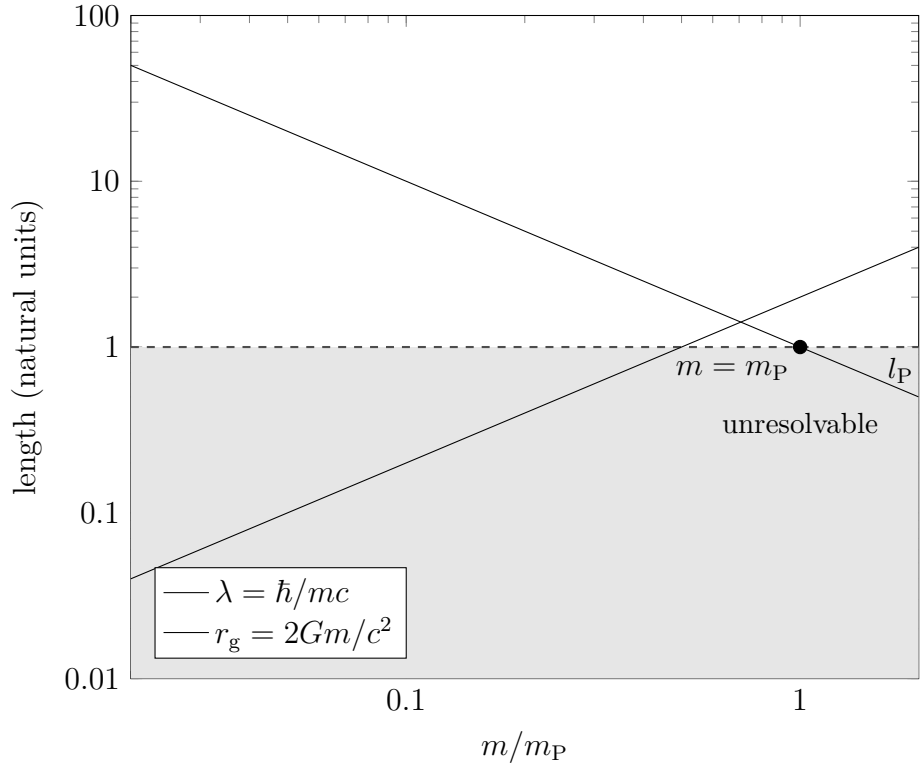
\begin{figure}[t]
\centering
\begin{tikzpicture}
\begin{axis}[
  width=0.75\textwidth,
  xlabel={$m/m_\mathrm{P}$},
  ylabel={length (natural units)},
  xmin=0.02,xmax=2, ymin=0.01,ymax=100,
  log ticks with fixed point,
  xmode=log,ymode=log,
  legend pos=south west,legend cell align=left]
\addplot[fill=gray!20,draw=none,domain=0.02:2,samples=100]
  {1} -- (2,0.01) -- (0.02,0.01) -- cycle;
\draw[dashed,thick,black!70] (axis cs:0.02,1) -- (axis cs:2,1);
  \node[below] at (axis cs:1.8,1) {\small $l_\mathrm{P}$};
\node at (axis cs:1.00,0.35) {\small unresolvable};
\addplot[domain=0.02:2,samples=200] {1/x};
\addlegendentry{$\lambda=\hbar/mc$}
\addplot[domain=0.02:2] {2*x};
\addlegendentry{$r_\mathrm{g}=2Gm/c^2$}
\addplot[only marks,mark=*,mark size=2.5pt] coordinates {(1,1)};
\node[anchor=north east] at (axis cs:1,0.95) {$m=m_\mathrm{P}$};
\end{axis}
\end{tikzpicture}
\caption{The particle--black hole confluence. The (reduced) Compton
wavelength $\lambda\propto m^{-1}$ and the gravitational radius
$r_\mathrm{g}\propto m$ cross at the Planck mass. The shaded region
below $l_\mathrm{P}$ is fundamentally unresolvable: distances smaller
than the Planck length cannot be measured in principle.}
\label{fig:pbh}
\end{figure}
One of
the most interesting consequences of having a minimal length is indeed
related to the dimensions of spacetime.
Too often we take for granted that we live in a three dimensional space.
We are also used to considering time as a fourth coordinate. However, the
number of dimensions is not an absolute concept. While we believe space
has three dimensions on the grounds of the independent directions along
which we can move, we tend to simplify the scenario when the motion is on
a plane, e.g.\ a chessboard, or on a straight line, e.g.\ a train on the
railway. We still believe space is three dimensional because, both in the
case of the chessboard and the train, we see directions in the full space.
But what would happen if we had no possibility of seeing the full space?
The study of the problem of the random walk can offer an interesting
answer to this question. A random walk, or drunkard's walk, consists in
predicting the direction of a drunk man attempting to walk along a
straight line. If the man is really drunk, he will walk randomly, and
rather than a straight line his path will be a curve on the plane.
Mathematically we can model it by considering a diffusive process, for
instance how the concentration of particles changes over time and
position. Another important process of this kind is heat diffusion. As a
start we consider a $D$-dimensional Euclidean space. We can later set $D=4$
but it is not necessary for now. We write the diffusion equation as
\begin{equation}
\frac{\partial}{\partial s}K(x,y;s)=\kappa\Delta K(x,y;s)
\end{equation}
where $s$ is the diffusion time and $\kappa$ the thermal diffusivity.
Here $K(x,y;s)$ governs the probability density of diffusion from the
point $x$ to the point $y$. In practice, for our random drunk walker,
$K(x,y;s)$ represents the probability density of finding the walker at
position $y$ at time $s$, given that the walker started at position $x$.
In the thermal case, $K$ governs the propagation of the temperature over
a certain material, which we assume for simplicity to be an unbounded,
homogeneous medium. For such a medium the diffusivity is a constant; for
simplicity we can set $\kappa=1$, which implies that the diffusion time
$s$ has the dimension of a length squared.

Following an analogy with thermal diffusion, we term $K$ the ``heat
kernel'' for every case where diffusion takes place. The initial condition
is
\begin{equation}
K(x,y;0)=\frac{1}{\sqrt{\det g_{ab}}}\,\delta^{(D)}(x-y)
\end{equation}
which means that the probability is peaked on $x$. In case of free
diffusion, namely no external drift, obstacles, boundaries, or sources, the
heat kernel is Gaussian
\begin{equation}
K(x,y;s)=\frac{e^{-\frac{(x-y)^2}{4s}}}{(4\pi s)^{D/2}}
\label{eq:heatk}
\end{equation}
From $K$ we can calculate
\begin{equation}
P_g(s)=\frac{\int d^{D}x\,\sqrt{\det g_{ab}}\,K(x,x;s)}
{\int d^{D}x\,\sqrt{\det g_{ab}}}
\end{equation}
which is called the average return probability, namely, the average
probability density for a random walker to return to its starting location
after diffusion time $s$. For the simple case of the walker, namely free
diffusion in flat space, $\det g_{ab}$ is constant and $P_g(s)=K(x,x;s)$.
This is useful because one can take the logarithmic derivative of $P_g(s)$
to get
\begin{equation}
-2\,\frac{\partial \ln P_g(s)}{\partial \ln s}
=
\text{number of spatial dimensions},
\label{eq:logder}
\end{equation}
namely $D$ in our case. The question is now to understand what happens in
a more general case. Suppose we have a space where there is a maximal
resolution, namely distances below a certain scale $\ell$ cannot be seen.
What would diffusion look like in this case? We can think about this
problem in terms of a porous medium, namely a solid with structures like
foams. For sure the very notion of point-like object can no longer be
valid on such a medium. Therefore, even if we do not know the exact
details of the heat equation over such a medium, we are sure that the
initial condition in terms of a Dirac delta must be modified. The cheapest
of the class of possible modifications is to replace the delta with a
Gaussian having a width $\ell$.
\begin{equation}
K(x,y;0)=\left(\frac{1}{4\pi \ell^{2}}\right)^{D/2}\,
e^{-\frac{(x-y)^2}{4\ell^{2}}}.
\end{equation}
This modification affects the solution. Eq.~\eqref{eq:heatk} becomes
\begin{equation}
K(x,y;s)\longrightarrow K_\ell(x,y;s)
=\frac{e^{-\frac{(x-y)^2}{ 4(s+\ell^{2}) }}}
{\left[4\pi(s+\ell^{2})\right]^{D/2} }
\end{equation}
and the logarithmic derivative of the average return probability
\eqref{eq:logder} gives another value of the spacetime
dimension~\cite{ModestoNicolini2010}
\begin{equation}
\mathbb{D}(s)=\frac{s}{s+\ell^{2}}\,D.
\end{equation}
This single formula changes the whole picture of spacetime dimension.

First, the dimension is no longer a constant: $\mathbb{D}(s)$ depends on
the scale, i.e.\ on the energy at which we observe the spacetime. The
conventional integer value $D$ is recovered only in the infrared limit
$s\gg\ell^2$, where the formula reduces to $\mathbb{D}\to D$. In this
sense $D$, which we may term the topological dimension, is just the
low energy limit of a scale dependent quantity.

Second, the dimension can take non-integer values. In the high energy
regime, $s\sim\ell^2$, the spacetime is effectively a fractal. The smooth
manifold of general relativity is not the ultimate substrate: it is the
condensate phase, the ordered outcome of a phase transition; the fractal
phase is its disordered precursor.

Third, the threshold of the transition is set by the parameter $\ell$
itself, the Planck length. In the present example $\ell$ is the minimal
length emerging from noncommutative geometry, a remnant of the
matter-gravity decoupling limit which leaves noncommutative behavior for
the end points of open strings on D-branes~\cite{SeibergWitten1999};
however, independent quantum gravity approaches, like causal dynamical
triangulations, have led to the same result~\cite{Ambjorn2005}.

Fourth, the dimension decreases as the energy increases: the shorter the
diffusion probe, the more the porous structure of space obstructs the
walker, and the fewer directions remain visible.

The miracle occurs at $s=\ell^2$: for $D=4$, the formula gives
$\mathbb{D}=2$~\cite{tHooft1993}. This is exactly the spontaneous
dimensional reduction to two dimensions at the Planck scale conjectured by
't Hooft in his original scenario for quantum gravity. The consequence is
striking. The gravitational coupling constant scales with the number of
dimensions as
\begin{equation}
G\sim \ell^{(\mathbb{D}-2)}.
\end{equation}
For $\mathbb{D}=2$, $G$ is dimensionless and gravity becomes a
renormalizable theory, just like QED. In other words, the notorious
non-renormalizability of gravity is a spurious effect of the classical,
condensate limit: quantum gravity is intrinsically two dimensional, and
renormalizable.

The above results are already formidable by themselves, but the story is
not over. Planckian fractalization effects can have a leftover at lower
energy by transmuting the ordinary particle states of matter into some
fractal stuff, termed ``unparticles''~\cite{Georgi2007a,Georgi2007b},
which captures some properties of the previously introduced HEIDI
model~\cite{vanderBij2006,Hill1987}. The formalism requires the existence
of a fixed point of the renormalization group, where perturbation theory
is possible for some, yet undetected, fields. Such fields, known as
Banks-Zaks (BZ) fields~\cite{BanksZaks1982}, interact with the Standard
Model at energy $M_\mathrm{BZ}$ between the 10 TeV and the Planck scale,
but at a lower energy scale $\Lambda_\mathrm{U}>10$ TeV they develop
scale invariance properties. In particular their particle number depends
on a continuous parameter $d_\mathrm{U}$, which is the unparticle scaling
dimension. This becomes evident by observing that the Feynman propagator
of the unparticle field
\begin{equation}
D_{\mathrm{U}}(x,x')
=
\frac{A_{d_\mathrm{U}}}{2\pi\left(\Lambda_\mathrm{U}^2\right)^{d_\mathrm{U}-1}}
\int_{0}^{\infty}
d m^2\,
\left(m^2\right)^{d_\mathrm{U}-2}
D(x,x';m^2),
\end{equation}
where
\[
A_{d_\mathrm{U}}
=
\frac{16\pi^{5/2}}{(2\pi)^{2d_\mathrm{U}}}
\frac{\Gamma\left(d_\mathrm{U}+\frac{1}{2}\right)}
{\Gamma(d_\mathrm{U}-1)\Gamma(2d_\mathrm{U})}.
\]
is a continuous superposition of Feynman propagators of fixed mass $m$.
Unparticles can affect a variety of physical phenomena, including testable
signatures in particle accelerator experiments~\cite{Cheung2007}, the
anomalous magnetic moment of the electron~\cite{Liao2007}, the Newtonian
potential and black holes~\cite{Goldberg2008,Mureika2008,Mureika2009,
Gaete2010,MureikaSpallucci2010}, the Hydrogen atom energy
levels~\cite{Wondrak2016}, and cosmological and astrophysical processes
such as Big Bang nucleosynthesis, stellar energy loss, supernova cooling,
and cosmological energy-density bounds~\cite{Davoudiasl2007,Freitas2007}.

Unparticles bring corrections that depend on the scaling dimension
$d_\mathrm{U}$, leading to a fractalization effect, such as the case of
the Casimir capacitor plates which are no longer perceived as two
dimensional surfaces~\cite{NicoliniSpallucci2011,Frassino2017}. Unparticles
explain charge anomalies in superconductors, such as the observed
violation of Luttinger's theorem for the transport of electrons in
cuprates~\cite{LeBlanc2015,Karch2016}.

Unparticles, however, are a ghost of fractalization surviving in ordinary
matter. The most direct evidence for the spacetime condensate comes
instead from the way objects propagate on it, starting from the path of a
single particle~\cite{AbbottWise1981,NicoliniNiedner2011}.

\begin{figure}[t]
\centering
\begin{tikzpicture}[line cap=round]
\draw[very thick,
      l-system={Koch, axiom=F, step=0.55cm, angle=60, order=2}]
  lindenmayer system;
\draw[thick,red!70!black] (0.83,-0.24) circle (0.75);
\node[below] at (2.5,-1.5) {\small resolution $\Delta x$};

\draw[dashed] (1.4,0.3) -- (3.9,2.2);

\begin{scope}[shift={(4.8,2.9)}]
  \clip (0,0) circle (1.35);
  \draw[very thick,
        l-system={Koch, axiom=F, step=0.28cm, angle=60, order=2}]
    lindenmayer system;
  \draw[thick,red!70!black] (0.42,-0.12) circle (0.55);
\end{scope}
\draw[thick] (4.8,2.9) circle (1.35);
\draw[thick] (3.85,2.35) -- (3.5,2.1);             
\node[below] at (4.8,1.4) {\small $\Delta x/3$};

\draw[dashed] (6.0,2.9) -- (7.3,2.9);

\begin{scope}[shift={(8.4,2.9)}]
  \clip (0,0) circle (1.35);
  \draw[very thick,
        l-system={Koch, axiom=F, step=0.093cm, angle=60, order=3}]
    lindenmayer system;
\end{scope}
\draw[thick] (8.4,2.9) circle (1.35);
\draw[thick] (7.45,2.35) -- (7.1,2.1);             
\node[below] at (8.4,1.4) {\small $\Delta x/9$};
\end{tikzpicture}
\caption{Self-similar structure of the Koch curve. The piece
highlighted in the red circle looks smooth at resolution
$\Delta x$; magnified by three ($\Delta x/3$), it reveals the
zigzag of the next iteration. Magnifying again ($\Delta x/9$)
resolves the next level --- and so on ad infinitum. The measured
length grows by a factor $4/3$ at every step without converging:
the curve has no definite length, only a Hausdorff dimension
$D_\mathrm{H}=\ln4/\ln3$.}
\label{fig:koch}
\end{figure}
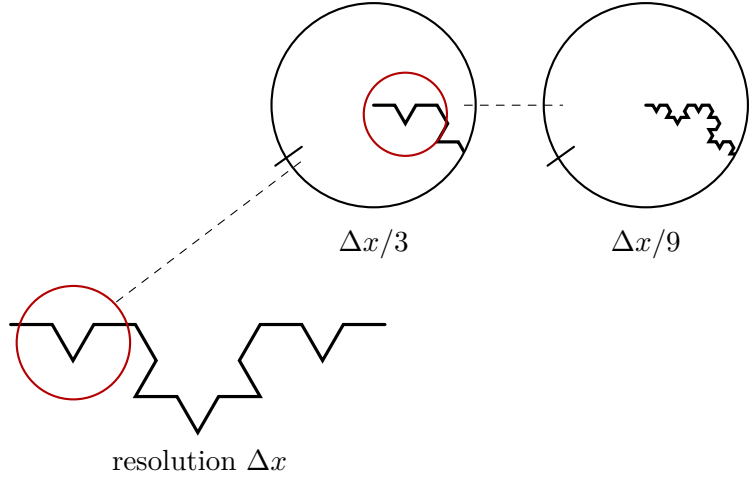

A classical path, namely a world line, is a one-dimensional object. More
complicated is the case of a quantum particle, because it is not properly
possible to speak of a quantum trajectory: due to uncertainty relations,
either one determines the position but neglects the momentum, or the
other way around. Therefore the best one can do is to consider average
values to keep speaking in terms of a path. Of course, we cannot expect
the dimension of what we improperly call ``quantum path'' to be one, as in
the case of the classical world line. Something has to change. It is
reasonable to expect that we cannot even speak of the dimension of a path
in simple terms. We probably need a new notion to address this question.
The answer is the Hausdorff dimension, $D_\mathrm{H}$. For a fractal curve
there is no unambiguous definition of the length, since it depends on the
resolution at which we observe the fractal. The higher the resolution, the
larger is the number of fractal wiggles that show up, therefore the length
increases. Hausdorff, however, revisited the concept of length of a
fractal by demanding it to be independent of the resolution. He proposed
\begin{equation}
L_\mathrm{H}=l_0\Delta x^{D_\mathrm{H}-1}
\label{eq:hlength}
\end{equation}
as a resolution independent length. Here $D_\mathrm{H}$ is a real number
and $\Delta x$ represents the minimal length that can be resolved on the
curve. For a non-fractal line, it is evident that the size of $\Delta x$
does not affect its length, because one can simply count how many
$\Delta x$ are necessary to cover the full curve. The smaller $\Delta x$,
the higher is the number of $\Delta x$ needed to cover the full length,
but the result is unambiguous, namely $l_0$. For this reason,
\eqref{eq:hlength} works and consistently gives $L_\mathrm{H}=l_0$ for
$D_\mathrm{H}=1$. For a fractal, however, it is less trivial to guarantee
the independence of $L_\mathrm{H}$ on $\Delta x$. For the Koch curve (cf.\ Fig.~\ref{fig:koch}), as
$\Delta x$ decreases by two thirds, the fractal length increases by one
third. In practice, if we require the Hausdorff length to be invariant
\begin{equation}
L_\mathrm{H}=L^\prime_\mathrm{H}
\Rightarrow l_0\Delta x^{D_\mathrm{H}-1}
=(l_0^\prime)(\Delta x^\prime)^{D_\mathrm{H}-1}
\end{equation}
the Hausdorff dimension must be non-integer, namely
$D_\mathrm{H}=\ln4/\ln3$.

One can repeat the same reasoning by considering the average distance
covered by a quantum particle in a time $T=N\Delta t$,
\begin{equation}
\langle l \rangle = N \Delta l
\end{equation}
where
\begin{equation}
\Delta l = \langle \psi| \hat{U}^\dagger (\Delta t)\,| \hat{\textbf{x}}|
\hat{U}(\Delta t) | \psi \rangle
\label{eq:ppuncert}
\end{equation}
with $\hat{U}(t)$ the evolution operator, $\hat{\textbf{x}}$ the position
operator and $\psi$ the wave function. Calculations show that
\begin{equation}
\langle \Delta l \rangle \propto
\frac{\hbar \Delta t}{m \Delta x}.
\label{eq:deltal}
\end{equation}
By applying the same reasoning as for the Koch curve: for large $\Delta x$
we have $\langle \Delta l \rangle\approx 0$ and $D_\mathrm{H}=1$.
Conversely, for fine resolution, namely small $\Delta x$, we have
$\langle \Delta l \rangle\sim (\Delta x)^{-1}$, which implies
$D_\mathrm{H}=2$, namely a fractal curve of dimension two. In practice
this means that, in the excited quantum state, the particle trajectory
increases the dimension of its world line: it is somehow thicker.

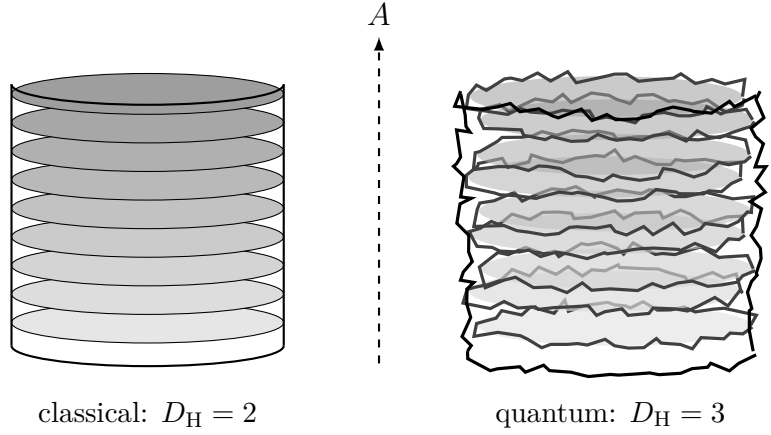
\begin{figure}[t]
\centering
\pgfmathsetseed{2026}   
\begin{tikzpicture}[scale=0.9]
\begin{scope}[shift={(-3.4,0)}]
  \foreach \i/\shade in {0/20, 1/27, 2/34, 3/41, 4/48, 5/55, 6/62, 7/69, 8/76}{
    \fill[gray!\shade, draw=black] (0, 0.42*\i) ellipse (2.0 and 0.30);
  }
  \draw[thick] (-2.0,-0.34) -- (-2.0,3.50);
  \draw[thick] ( 2.0,-0.34) -- ( 2.0,3.50);
  \draw[thick] (-2.0, 3.50) arc (180:360:2.0 and 0.30);
  \draw[thick] (-2.0,-0.34) arc (180:360:2.0 and 0.30);
  \node at (0,-1.35) {\small classical: $D_\mathrm{H}=2$};
\end{scope}
\begin{scope}[shift={(3.4,0)}]
  \foreach \i/\shade/\op/\dx/\dy in
    {0/20/0.70/ 0.06/-0.04, 1/27/0.68/-0.08/ 0.04,
     2/34/0.66/ 0.10/-0.05, 3/41/0.64/-0.05/ 0.05,
     4/48/0.62/ 0.07/-0.04, 5/55/0.60/-0.09/ 0.04,
     6/62/0.58/ 0.04/-0.05, 7/69/0.56/ 0.09/ 0.04,
     8/76/0.54/-0.06/-0.04}{
    \fill[gray!\shade, draw=none, fill opacity=\op]
      (\dx, 0.42*\i+\dy) ellipse (2.0 and 0.30);
    \draw[very thick, draw=black!75, fill=none,
          decoration={random steps, segment length=5pt, amplitude=2.5pt},
          decorate]
      (\dx, 0.42*\i+\dy) ellipse (2.0 and 0.30);
  }
  \draw[very thick, decoration={random steps, segment length=4pt, amplitude=3.5pt},
        decorate] (-2.10,-0.42) -- (-2.20,3.36);
  \draw[very thick, decoration={random steps, segment length=4pt, amplitude=3.5pt},
        decorate] ( 2.10,-0.42) -- ( 2.20,3.40);
  \draw[very thick, decoration={random steps, segment length=5pt, amplitude=2.5pt},
        decorate] (-2.10,-0.42) arc (180:360:2.10 and 0.34);
  \draw[very thick, decoration={random steps, segment length=5pt, amplitude=2.5pt},
        decorate] (-2.20,3.40) arc (180:360:2.15 and 0.34);
  \node at (0,-1.35) {\small quantum: $D_\mathrm{H}=3$};
\end{scope}
\draw[-latex, dashed, thick] (0,-0.6) -- (0,4.2) node[above] {$A$};
\end{tikzpicture}
\caption{World-sheet of a quantized string as a foliation of closed loops
along the area parameter $A$. Left: classical stack of smooth loops; the
cylinder mantle is a resolvable two-dimensional sheet ($D_\mathrm{H}=2$).
Right: quantum fuzziness thickens the loops and the mantle into a
three-dimensional fractal object ($D_\mathrm{H}=3$).}
\label{fig:stringstack}
\end{figure}

At this point one may be tempted to ask what happens if the classical
object is no longer a point-like object but an extended one, for instance
a $p$-brane. Even if we do not have a full answer to this question, it is
instructive to explore what happens in the case $p=1$, namely the string.
We expect the world-volume swept by the string to be a two-dimensional
world-sheet. Intriguing, on the other hand, is the case of the quantum
string and the related concept of Hausdorff dimension for it. To this
purpose it is instructive to recall that the conventional string
quantization from the Nambu-Goto action is not the only possibility. For
our purposes it is far more instructive to follow Eguchi's quantization
scheme in terms of areal functionals~\cite{Eguchi1980}, which allows for a
transparent quantization of the string, as an object flying in target
space, along a one-to-one correspondence to the conventional procedure for
the particle case. In practice, rather than performing an oscillatory
expansion, the closed string is quantized as a whole, by identifying a
time-like dynamical variable and a space-like dynamical variable in terms
of the world-sheet area $A$ and the internal area enclosed by a specific
string loop configuration $C$. The spatial part is quantized and $A$ acts
as an evolution parameter. The quantization proceeds by adopting the
Schr\"odinger picture, where a wave function $\Psi[C, A]$ solves a
Schr\"odinger-like equation
\begin{equation}
-\frac{1}{4m^2}
\left(\oint_C dl(s)\right)^{-1}
\oint_C dl(s)\,
\frac{\delta^2\Psi[C;A]}
{\delta\sigma^{\mu\nu}(s)\,\delta\sigma_{\mu\nu}(s)}
=
i\,\frac{\partial\Psi[C;A]}{\partial A},
\label{eq:schrstring}
\end{equation}
descending from the correspondence
\begin{equation}
P_{\mu\nu}(s)\longrightarrow
\frac{i}{\sqrt{x'^2(s)}}
\frac{\delta}{\delta \sigma^{\mu\nu}(s)}
\end{equation}
\begin{equation}
H\longrightarrow -i\frac{\partial}{\partial A},
\end{equation}
where $\sigma^{\mu\nu}$
\begin{equation}
\sigma^{\mu\nu}(C)
\equiv
\oint_C x^\mu\,dx^\nu
\end{equation}
is the area element of the loop $C$, $x^\mu=x^\mu(s)$ is the world-sheet
coordinate, $dl(s) \equiv ds\,\sqrt{x'^{\,2}(s)}$ is the invariant element
of string length, $s$ is the affine parameter labeling the loop coordinate
and $m^2 = 1/(2\pi\alpha')$ is defined in terms of the string tension.
The string wave functional $\Psi[C, A]$ gives the probability amplitude of
finding the string loop $C$ with area elements $\sigma^{\mu\nu}(C)$ as the
only boundary of the two-surface $A$. We also note that the tensor
$\sigma^{\mu\nu}(C)$ corresponds to the string position, or better to the
$p=1$-brane analog of the particle position at a point $P$, namely
\begin{equation}
x^{\mu}(P)\longrightarrow \sigma^{\mu\nu}(C).
\end{equation}
Similarly to the quantum particle case, for an $A$-independent potential
one writes $\Psi$ as the product of functions depending on $C$ and $A$,
\begin{equation}
\Psi(C; A) = \Psi(C)\exp(-iEA).
\end{equation}
It is also possible to faithfully repeat the procedure of the calculation
of the Hausdorff dimension as in the particle case, by considering a
Gaussian packet built with a superposition of monochromatic string wave
trains. There are of course some caveats. First, the world sheet has to be
``foliated'' into a stack of closed lines labeled by $A$. In view of the
discretization of the evolution in $N$ steps, we approximate the string
stack as a discrete set of loops
$x^\mu_n(s)=x^\mu(s; n\Delta A)$.
Second, in the quantum regime, the world sheet is affected by graininess
that increases with increasing resolution, similarly to what is seen in
the case of the Koch curve and the quantum particle trajectory. As a
result, the string stack (cf.\ Fig.~\ref{fig:stringstack}) acquires a certain thickness when quantum
fluctuations are visible.

Therefore the Gaussian string packet displays a width that depends on
$\Delta\sigma$, a parameter which represents the position uncertainty in
loop space, corresponding to an uncertainty in the physical shape of the
loop. This is the analog of the resolution in the particle case,
\begin{equation}
\Delta x\longrightarrow \Delta \sigma.
\end{equation}
At this point we can calculate the average of the ``surface position'' of
the quantized string
\begin{equation}
\langle S\rangle = N\langle \Delta S\rangle
\end{equation}
where
\begin{equation}
\langle \Delta S\rangle \equiv
\left[
\int [d\sigma]\,
\sigma^{\mu\nu}(C)\sigma_{\mu\nu}(C)\,
\left|\Psi(C;\Delta A)\right|^2
\right]^{1/2}
\label{eq:surfuncert}
\end{equation}
is the stringy analog of \eqref{eq:ppuncert}. The quantity $\langle
S\rangle$ diverges in the limit $\Delta\sigma\to 0$, therefore one has to
introduce the following Hausdorff measure
\begin{equation}
S_\mathrm{H} = N\langle \Delta S\rangle(\Delta\sigma)^{D_\mathrm{H}-2},
\end{equation}
where $D_\mathrm{H}$ is determined by the requirement that $S_\mathrm{H}$
be independent of the resolution $\Delta\sigma$. Explicit calculations of
\eqref{eq:surfuncert} give
\begin{equation}
\langle \Delta S \rangle \propto
\frac{\Delta A}{4m^2 \Delta \sigma}.
\label{eq:deltaS}
\end{equation}
The above result implies that for $\Delta\sigma \gg \Delta A$ the measure
$S_\mathrm{H}\sim (\Delta\sigma)^{D_\mathrm{H}-2}$, which requires
$D_\mathrm{H}=2$. This is the classical case, where the roughness of the
world sheet is not visible. Conversely, for $\Delta\sigma \ll \Delta A$,
the measure
$S_\mathrm{H}\sim \frac{\Delta A}{\Delta \sigma}(\Delta\sigma)^{D_\mathrm{H}-2}$
requires $D_\mathrm{H}=3$. This confirms our idea about the thickness of
the string world-sheet, which becomes a three-dimensional fractal object
due to fuzziness, as a consequence of quantum
fluctuations~\cite{Ansoldi1997}.

The consequences of the above calculations go beyond the mere geometric
extension from the $p=0$ (particle) to the $p=1$ (string) case. As a
start we notice that the quantum of action is replaced by a quantum of
length
\begin{equation}
\hbar\longrightarrow \sqrt{\alpha^\prime}
\end{equation}
when comparing \eqref{eq:deltal} with \eqref{eq:deltaS}. This means that
even in the absence of conventional quantum mechanical effects, i.e.\
$\hbar=0$, it is still possible to have a quantum phase for the
world-sheet, namely a quantum geometry phase.

As a second remark, the full procedure carried out in loop space does not
require the knowledge of a pre-assigned spacetime. This feature is
actually the aforementioned ``emergence'' of spacetime, but it also
offers a possible solution to the problem of background independence,
when dealing with gravity quantization within the string theory
framework.

As a third point, spacetime emergence results from the process of
``$p$-brane--'' or more specifically ``string--condensation''. The
opposite process is the ``spacetime boiling'', which transforms a Riemann
geometry into a fractal. What fully captures our central claim -- the
spacetime is a condensate itself -- is that the Ginzburg-Landau (GL)
theory for superconductors can describe string condensation similarly to
the formation of superconducting phases below the critical
temperature~\cite{Ansoldi1999}.

To this purpose, the first step is to introduce the analog of the
temperature. From a field-theoretic perspective, the Wick rotation
$t=-i\tau$,
\begin{equation}
0 \leq \tau \leq \beta,
\qquad
\beta=\frac{1}{k_B T},
\end{equation}
is a customary method to calculate Green's functions of non-pure, i.e.\
thermal, states. Similarly here, we Wick rotate the evolution parameter
\begin{equation}
iA \longrightarrow a \sim \beta
= \frac{1}{k_B T},
\end{equation}
to obtain the analog of the temperature. This formally allows for a
statistical description of the string vacuum fluctuations, in the sense
of finite-area loop quantum mechanics, as summarized in the dictionary
\eqref{tab:identif}.
\begin{eqnarray}
\begin{array}{c|c}
\text{Particle field theory}
&
\text{Loop-space theory}
\\[4pt]
\hline
\lvert x\rangle
&
\lvert C\rangle
\\[4pt]
\psi(x)
&
\Psi[C]
\\[4pt]
\text{particle-number state}
&
\text{string/loop-number state}
\\[4pt]
\langle A_\mu(x)\rangle
&
\langle A_{\mu\nu}(x)\rangle
\end{array}
\label{tab:identif}
\end{eqnarray}
For the fields
\begin{equation}
\Psi\equiv\Psi(C,a),
\qquad
\Psi^*\equiv\Psi^*(C,a),
\end{equation}
we introduce the following GL Lagrangian
\begin{eqnarray}
\label{eq:GLLagr}
\mathcal{L}(\Psi,\Psi^*)
&=&
\underbrace{\Psi^*\frac{\partial\Psi}{\partial a}}
_{\text{area-evolution}}
-
\underbrace{
\frac{1}{4m^2}
\left(\oint_C dl(s)\right)^{-1}
\oint_C dl(s)\,
\left|
\left(
\frac{\delta}{\delta\sigma^{\mu\nu}(s)}
-igA_{\mu\nu}
\right)\Psi
\right|^2
}_{\text{loop-space kinetic}}
\\[0.5em]
&&
-
\underbrace{V(|\Psi|^2)}_{\text{GL potential}}
-
\underbrace{
\frac{1}{2\cdot3!}
H_{\lambda\mu\nu}H^{\lambda\mu\nu}
}_{\text{two-form field strength}} .\nonumber
\end{eqnarray}
The Lagrangian \eqref{eq:GLLagr} looks daunting, but at the end of the
day each term can be explained in plain words.

The term $\Psi^*\frac{\partial\Psi}{\partial a}$ is unusual in a GL
theory. It measures how the string field $\Psi$ changes as $a$ changes.
It may be interpreted as a first-order areal kinetic term.

The second term is more conventional. Apart from the rank of the tensors,
\begin{equation}
\nabla
\quad\longrightarrow\quad
\frac{\delta}{\delta\sigma^{\mu\nu}(s)},
\end{equation}
\begin{equation}
q\mathbf{A} \quad\longrightarrow\quad gA_{\mu\nu}(x),
\end{equation}
it coincides with the gauge-covariant kinetic term in ordinary
Ginzburg-Landau theory. Here $A_{\mu\nu}$ is the Kalb-Ramond
antisymmetric gauge potential and the antisymmetric tensor field
strength is
\begin{equation}
H_{\lambda\mu\nu}(x)
=
\partial_{[\lambda}A_{\mu\nu]}(x),
\end{equation}
where the square brackets indicate antisymmetrization.

Finally the key term is the catastrophic potential \eqref{eq:glpotential},
which has the conventional Ginzburg-Landau form. It is a potential one
encounters also in other phase transitions, or in other contexts like
Higgs mechanisms and in mechanical systems like Zeeman's catastrophe
machine~\cite{ZeemanCatastrophe}:
\begin{equation}
V(|\Psi|^2)
=
\underbrace{
\mu_0^2\left(\frac{a_\mathrm{c}}{a}-1\right)|\Psi|^2
}_{\text{quadratic term}}
+
\underbrace{
\frac{b}{4}|\Psi|^4
}_{\text{quartic term}} \qquad b>0.
\label{eq:glpotential}
\end{equation}
To simplify the discussion we consider the free case, by setting
$A_{\mu\nu}=0$. The Lagrangian \eqref{eq:GLLagr} then has basically the
kinetic term and the above potential,
where $a_\mathrm{c}$ is a critical loop area: for $a\leq a_\mathrm{c}$ the
potential energy is minimized by the ordinary vacuum $\Psi[C]=0$, while
for $a>a_\mathrm{c}$ strings condense into a superconducting vacuum
\begin{equation}
|\Psi|^2=-2\mu_0^2\left(\frac{a_\mathrm{c}}{a}-1\right)/b.
\end{equation}
We can draw the following scenario: the spacetime is destroyed in the
disordered pre-geometry phase for
\begin{equation}
 T\geq T_\mathrm{c}\Longleftrightarrow a\leq a_\mathrm{c}.
\end{equation}
As the temperature decreases, long range correlations form and an
ordered phase shows up, resulting in a superconducting geometric
condensate: the Riemann manifold.

One of the reasons why string theory is a powerful framework is that it
has one free parameter only. At this point, this leads us to consider
the possibility that $a\sim \alpha^\prime\sim G \sim l_\mathrm{P}^2$,
with the string length and the Newton constant being the only parameters
available on the market for a quantum gravity calculation. We can dig
further in this direction by assuming an even more radical view: gravity
is not a fundamental interaction but a phenomenon resulting from quantum
matter. From this perspective the loop space string approach is
consistent with Sakharov's idea about induced gravity. In other words,
the Einstein-Hilbert action would result as the effective action
emerging from matter fields at one
loop~\cite{Sakharov1968,BirrellDavies1982},
\begin{equation}
S_\mathrm{EH}\sim W\sim \mathrm{Tr}\log (- G_\mathrm{F})
\qquad
\Longrightarrow
\qquad
G_{\mu\nu}
=
\frac{8\pi G}{c^4}
\left\langle \hat{T}_{\mu\nu}\right\rangle ,
\end{equation}
where the Feynman propagator has to be interpreted as an operator acting
on states $|x\rangle$~\cite{BirrellDavies1982}. The last equation is
nothing but the semiclassical Einstein equation: the quantum stress
tensor of matter, derived from $W$, determines the Einstein tensor. In
Sakharov's words, the elastic Riemann condensate undergoes a phase
transition to a Planckian foam of fractals, and the other way around
(see~\cite{Ord1983,Cannata1988,Nottale1989,Nottale1992} for further
references about fractality and spacetime).
Interestingly, at the classical level, the non-fundamental nature of
gravity is connected to the Equivalence Principle. Feynman, in
particular, has interpreted gravity as a pseudo-force, rather than an
interaction~\footnote{R.\,P. Feynman, R.\,B. Leighton and M.\,Sands,
\emph{The Feynman Lectures on Physics}, Vol.~I, Sec.~12-5.}.

As a further piece of evidence in favor of the interpretation of
spacetime as a condensate, we mention the recent developments within the
paradigm of ``corpuscular gravity'', in which black holes are
re-interpreted as a new phase of matter complementary to the conventional
concept of
particle~\cite{Casadio2014,Casadio2014b,Casadio2015,Casadio2016,
Spallucci2016,Spallucci2017,DvaliGomez2013,DvaliGomez2014}.

As a premise, we recall that, as initially noted by Aurilia and
Spallucci~\cite{AuriliaSpallucci2013} and more recently by
Carr~\cite{Carr2016,Carr2015}, Dvali~\cite{Dvali2011} and their
collaborators, gravity does not admit a Wilsonian ultraviolet
completion: unlike other non-renormalizable effective theories, such
as the Fermi theory, it is not the low energy limit of a deeper theory
awaiting discovery in the ultraviolet. Gravity simply does not have
one. It is self-complete: as shown in Fig.~\ref{fig:pbh}, it provides
its own cutoff at the Planck energy and classicalizes beyond it ---
more energy makes the object bigger, not smaller. The gravitational
radius $r_\mathrm{g}(M,G,c)$ grows with $M$; the Compton wavelength
$\lambda(M,\hbar,c)\sim 1/M$ shrinks with it. Quantum gravity is just
their confluence, the point $m=m_\mathrm{P}(\hbar,c, G)$: there, and only there,
gravity and quantum mechanics meet.

Given this scenario, it is natural to ask if the classical regime of gravity
can emerge from some quantum theory. In this sense corpuscular gravity
is an umbrella term which covers any attempt to model the emergence of
black holes from a quantum pregeometric framework. To put it simply, one
should come up with some quantum system which, in its classical limit,
describes known black hole properties, without invoking any notion of
space or geometry. In this sense, the spacetime can only emerge: it is
never pre-assigned or given a priori.

Within the black hole quantum $N$-portrait, the degree of classicality
is controlled by the parameter $N$, namely the occupation number of
gravitons in a gravitational field~\cite{DvaliGomez2013}. 
The main idea of the whole construction is simple: a black hole is a
Bose--Einstein condensate of gravitons at maximal packing, and it always
sits exactly at the critical point $\alpha N=1$ of a quantum phase
transition. Everything that follows is the path leading to this claim.

\begin{figure}[t]
\centering
\begin{tikzpicture}[scale=1.0,line cap=round]
\begin{scope}
  \draw[very thick] (0,0) circle (1.6);
  \foreach \yy in {-1.2,-0.8,...,1.2}{%
    \foreach \xx in {-1.2,-0.8,...,1.2}{%
      \pgfmathparse{int(\xx*\xx+\yy*\yy < 2.1)}%
      \ifnum\pgfmathresult=1
        \fill[gray!55] (\xx,\yy) circle (0.13);
      \fi
    }%
  }%
  \foreach \yy in {-1.0,-0.6,...,1.0}{%
    \foreach \xx in {-1.0,-0.6,...,1.0}{%
      \pgfmathparse{int((\xx+0.2)*(\xx+0.2)+\yy*\yy < 2.1)}%
      \ifnum\pgfmathresult=1
        \fill[gray!55] (\xx+0.2,\yy) circle (0.13);
      \fi
    }%
  }%
  \fill[gray!55] (-1.85,1.25) circle (0.13);
  \draw[-latex,thick,red!70!black] (-1.6,1.1) -- (-2.6,1.65);
  \node[align=center] at (-2.3,2.15)
    {\small quantum depletion\\[-2pt]\small (Hawking radiation)};
  \draw[-latex,thick] (2.75,-1.15) -- (1.35,-0.55);
  \node[align=center] at (2.55,-1.6) {\small accretion};
  \node at (0,-2.15)
    {\small $N\sim(M/m_\mathrm{P})^{2}$,\quad
     $\lambda\sim\sqrt{N}\,l_\mathrm{P}$,\quad
     $\alpha N=1$};
\end{scope}
\begin{scope}[shift={(3.8,-2.8)}]
  \fill[gray!30] (0,1.6) rectangle (7.5,5.0);
  \fill[gray!10] (0,0) rectangle (7.5,1.6);
  
  \draw[-latex,thick] (0,0) -- (7.8,0) node[right] {$N$};
  \draw[-latex,thick] (0,0) -- (0,5.3) node[above] {$\alpha N$};
  
  \draw[dashed,very thick] (0,1.6) -- (7.5,1.6);
  \node[anchor=west,align=left,yshift=0.1cm] at (0.2,1.6)
    {\small critical point: $\alpha N=1$};
  
  \node[anchor=west,align=left,yshift=-0.3cm] at (0.2,3.8)
    {\small strongly correlated phase\\[-2pt]\small (collapse)};
  \node[anchor=west,align=left,yshift=0.0cm] at (0.2,0.5)
    {\small weakly correlated phase\\[-2pt]\small (dilute graviton gas)};
  
  \fill (5.5,1.6) circle (2.8pt);
  \node[above,align=center,yshift=0.1cm] at (5.5,1.6) 
    {\small BH};
  
  \draw[-latex,thick] (5.2,1.6) -- (3.5,1.6);
  \node[below,align=center,yshift=0.0cm] at (4.3,1.5)
    {\small evaporation (depletion)
    };
  
  \draw[-latex,thick] (5.8,1.6) -- (7.2,1.6);
  \node[above,align=center,yshift=0.0cm] at (6.8,1.7)
    {\small accretion};
\end{scope}
\end{tikzpicture}
\caption{The black hole as a self-sustained graviton condensate.
Left: the black hole is a Bose--Einstein condensate of $N$ weakly
interacting gravitons at maximal packing. The condensate is leaky:
individual gravitons escape (quantum depletion, i.e.\ Hawking
radiation), while others fall in (accretion).
Right: the condensate always sits on the critical line $\alpha N=1$
between the weakly correlated phase ($\alpha N<1$, dilute gas) and the
strongly correlated one ($\alpha N>1$, collapse). Depletion
$N\to N-1$ does not push the black hole below the line: since the
condensate shrinks, the coupling rises as $\alpha=1/(N-1)$, and the
state glides along the critical point at every step of the
evaporation cascade.}
\label{fig:bhcondensate}
\end{figure}
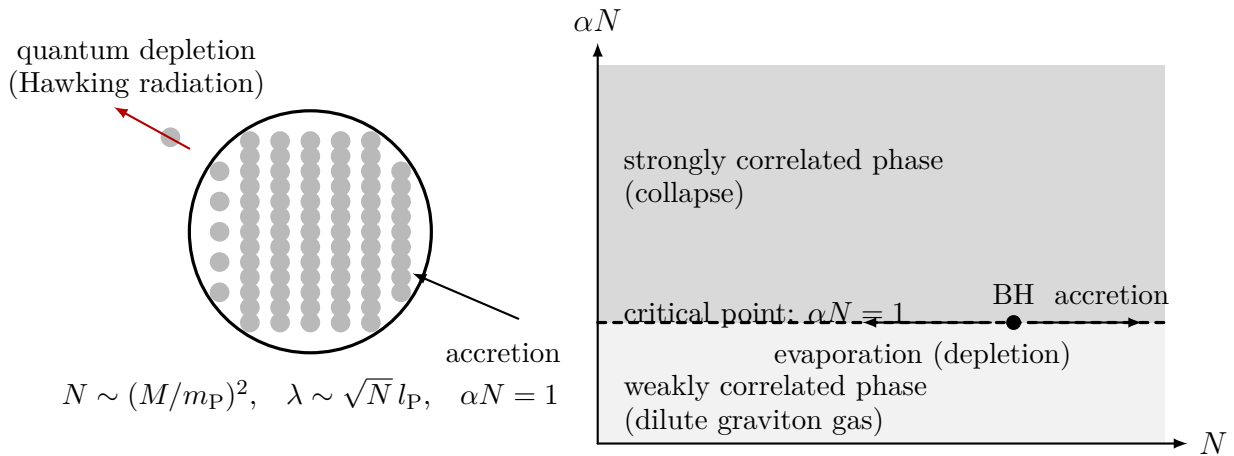

The higher $N$, the more classical is the object which generates the field. The
parameter works also for the boson occupation number in an arbitrary
classical field, but for gravity it reads ($c=1$)
\begin{equation}
N\equiv\frac{Mr_\mathrm{g}}{\hbar}.
\end{equation}
In view of introducing black holes as the state of maximum graviton
packing (cf.\ Fig.~\ref{fig:bhcondensate}), namely a Bose-Einstein condensate of gravitons, it is
instructive to introduce the interaction intensity parameter\footnote{Not to be confused with the deformation parameter in \eqref{eq:gup}.}
\begin{equation}
\alpha\equiv \frac{\hbar G}{\lambda^2}
\label{eq:alpha}
\end{equation}
where $\lambda$ is the Compton wavelength of such gravitons. Let us
consider a microscopic, particle-like source of mass $M\ll m_\mathrm{P}$
and size $R\sim 1/M$: the source size is then set by its own Compton
wavelength, and consistently $R\gg r_\mathrm{g}$. The graviton occupation
number of its gravitational field reads
\begin{equation}
N\equiv\frac{Mr_\mathrm{g}}{\hbar}
=\frac{r_\mathrm{g}}{R}
\sim\left(\frac{M}{m_\mathrm{P}}\right)^{2}
\ll 1 .
\label{eq:coupling}
\end{equation}
 On the other hand, if the source had
collapsed into a black hole $R\sim r_\mathrm{g}$, we would have found
\begin{equation}
M\sim  \frac{r_\mathrm{g}}{G}\Longrightarrow N_\mathrm{BH}\sim
\frac{r_\mathrm{g}^2}{l_\mathrm{P}^2}\sim \frac{M^2}{m_\mathrm{P}^2}\gg 1.
\label{eq:nbh}
\end{equation}
From this perspective black holes are the most classical objects on the
market, because they contain the maximum number of quanta for a given
size. Any increase of $N$ implies an increase of the black hole size,
since for black holes the graviton density is saturated. Black holes
are also the simplest objects, since their physics is controlled by one
parameter only, $N$. This is not a surprise since $N$ is universal,
namely it does not depend on the details of the collapse of the source,
nor on the number or type of particles involved, be it a gravitational
collapse or a collision in the deep ultraviolet energy regime.

The parameter $N$ is very significant since it has another meaning in
terms of energy, namely
\begin{equation}
N = \frac{\text{total gravitational energy in the field}}
         {\text{energy of a single graviton}}.
\end{equation}
We can see this by considering the ratio between the gravitational
potential energy (up to a sign)
\begin{equation}
E_{\mathrm{g}} \sim \frac{M r_{\mathrm{g}}}{R}
\end{equation}
as total energy, and the graviton energy $\hbar R^{-1}$. As long as
$R\gg r_\mathrm{g}$, we get $E_{\mathrm{g}}< M$. Conversely, for
$R\sim r_\mathrm{g}$, we find $E_{\mathrm{g}}\sim M$. Indeed, black
holes are stable states of self-gravitating maximally packed gravitons,
which motivates their interpretation in terms of a Bose-Einstein
condensate. As a further remark, we can express $N$ for black holes,
in \eqref{eq:nbh}, as
\begin{equation}
N=N_\mathrm{BH}\sim  \frac{\lambda^2}{l_\mathrm{P}^2}=\alpha^{-1}
\Longrightarrow \alpha=\alpha_\mathrm{BH}\equiv\frac{1}{N}\ll 1,
\end{equation}
being $\lambda\sim r_\mathrm{g}$. This means that black holes can be
more precisely defined as Bose-Einstein condensates of weakly
interacting gravitons having Compton wavelength
$\lambda\sim \sqrt{N}l_\mathrm{P}$. To see this clearly, we establish a
more concrete contact with known black hole properties. The black hole
mass corresponds to $N \lambda^{-1}$, namely $M=\sqrt{N}m_\mathrm{P}$.
The black hole area goes like $r_\mathrm{g}^2\sim N l_\mathrm{P}^2$,
confirming the holographic interpretation of the pixelation of the event
horizon in plaquettes of area $l_\mathrm{P}^2$. The celebrated area law
for the Bekenstein-Hawking entropy would reduce to $S\sim N$. This is
compelling theoretical support for \eqref{eq:degfreed} and
\eqref{eq:arealaw}, at the basis of the emergent nature of gravity.
More importantly, it is possible to describe black holes beyond the
classical limit.

We can start by considering a generic source of $N$ gravitons with
wavelength $\lambda\gg r_\mathrm{g}$. In this regime we can treat
gravity as linear and use the Newtonian approximation. The potential
energy between a single graviton and one of the other gravitons of the
source will be proportional to $G (\hbar \lambda^{-1})(\hbar
\bar{\lambda}^{-1})\sim \hbar\alpha$, where $\lambda\sim
\bar{\lambda}$ are the wavelengths of the two gravitons. To get the
energy between the graviton and the full source, we just have to sum up
the interaction energy between two gravitons over all possible
pairings. Since all such interaction energies between all possible
graviton pairs are approximately equal, the resulting binding energy
will be $N$ times the single interaction energy, namely
\begin{equation}
V(r)\big|_{r\gtrsim\lambda}
  \sim N\left( \hbar\alpha\frac{1}{r}\right)
\end{equation}
where $r$ is the interaction distance. Such an energy is peaked at
$r=\lambda$. Therefore the probe graviton can escape the attraction
from the source if its energy exceeds the maximal depth of the
potential well, namely
\begin{equation}
E_{\mathrm{escape}}
  \equiv \frac{\hbar}{\lambda_{\mathrm{escape}}}
   = \hbar\alpha N\frac{1}{\lambda}.
\label{eq:escen}
\end{equation}
When the black hole forms, all the quantities involved in
\eqref{eq:escen} saturate, namely
$\lambda_{\mathrm{escape}}=\sqrt{N}l_\mathrm{P}$,
\begin{equation}
V_\mathrm{BH} = \frac{\hbar}{\sqrt{N}\,l_\mathrm{P}}.
\end{equation}
This means that the condensate is actually self-sustained but leaky:
there is always a graviton with enough energy to escape from it. In
practice, the black hole condensate is like a glass full of water: any
thermal noise will cause a drop to leak out of the glass. This
phenomenon, known as quantum depletion, is actually instrumental to
understanding the black hole Hawking radiation. We just calculate the
rate at which gravitons can escape: the graviton--graviton scattering probability $\alpha_\mathrm{BH}^2$, times the number of pairings $N^2$, times the escape frequency $E_\mathrm{escape}/\hbar$:
\begin{equation}
\frac{dN}{dt}=- \frac{V_\mathrm{BH}}{\hbar}
= -\frac{1}{\sqrt{N}\,l_\mathrm{P}}.
\label{eq:dndt}
\end{equation}
To obtain the loss of mass per unit of time, we have to multiply
\eqref{eq:dndt} times the energy of each leaking graviton:
\begin{equation}
\frac{dM}{dt}= \frac{dN}{dt}\,\frac{\hbar}{\lambda_{\mathrm{escape}}}
= -\frac{V_\mathrm{BH}}{\lambda_{\mathrm{escape}}}
= -\frac{\hbar}{N l_\mathrm{P}^2}.
\end{equation}
The application of the Stefan-Boltzmann law for black holes
\begin{equation}
\frac{dM}{dt}=-\frac{T^2}{\hbar}
\end{equation}
leads to the identification of the temperature for the condensate of
gravitons
\begin{equation}
T=\frac{\hbar}{\sqrt{N}\,l_\mathrm{P}},
\end{equation}
consistent with Hawking's original calculation~\cite{Hawking1975}, up to
numerical factors. The Hawking radiation, however, poses a problem for
the interpretation of a black hole in terms of a condensate. If the
black hole decays, the occupation number will decrease, for instance as
\begin{equation}
N\longrightarrow N-1.
\end{equation}
In a condensate this means that the collective forces acting on the
probe boson of the condensate decrease. In practice, the condensate
would decay back to a uniform, unstable, subcritical phase. Consequently, 
this mechanism would imply that the black hole evaporates
off and leaves a horizonless, soliton-like remnant star behind. While
the destiny of an evaporating black hole has been discussed by many
authors~\cite{Barrau2014,AuriliaSpallucci2013}, the problematic aspect
of the quantum $N$-portrait would be the nonvanishing probability for a
black hole of any mass to switch to a remnant star, and not exclusively
at the end of the evaporation.

To clarify the issue we recall that for a condensate described by the
Gross--Pitaevskii field $\Psi(\mathbf{x},t)$, the energy functional
reads
\begin{equation}
E[\Psi]
=
\int d^3x
\left[
\hbar R_0|\nabla\Psi|^2
-g|\Psi|^4
\right],
\end{equation}
where $R_0$ is a length scale and $g>0$ denotes an attractive
interaction. For a configuration in which particles are localized within
a region of size \(R\) we have $ |\Psi|^2\sim N/R^3$. Therefore, the
energy scales as
\begin{equation}
E(R)
\sim
\hbar R_0\frac{N}{R^2}
-
g\frac{N^2}{R^3}.
\label{eq:eL}
\end{equation}
The first term is the positive quantum-pressure, or kinetic-energy,
contribution, whereas the second term is the negative
attractive-interaction contribution. The two terms are crucial to define
the intensity of an interaction, namely the parameter $\alpha$, which
for the special case of gravity has been introduced in \eqref{eq:alpha}.
In general, $\alpha$ is the relative value per unit of particle of the
interaction energy versus kinetic energy. In formulas
\begin{equation}
\alpha  \equiv\frac{1}{N}\left(\frac{\textrm{interaction energy}}
{\textrm{quantum pressure}}\right).
\end{equation}
Since we are interested in a collective interaction in a condensate, we
will consider as before the product $\alpha N$ to understand whether or
not the condensate is stable: the condition $\alpha N<1$ indicates a
sub-critical, dilute, approximately stable phase; $\alpha N>1$
corresponds to a collapsing, unstable supercritical phase; while the
condition $\alpha N=1$ represents a marginally stable critical phase
between the other two phases. From \eqref{eq:eL}, we find
\begin{equation}
\alpha=\frac{g}{\hbar R_0 R}
\label{eq:gpalpha}
\end{equation}
The above equation allows one to write \eqref{eq:eL} as
\begin{equation}
E(R)= \hbar R_0 \frac{N}{R^2}\left[1-\alpha N\right].
\end{equation}
We take the derivative of the above expression to determine the
stationary points
\begin{equation}
0=\frac{dE}{dR}= \frac{\hbar R_0 N}{R^3} \left[3\alpha N-2\right]
\Longrightarrow \alpha=\alpha_\mathrm{c}\equiv\frac{2}{3}\frac{1}{N}
\approx \frac{1}{N}
\label{eq:statpoint}
\end{equation}
As a result, for the Gross--Pitaevskii model, the critical phase
$\alpha N=1$ occurs in the vicinity of the stationary point of the
energy \eqref{eq:eL}. Such a stationary point is a local maximum. From
\eqref{eq:gpalpha} and \eqref{eq:statpoint}, we obtain
\begin{equation}
R_\mathrm{c}=\frac{3gN}{2\hbar R_0}.
\end{equation}
By increasing $N$, the whole condensate is in the supercritical, compressing phase; conversely,  by
decreasing $N$, the condensate is in the diluted, subcritical phase. 
Therefore the condition $\alpha
N=1$ marks the tip of the barrier which separates the dilute from the
collapsing regime.

From \eqref{eq:eL}, if the size \(R\) is kept fixed, then \(\alpha\)
would be a constant and the collective coupling would decrease,
\begin{equation}
\alpha N\longrightarrow \alpha(N-1)<\alpha N,
\end{equation}
in case of depletion. This would move the system slightly to the
subcritical side, implying a violation of the black hole condition
$\alpha N= 1$, were black holes described in terms of the
Gross--Pitaevskii model.

For gravity, however, the intensity of the interaction becomes stronger
and compensates the depletion. This is evident from the fact that for a
black hole $R\sim r_\mathrm{g}\sim \sqrt{N}l_\mathrm{P}$. If, due to
depletion, the black hole shrinks, the value of
\begin{equation}
\alpha=\frac{l_\mathrm{P}^2}{R^2},
\end{equation}
is no longer constant, but increases because $R$ depends on $N$, in
marked contrast to the above Gross--Pitaevskii model. As a result
\begin{equation}
\alpha=\frac1N \longrightarrow \alpha'=\frac1{N-1}.
\end{equation}
Consequently,
\begin{equation}
\alpha'(N-1)
=
\frac1{N-1}(N-1)
=1.
\end{equation}
Thus the new state is again critical:
\begin{equation}
(N,R(N))
\longrightarrow
(N-1,R(N-1)).
\end{equation}
Repeated emission gives the sequence
\begin{equation}
(N_0,R_0)
\longrightarrow
(N_0-1,R_{-1})
\longrightarrow
(N_0-2,R_{-2})
\longrightarrow\cdots,
\end{equation}
where
\begin{equation}
R_{-k}=l_\mathrm{P}\sqrt{N_0-k}.
\end{equation}
In practice, the black hole, by leaking, jumps from a configuration at
the critical point to another one. The same holds for the opposite
process, namely when the graviton occupation number increases,
\begin{equation}
(N_0,R_0)
\longrightarrow
(N_0+1,R_{1})
\longrightarrow
(N_0+2,R_{2})
\longrightarrow\cdots,
\end{equation}
corresponding to the case in which the black hole accretes mass. To
make the above properties concrete, we have to write a Lagrangian in
which the kinetic pressure compensates the leak. Dvali \& G\'omez have
proposed a Ginzburg-Landau like expression for the collective variable
\(N\)~\cite{DvaliGomez2014}:
\begin{equation}
\mathcal{L}_{\mathrm{GL}}
=
(\dot N)^2
+
\frac{1}{N l_\mathrm{P}^2}
+
l_\mathrm{P}^{-2}\mathcal{O}\!\left(\frac1{N^2}\right).
\label{eq:dgaction}
\end{equation}
Here the first term, the kinetic pressure, balances the second term,
the interaction. Indeed, in agreement with the depletion rate
\eqref{eq:dndt},
\begin{equation}
\dot N\sim-\frac1{\sqrt N l_\mathrm{P}},
\end{equation}
so that $(\dot N)^2\sim 1/(N l_\mathrm{P}^2)$: the black hole sits at
the critical point at each step of the evaporation cascade.

In conclusion, black holes are self-similar, critical graviton
condensates (cf.\ Fig.~\ref{fig:bhcondensate}).

\section{Conclusions}

We have provided evidence in favor of a fruitful relation between quantum gravity and condensed matter physics, 
and we have done so at three depths: from accurate testbeds, through analog and dual systems, to the one-to-one identification of spacetime with a condensate itself.

The deepest of these levels is the most promising, because it is
connected to the very nature of gravity as an emergent phenomenon, a
vast research area to which many authors are
contributing~\cite{Jacobson1995,Padmanabhan2002,Padmanabhan2010,
Bianconi2025}. Despite the progress, there are many open issues one
should think about. For instance, the macroquantumness, namely the
expected large scale effects that a spacetime condensate should
exhibit. In the past, Dvali \& Gomez~\cite{DvaliGomez2012}, as well as
Giddings with similar arguments, have addressed this
question~\cite{Giddings2017}. We also have a spacetime metric -- the
holographic screen metric~\cite{NicoliniSpallucci2014} -- which
interpolates the transition from the black hole condensate to the
Schwarzschild metric and allows us to compute macroquantumness
concretely in an astrophysical setting.

Additionally, the full reasoning of the quantum $N$-portrait
formulation is rooted in the nature of the phase transition at the
Planck scale, where gravity classicalizes and any massive object which
generates gravity turns into a condensate. To some extent, this is the
crux of the question of any quantum gravity
formulation~\cite{Nicolini2025,Nicolini2018}. The self-similarity
properties of black holes at the critical point are a further piece of
evidence of the non-local, fractal nature of the
pre-geometry~\cite{Kleinert1987}. In this regime, higher dimensional
structures like $p$-branes are expected to play a role, and it is not
excluded that a new mathematical formalism must be invoked, like
polyvectors within the Clifford algebra
paradigm~\cite{AuriliaAnsoldiSpallucci2002}.

From a more technical viewpoint, there are further issues we should
explore in the future. A fundamental problem is the relation between
the quantum string and the black hole condensate, in particular between
the actions \eqref{eq:GLLagr} and \eqref{eq:dgaction}. More in
general, we are interested in the relation between string theory, or
any quantum gravity formulation~\cite{Nicolai2014}, and the recent
proposals of quantum mechanical descriptions of black
holes~\cite{Casadio2016,Spallucci2017}.

In conclusion, the question of all the questions in physics dates back
to the time of Leucippus and
Democritus,\footnote{The atomists, Leucippus and Democritus: fragments,
a text and translation with a commentary by C.C.W.~Taylor, University
of Toronto Press, 1999, ISBN 0-8020-4390-9, pp.~157--158.} and
Kant,\footnote{Critique of Pure Reason (German: Kritik der reinen
Vernunft; 1781; second edition 1787).} namely Boltzmann's
problem:\footnote{Ludwig Boltzmann: Nochmals \"uber die Atomistik. In:
Popul\"are Schriften, pp.~158--161.} How can we describe the
macroscopic behavior in terms of microscopic fundamental entities? It
is hard to predict whether particle accelerator based physics can say
something significant in the future, or is destined to be surpassed by
other investigation methods. Be that as it may, in the most optimistic
scenario, particle physics alone will likely not provide any answer.
The future of research in fundamental physics will require a collective
effort and a coordination of investigations via a variety of research
methods, of which particle accelerators are just one --- and not the
leading and only one as in the
past.\footnote{M. Schott: ``Particle Physics Is Not Over: It's Just
Getting Harder'', Physics Colloquium, Goethe-Universit\"at Frankfurt,
15 July 2026.}

We have indeed growing evidence that condensed matter physics will
become increasingly important, not just as a niche topic, and not only
for its technological applications in a variety of sectors, including
information theory and energy, but as a key element to understand
fundamental physics.

Nobody can actually predict the future, nor say which research lines
will succeed. What we can say is where the next measurement is most
likely to come from. The bridge between condensed matter physics and
gravity is built; the remaining task is to cross it.

Godot never came. Quantum gravity hides not in colliders, 
but in the lattice beneath our feet.

\section{My memories of Guido Barbiellini}

I first met Guido Barbiellini in the early 1990s. At the time, he was a
professor at the University of Trieste and I was a physics
undergraduate student. He was a very well established, highly regarded
scientist, but I did not have much interaction with him in the
beginning. He was teaching physics in the engineering school and
conducted research in experimental particle physics and astroparticle
physics. Being a student majoring in theoretical physics, also in
preparation of my Laurea degree thesis, I spent most of my time at the
Miramare Campus, not the main University Campus where Barbiellini had
his office. Nevertheless, through the years I developed a clear image
of his persona and his scientific profile. What was striking in
relation to the rest of the faculty was Barbiellini's international
agility. His experience of life, his anecdotal stories about Fermi and
other scientists, his contacts with groups at CERN in Geneva were a
gold mine for me and for any student or young scientist. On the pretty
personal level, Barbiellini was an elegant person, with a charisma
that seemed soft but in reality concealed self-confidence and some sort
of fire fueling his instinct of battling for the right causes. His
elegance was transparent also in some of the sentences he kept
repeating:
``Tu hai perfettamente ragione, ma per dirlo dovresti avere un certo
\textit{modo}''. In English: ``You are perfectly right, but to say it,
you should have a certain manner.''
I did not understand what ``modo'' (way) would exactly mean in that
context, but it was clear that social behavior and acceptance were for
him high stakes, at a level comparable to what one can find in the
British culture.

Another of his words of wisdom was the following: ``Inutile sognare di
avere collaboratori migliori, \`e meglio imparare a lavorare con quelli
che si hanno''. In English: ``There is no use dreaming of better
collaborators; it is better to learn to work with those you have.''
This was a sort of guideline for the management of manpower and
self-motivation at the same time, but it also suggests that Guido was
nostalgic for the golden time he had spent at CERN.

On the more scientific level, Barbiellini had a global mindset and was
interested in several research fields, including some aspects of
theoretical physics. For instance, he was interested in testing with
high energy gamma rays~\cite{Ackermann2009,Abdo2009} any violation of
the uncertainty principle emerging from quantum gravity modified
commutators like \eqref{eq:gup} or similar
ones~\cite{AmelinoCamelia1998}.
This topic was actually the point of closest contact I had with Guido
Barbiellini. In Spring 2003 there were several meetings to discuss my
enrollment as a postdoctoral researcher in the group of Guido and his
colleague Gianrossano Giannini, in order to investigate quantum gravity
signatures in high energy astrophysical events. 
In the end, despite the attractive opportunities, the deal never
materialized. The contact with Barbiellini was nevertheless very
instructive, because it helped me to understand what I wanted and who
I am: I did not want to be a member of a large collaboration, and I
preferred to follow my own research line. More importantly, the episode
is a piece of evidence that Guido was able to see the good, the talent
in people: he certainly saw that in me, irrespective of research
field, community habits or academic logic of any kind.

The contacts with Guido did not end there. In January 2007, I started
the first of my long stays in the U.S. and I was hosted for about
three months at Northeastern University in Boston. There, I met Guido's
son Bernardo, who since then has become a colleague, collaborator and
friend. Following his father's footsteps, Bernardo has become an
all-round physicist capable of performing research in a variety of
sectors.

Life is in any case unpredictable. It makes me think, while writing
this paper, that as of today, as a professor in Trieste, I teach
introductory physics in the school of engineering. I feel honored to be
Guido's successor in his teaching post.

\section*{Acknowledgments}
The author acknowledges institutional support from the GNFM (Italian National Group for Mathematical Physics) and the ``Iniziativa Specifica FLAG'' of the INFN (Italian National Institute for Nuclear Physics). Gratitude is extended to the editors of the Special Issue and the journal Condensed Matter for the invitation and patience during the review process.


\begin{thebibliography}{156}%
\makeatletter
\providecommand \@ifxundefined [1]{%
 \@ifx{#1\undefined}
}%
\providecommand \@ifnum [1]{%
 \ifnum #1\expandafter \@firstoftwo
 \else \expandafter \@secondoftwo
 \fi
}%
\providecommand \@ifx [1]{%
 \ifx #1\expandafter \@firstoftwo
 \else \expandafter \@secondoftwo
 \fi
}%
\providecommand \natexlab [1]{#1}%
\providecommand \enquote  [1]{``#1''}%
\providecommand \bibnamefont  [1]{#1}%
\providecommand \bibfnamefont [1]{#1}%
\providecommand \citenamefont [1]{#1}%
\providecommand \href@noop [0]{\@secondoftwo}%
\providecommand \href [0]{\begingroup \@sanitize@url \@href}%
\providecommand \@href[1]{\@@startlink{#1}\@@href}%
\providecommand \@@href[1]{\endgroup#1\@@endlink}%
\providecommand \@sanitize@url [0]{\catcode `\\12\catcode `\$12\catcode
  `\&12\catcode `\#12\catcode `\^12\catcode `\_12\catcode `\%12\relax}%
\providecommand \@@startlink[1]{}%
\providecommand \@@endlink[0]{}%
\providecommand \url  [0]{\begingroup\@sanitize@url \@url }%
\providecommand \@url [1]{\endgroup\@href {#1}{\urlprefix }}%
\providecommand \urlprefix  [0]{URL }%
\providecommand \Eprint [0]{\href }%
\providecommand \doibase [0]{http://dx.doi.org/}%
\providecommand \selectlanguage [0]{\@gobble}%
\providecommand \bibinfo  [0]{\@secondoftwo}%
\providecommand \bibfield  [0]{\@secondoftwo}%
\providecommand \translation [1]{[#1]}%
\providecommand \BibitemOpen [0]{}%
\providecommand \bibitemStop [0]{}%
\providecommand \bibitemNoStop [0]{.\EOS\space}%
\providecommand \EOS [0]{\spacefactor3000\relax}%
\providecommand \BibitemShut  [1]{\csname bibitem#1\endcsname}%
\let\auto@bib@innerbib\@empty


\bibitem [{\citenamefont {Aubert}\ \emph {et~al.}(1974)\citenamefont {Aubert}
  \emph {et~al.}}]{Aubert:1974js}%
  \BibitemOpen
  \bibfield  {author} {\bibinfo {author} {\bibfnamefont {J.~J.}\ \bibnamefont
  {Aubert}} \emph {et~al.},\ }\href@noop {} {\bibfield  {journal} {\bibinfo
  {journal} {Phys. Rev. Lett.}\ }\textbf {\bibinfo {volume} {33}},\ \bibinfo
  {pages} {1404} (\bibinfo {year} {1974})}\BibitemShut {NoStop}%
\bibitem [{\citenamefont {Augustin}\ \emph {et~al.}(1974)\citenamefont
  {Augustin} \emph {et~al.}}]{Augustin:1974xv}%
  \BibitemOpen
  \bibfield  {author} {\bibinfo {author} {\bibfnamefont {J.-E.}\ \bibnamefont
  {Augustin}} \emph {et~al.},\ }\href@noop {} {\bibfield  {journal} {\bibinfo
  {journal} {Phys. Rev. Lett.}\ }\textbf {\bibinfo {volume} {33}},\ \bibinfo
  {pages} {1406} (\bibinfo {year} {1974})}\BibitemShut {NoStop}%
\bibitem [{\citenamefont {Hawking}(1975)}]{Hawking1975}%
  \BibitemOpen
  \bibfield  {author} {\bibinfo {author} {\bibfnamefont {S.~W.}\ \bibnamefont
  {Hawking}},\ }\href@noop {} {\bibfield  {journal} {\bibinfo  {journal}
  {Commun. Math. Phys.}\ }\textbf {\bibinfo {volume} {43}},\ \bibinfo
  {pages} {199} (\bibinfo {year} {1975})}\BibitemShut {NoStop}%
\bibitem [{\citenamefont {Nicolini}(2025)}]{Nicolini2025}%
  \BibitemOpen
  \bibfield  {author} {\bibinfo {author} {\bibfnamefont {P.}~\bibnamefont
  {Nicolini}},\ }\href {\doibase 10.1007/978-3-031-76066-2\_13} {\bibfield
  {journal} {\bibinfo  {journal} {Fundamental Theories of Physics}\ }\textbf
  {\bibinfo {volume} {219}},\ \bibinfo {pages} {275} (\bibinfo {year}
  {2025})}\BibitemShut {NoStop}%
\bibitem [{\citenamefont {Kramida}(2010)}]{Kramida2010}%
  \BibitemOpen
  \bibfield  {author} {\bibinfo {author} {\bibfnamefont {A.~E.}\ \bibnamefont
  {Kramida}},\ }\href@noop {} {\bibfield  {journal} {\bibinfo  {journal}
  {Atom. Data Nucl. Data Tables}\ }\textbf {\bibinfo {volume} {96}},\ \bibinfo
  {pages} {586} (\bibinfo {year} {2010})}\BibitemShut {NoStop}%
\bibitem [{\citenamefont {Antoniadis}(1990)}]{Antoniadis1990}%
  \BibitemOpen
  \bibfield  {author} {\bibinfo {author} {\bibfnamefont {I.}~\bibnamefont
  {Antoniadis}},\ }\href@noop {} {\bibfield  {journal} {\bibinfo  {journal}
  {Phys. Lett. B}\ }\textbf {\bibinfo {volume} {246}},\ \bibinfo {pages}
  {377} (\bibinfo {year} {1990})}\BibitemShut {NoStop}%
\bibitem [{\citenamefont {Arkani-Hamed}\ \emph {et~al.}(1998)\citenamefont
  {Arkani-Hamed}, \citenamefont {Dimopoulos},\ and\ \citenamefont
  {Dvali}}]{ADD1998}%
  \BibitemOpen
  \bibfield  {author} {\bibinfo {author} {\bibfnamefont {N.}~\bibnamefont
  {Arkani-Hamed}}, \bibinfo {author} {\bibfnamefont {S.}~\bibnamefont
  {Dimopoulos}}, \ and\ \bibinfo {author} {\bibfnamefont {G.}~\bibnamefont
  {Dvali}},\ }\href@noop {} {\bibfield  {journal} {\bibinfo  {journal}
  {Phys. Lett. B}\ }\textbf {\bibinfo {volume} {429}},\ \bibinfo {pages}
  {263} (\bibinfo {year} {1998})}\BibitemShut {NoStop}%
\bibitem [{\citenamefont {Arkani-Hamed}\ \emph {et~al.}(1999)\citenamefont
  {Arkani-Hamed}, \citenamefont {Dimopoulos},\ and\ \citenamefont
  {Dvali}}]{ADD1999}%
  \BibitemOpen
  \bibfield  {author} {\bibinfo {author} {\bibfnamefont {N.}~\bibnamefont
  {Arkani-Hamed}}, \bibinfo {author} {\bibfnamefont {S.}~\bibnamefont
  {Dimopoulos}}, \ and\ \bibinfo {author} {\bibfnamefont {G.}~\bibnamefont
  {Dvali}},\ }\href@noop {} {\bibfield  {journal} {\bibinfo  {journal}
  {Phys. Rev. D}\ }\textbf {\bibinfo {volume} {59}},\ \bibinfo {pages}
  {086004} (\bibinfo {year} {1999})}\BibitemShut {NoStop}%
\bibitem [{\citenamefont {Antoniadis}\ \emph {et~al.}(1998)\citenamefont
  {Antoniadis}, \citenamefont {Arkani-Hamed}, \citenamefont {Dimopoulos},\ and\
  \citenamefont {Dvali}}]{AADD1998}%
  \BibitemOpen
  \bibfield  {author} {\bibinfo {author} {\bibfnamefont {I.}~\bibnamefont
  {Antoniadis}}, \bibinfo {author} {\bibfnamefont {N.}~\bibnamefont
  {Arkani-Hamed}}, \bibinfo {author} {\bibfnamefont {S.}~\bibnamefont
  {Dimopoulos}}, \ and\ \bibinfo {author} {\bibfnamefont {G.}~\bibnamefont
  {Dvali}},\ }\href@noop {} {\bibfield  {journal} {\bibinfo  {journal}
  {Phys. Lett. B}\ }\textbf {\bibinfo {volume} {436}},\ \bibinfo {pages}
  {257} (\bibinfo {year} {1998})}\BibitemShut {NoStop}%
\bibitem [{\citenamefont {Randall}\ and\ \citenamefont
  {Sundrum}(1999{\natexlab{a}})}]{Randall1999a}%
  \BibitemOpen
  \bibfield  {author} {\bibinfo {author} {\bibfnamefont {L.}~\bibnamefont
  {Randall}}\ and\ \bibinfo {author} {\bibfnamefont {R.}~\bibnamefont
  {Sundrum}},\ }\href@noop {} {\bibfield  {journal} {\bibinfo  {journal}
  {Phys. Rev. Lett.}\ }\textbf {\bibinfo {volume} {83}},\ \bibinfo {pages}
  {3370} (\bibinfo {year} {1999}{\natexlab{a}})}\BibitemShut {NoStop}%
\bibitem [{\citenamefont {Randall}\ and\ \citenamefont
  {Sundrum}(1999{\natexlab{b}})}]{Randall1999b}%
  \BibitemOpen
  \bibfield  {author} {\bibinfo {author} {\bibfnamefont {L.}~\bibnamefont
  {Randall}}\ and\ \bibinfo {author} {\bibfnamefont {R.}~\bibnamefont
  {Sundrum}},\ }\href@noop {} {\bibfield  {journal} {\bibinfo  {journal}
  {Phys. Rev. Lett.}\ }\textbf {\bibinfo {volume} {83}},\ \bibinfo {pages}
  {4690} (\bibinfo {year} {1999}{\natexlab{b}})}\BibitemShut {NoStop}%
\bibitem [{\citenamefont {Appelquist}\ \emph {et~al.}(2001)\citenamefont
  {Appelquist}, \citenamefont {Cheng},\ and\ \citenamefont
  {Dobrescu}}]{Appelquist2001}%
  \BibitemOpen
  \bibfield  {author} {\bibinfo {author} {\bibfnamefont {T.}~\bibnamefont
  {Appelquist}}, \bibinfo {author} {\bibfnamefont {H.-C.}\ \bibnamefont
  {Cheng}}, \ and\ \bibinfo {author} {\bibfnamefont {B.~A.}\ \bibnamefont
  {Dobrescu}},\ }\href@noop {} {\bibfield  {journal} {\bibinfo  {journal}
  {Phys. Rev. D}\ }\textbf {\bibinfo {volume} {64}},\ \bibinfo {pages}
  {035002} (\bibinfo {year} {2001})}\BibitemShut {NoStop}%
\bibitem [{\citenamefont {Dimopoulos}\ and\ \citenamefont
  {Landsberg}(2001)}]{Dimopoulos2001}%
  \BibitemOpen
  \bibfield  {author} {\bibinfo {author} {\bibfnamefont {S.}~\bibnamefont
  {Dimopoulos}}\ and\ \bibinfo {author} {\bibfnamefont {G.}~\bibnamefont
  {Landsberg}},\ }\href@noop {} {\bibfield  {journal} {\bibinfo  {journal}
  {Phys. Rev. Lett.}\ }\textbf {\bibinfo {volume} {87}},\ \bibinfo {pages}
  {161602} (\bibinfo {year} {2001})}\BibitemShut {NoStop}%
\bibitem [{\citenamefont {Giddings}\ and\ \citenamefont
  {Thomas}(2002)}]{Giddings2002}%
  \BibitemOpen
  \bibfield  {author} {\bibinfo {author} {\bibfnamefont {S.~B.}\ \bibnamefont
  {Giddings}}\ and\ \bibinfo {author} {\bibfnamefont {S.}~\bibnamefont
  {Thomas}},\ }\href@noop {} {\bibfield  {journal} {\bibinfo  {journal}
  {Phys. Rev. D}\ }\textbf {\bibinfo {volume} {65}},\ \bibinfo {pages}
  {056010} (\bibinfo {year} {2002})}\BibitemShut {NoStop}%
\bibitem [{\citenamefont {Banks}\ and\ \citenamefont
  {Fischler}(1999)}]{Banks1999}%
  \BibitemOpen
  \bibfield  {author} {\bibinfo {author} {\bibfnamefont {T.}~\bibnamefont
  {Banks}}\ and\ \bibinfo {author} {\bibfnamefont {W.}~\bibnamefont
  {Fischler}},\ }\href@noop {} {\bibfield  {journal} {\bibinfo  {journal}
  {arXiv e-prints}\ } (\bibinfo {year} {1999})},\ \Eprint
  {http://arxiv.org/abs/hep-th/9906038}{hep-th/9906038}\BibitemShut {NoStop}%
\bibitem [{\citenamefont {Br{\"u}ning}\ and\ \citenamefont
  {Zerlauth}(2025)}]{LHC2025}%
  \BibitemOpen
  \bibfield  {author} {\bibinfo {author} {\bibfnamefont {O.}~\bibnamefont
  {Br{\"u}ning}}\ and\ \bibinfo {author} {\bibfnamefont {M.}~\bibnamefont
  {Zerlauth}},\ }\href {\doibase 10.48550/arXiv.2505.03535} {\bibfield
  {journal} {\bibinfo  {journal} {arXiv e-prints}\ } (\bibinfo {year} {2025}),\
  10.48550/arXiv.2505.03535},\ \Eprint
  {http://arxiv.org/abs/2505.03535}{2505.03535 [physics.acc-ph]}\BibitemShut
  {NoStop}%
\bibitem [{\citenamefont {Casanova}\ and\ \citenamefont
  {Spallucci}(2005)}]{Casanova2005}%
  \BibitemOpen
  \bibfield  {author} {\bibinfo {author} {\bibfnamefont {J.}~\bibnamefont
  {Casanova}}\ and\ \bibinfo {author} {\bibfnamefont {E.}~\bibnamefont
  {Spallucci}},\ }\href@noop {} {\bibfield  {journal} {\bibinfo  {journal}
  {arXiv e-prints}\ } (\bibinfo {year} {2005})},\ \Eprint
  {http://arxiv.org/abs/hep-ph/0512063}{hep-ph/0512063}\BibitemShut {NoStop}%
\bibitem [{\citenamefont {Bleicher}\ and\ \citenamefont
  {Nicolini}(2010)}]{Bleicher2010}%
  \BibitemOpen
  \bibfield  {author} {\bibinfo {author} {\bibfnamefont {M.}~\bibnamefont
  {Bleicher}}\ and\ \bibinfo {author} {\bibfnamefont {P.}~\bibnamefont
  {Nicolini}},\ }\href {\doibase 10.1088/1742-6596/237/1/012008} {\bibfield
  {journal} {\bibinfo  {journal} {J. Phys.: Conf. Ser.}\ }\textbf
  {\bibinfo {volume} {237}},\ \bibinfo {pages} {012008} (\bibinfo {year}
  {2010})},\ \Eprint {http://arxiv.org/abs/1001.2211}{1001.2211
  [hep-ph]}\BibitemShut {NoStop}%
\bibitem [{\citenamefont {Heckler}(1997)}]{Heckler1997}%
  \BibitemOpen
  \bibfield  {author} {\bibinfo {author} {\bibfnamefont {A.~F.}\ \bibnamefont
  {Heckler}},\ }\href {\doibase 10.1103/PhysRevD.55.480} {\bibfield  {journal}
  {\bibinfo  {journal} {Phys. Rev. D}\ }\textbf {\bibinfo {volume} {55}},\
  \bibinfo {pages} {480} (\bibinfo {year} {1997})},\ \Eprint
  {http://arxiv.org/abs/astro-ph/9601029}{astro-ph/9601029}\BibitemShut
  {NoStop}%
\bibitem [{\citenamefont {Hayrapetyan}\ \emph {et~al.}(2026)\citenamefont
  {Hayrapetyan} \emph {et~al.}}]{CMS2026}%
  \BibitemOpen
  \bibfield  {author} {\bibinfo {author} {\bibfnamefont {A.}~\bibnamefont
  {Hayrapetyan}} \emph {et~al.} (\bibinfo {collaboration} {CMS}),\ }\href
  {\doibase 10.1007/JHEP08(2026)098} {\bibfield  {journal} {\bibinfo  {journal}
  {JHEP}\ }\textbf {\bibinfo {volume} {08}},\ \bibinfo {pages} {098} (\bibinfo
  {year} {2026})},\ \Eprint
  {http://arxiv.org/abs/2604.10732}{2604.10732 [hep-ex]}\BibitemShut {NoStop}%
\bibitem [{\citenamefont {Nicolini}\ \emph {et~al.}(2006)\citenamefont
  {Nicolini}, \citenamefont {Smailagic},\ and\ \citenamefont
  {Spallucci}}]{Nicolini:2005vd}%
  \BibitemOpen
  \bibfield  {author} {\bibinfo {author} {\bibfnamefont {P.}~\bibnamefont
  {Nicolini}}, \bibinfo {author} {\bibfnamefont {A.}~\bibnamefont {Smailagic}},
  \ and\ \bibinfo {author} {\bibfnamefont {E.}~\bibnamefont {Spallucci}},\
  }\href {\doibase 10.1016/j.physletb.2005.11.004} {\bibfield  {journal}
  {\bibinfo  {journal} {Phys. Lett. B}\ }\textbf {\bibinfo {volume} {632}},\
  \bibinfo {pages} {547} (\bibinfo {year} {2006})},\ \Eprint
  {http://arxiv.org/abs/gr-qc/0510112}{arXiv:gr-qc/0510112}\BibitemShut
  {NoStop}%
\bibitem [{\citenamefont {Nicolini}(2009)}]{Nicolini:2008aj}%
  \BibitemOpen
  \bibfield  {author} {\bibinfo {author} {\bibfnamefont {P.}~\bibnamefont
  {Nicolini}},\ }\href {\doibase 10.1142/S0217751X09043353} {\bibfield
  {journal} {\bibinfo  {journal} {Int. J. Mod. Phys. A}\ }\textbf
  {\bibinfo {volume} {24}},\ \bibinfo {pages} {1229} (\bibinfo {year}
  {2009})},\ \Eprint
  {http://arxiv.org/abs/0807.1939}{arXiv:0807.1939 [hep-th]}\BibitemShut
  {NoStop}%
\bibitem [{\citenamefont {Rizzo}(2006)}]{Rizzo:2006zb}%
  \BibitemOpen
  \bibfield  {author} {\bibinfo {author} {\bibfnamefont {T.~G.}\ \bibnamefont
  {Rizzo}},\ }\href {\doibase 10.1088/1126-6708/2006/09/021} {\bibfield
  {journal} {\bibinfo  {journal} {JHEP}\ }\textbf {\bibinfo {volume} {09}},\
  \bibinfo {pages} {021} (\bibinfo {year} {2006})},\ \Eprint
  {http://arxiv.org/abs/hep-ph/0606051}{arXiv:hep-ph/0606051}\BibitemShut
  {NoStop}%
\bibitem [{\citenamefont {Nicolini}\ and\ \citenamefont
  {Winstanley}(2011)}]{Nicolini:2011nz}%
  \BibitemOpen
  \bibfield  {author} {\bibinfo {author} {\bibfnamefont {P.}~\bibnamefont
  {Nicolini}}\ and\ \bibinfo {author} {\bibfnamefont {E.}~\bibnamefont
  {Winstanley}},\ }\href {\doibase 10.1007/JHEP11(2011)075} {\bibfield
  {journal} {\bibinfo  {journal} {JHEP}\ }\textbf {\bibinfo {volume}
  {11}},\ \bibinfo {pages} {075} (\bibinfo {year} {2011})},\ \Eprint
  {http://arxiv.org/abs/1108.4419}{arXiv:1108.4419 [hep-ph]}\BibitemShut
  {NoStop}%
\bibitem [{\citenamefont {Gingrich}(2010)}]{Gingrich:2010ed}%
  \BibitemOpen
  \bibfield  {author} {\bibinfo {author} {\bibfnamefont {D.~M.}\ \bibnamefont
  {Gingrich}},\ }\href {\doibase 10.1007/JHEP05(2010)022} {\bibfield  {journal}
  {\bibinfo  {journal} {JHEP}\ }\textbf {\bibinfo {volume} {05}},\ \bibinfo
  {pages} {022} (\bibinfo {year} {2010})},\ \Eprint
  {http://arxiv.org/abs/1003.1798}{arXiv:1003.1798 [hep-ph]}\BibitemShut
  {NoStop}%
\bibitem [{\citenamefont {Mureika}\ \emph {et~al.}(2012)\citenamefont
  {Mureika}, \citenamefont {Nicolini},\ and\ \citenamefont
  {Spallucci}}]{Mureika2012}%
  \BibitemOpen
  \bibfield  {author} {\bibinfo {author} {\bibfnamefont {J.}~\bibnamefont
  {Mureika}}, \bibinfo {author} {\bibfnamefont {P.}~\bibnamefont {Nicolini}}, \
  and\ \bibinfo {author} {\bibfnamefont {E.}~\bibnamefont {Spallucci}},\ }\href
  {\doibase 10.1103/PhysRevD.85.106007} {\bibfield  {journal} {\bibinfo
  {journal} {Phys. Rev. D}\ }\textbf {\bibinfo {volume} {85}},\ \bibinfo
  {pages} {106007} (\bibinfo {year} {2012})},\ \Eprint
  {http://arxiv.org/abs/1111.5830}{1111.5830 [hep-ph]}\BibitemShut {NoStop}%
\bibitem [{\citenamefont {Roser}\ \emph {et~al.}(2023)\citenamefont {Roser},
  \citenamefont {Brinkmann}, \citenamefont {Cousineau}, \citenamefont
  {Denisov}, \citenamefont {Gessner}, \citenamefont {Gourlay}, \citenamefont
  {Lebrun}, \citenamefont {Narain}, \citenamefont {Oide}, \citenamefont
  {Raubenheimer} \emph {et~al.}}]{Roser2023}%
  \BibitemOpen
  \bibfield  {author} {\bibinfo {author} {\bibfnamefont {T.}~\bibnamefont
  {Roser}}, \bibinfo {author} {\bibfnamefont {R.}~\bibnamefont {Brinkmann}},
  \bibinfo {author} {\bibfnamefont {S.}~\bibnamefont {Cousineau}}, \bibinfo
  {author} {\bibfnamefont {D.}~\bibnamefont {Denisov}}, \bibinfo {author}
  {\bibfnamefont {S.}~\bibnamefont {Gessner}}, \bibinfo {author} {\bibfnamefont
  {S.}~\bibnamefont {Gourlay}}, \bibinfo {author} {\bibfnamefont
  {P.}~\bibnamefont {Lebrun}}, \bibinfo {author} {\bibfnamefont
  {M.}~\bibnamefont {Narain}}, \bibinfo {author} {\bibfnamefont
  {K.}~\bibnamefont {Oide}}, \bibinfo {author} {\bibfnamefont
  {T.}~\bibnamefont {Raubenheimer}},  \emph {et~al.},\ }\href@noop {} {\bibfield
  {journal} {\bibinfo  {journal} {J. Instrum.}\ }\textbf {\bibinfo {volume}
  {18}},\ \bibinfo {pages} {P05018} (\bibinfo {year} {2023})},\ \Eprint
  {http://arxiv.org/abs/2208.06030}{2208.06030 [physics.acc-ph]}\BibitemShut
  {NoStop}%
\bibitem [{\citenamefont {Loeb}(2015)}]{Loeb2015}%
  \BibitemOpen
  \bibfield  {author} {\bibinfo {author} {\bibfnamefont {A.}~\bibnamefont
  {Loeb}},\ }\href@noop {} {\bibfield  {journal} {\bibinfo  {journal} {arXiv
  e-prints}\ } (\bibinfo {year} {2015})},\ \Eprint
  {http://arxiv.org/abs/1503.01509}{1503.01509 [astro-ph.HE]}\BibitemShut
  {NoStop}%
\bibitem [{\citenamefont {Casher}\ and\ \citenamefont
  {Nussinov}(1995)}]{Casher1995}%
  \BibitemOpen
  \bibfield  {author} {\bibinfo {author} {\bibfnamefont {A.}~\bibnamefont
  {Casher}}\ and\ \bibinfo {author} {\bibfnamefont {S.}~\bibnamefont
  {Nussinov}},\ }\href@noop {} {\bibfield  {journal} {\bibinfo  {journal}
  {arXiv e-prints}\ } (\bibinfo {year} {1995})},\ \Eprint
  {http://arxiv.org/abs/hep-ph/9510364}{hep-ph/9510364 [hep-ph]}\BibitemShut
  {NoStop}%
\bibitem [{\citenamefont {Loeb}(2023)}]{LoebMedium}%
  \BibitemOpen
  \bibfield  {author} {\bibinfo {author} {\bibfnamefont {A.}~\bibnamefont
  {Loeb}},\ }\href@noop {} {\enquote {\bibinfo {title} {Planck energy
  accelerators},}\ }\bibinfo {howpublished}
  {\url{https://avi-loeb.medium.com/planck-energy-accelerators-a966bddee59c}}
  (\bibinfo {year} {2023}),\ \bibinfo {note} {online essay, Medium}\BibitemShut
  {NoStop}%
\bibitem [{\citenamefont {Siegel}(2024)}]{Siegel2024}%
  \BibitemOpen
  \bibfield  {author} {\bibinfo {author} {\bibfnamefont {E.}~\bibnamefont
  {Siegel}},\ }\href@noop {} {\enquote {\bibinfo {title} {Ask {Ethan}: Could we
  build a collider bigger than {Earth}?}}\ }\bibinfo {howpublished}
  {\url{https://medium.com/starts-with-a-bang/ask-ethan-could-we-build-a-collider-bigger-than-earth-37a1dc1b6e7b}}
  (\bibinfo {year} {2024}),\ \bibinfo {note} {Starts With a Bang, August
  2024}\BibitemShut {NoStop}%
\bibitem [{\citenamefont {Danchev}\ \emph {et~al.}(2026)\citenamefont
  {Danchev}, \citenamefont {Dyer}, \citenamefont {Grau},\ and\ \citenamefont
  {Vazeille}}]{Danchev2026}%
  \BibitemOpen
  \bibfield  {author} {\bibinfo {author} {\bibfnamefont {V.}~\bibnamefont
  {Danchev}}, \bibinfo {author} {\bibfnamefont {A.}~\bibnamefont {Dyer}},
  \bibinfo {author} {\bibfnamefont {S.}~\bibnamefont {Grau}}, \ and\ \bibinfo
  {author} {\bibfnamefont {G.}~\bibnamefont {Vazeille}},\ }\href@noop {}
  {\bibfield  {journal} {\bibinfo  {journal} {arXiv e-prints}\ } (\bibinfo
  {year} {2026})},\ \Eprint {http://arxiv.org/abs/2605.08239}{2605.08239
  [physics.acc-ph]}\BibitemShut {NoStop}%
\bibitem [{\citenamefont {Bird}\ \emph {et~al.}(1995)\citenamefont {Bird} \emph
  {et~al.}}]{Bird1995}%
  \BibitemOpen
  \bibfield  {author} {\bibinfo {author} {\bibfnamefont {D.~J.}\ \bibnamefont
  {Bird}} \emph {et~al.},\ }\href {\doibase 10.1086/175344} {\bibfield
  {journal} {\bibinfo  {journal} {Astrophys. J.}\ }\textbf {\bibinfo {volume}
  {441}},\ \bibinfo {pages} {144} (\bibinfo {year} {1995})},\ \Eprint
  {http://arxiv.org/abs/astro-ph/9410067}{astro-ph/9410067}\BibitemShut
  {NoStop}%
\bibitem [{\citenamefont {Swordy}(2001)}]{Swordy2001}%
  \BibitemOpen
  \bibfield  {author} {\bibinfo {author} {\bibfnamefont {S.~P.}\ \bibnamefont
  {Swordy}},\ }\href@noop {} {\bibfield  {journal} {\bibinfo  {journal}
  {Space Sci. Rev.}\ }\textbf {\bibinfo {volume} {99}},\ \bibinfo {pages}
  {85} (\bibinfo {year} {2001})}\BibitemShut {NoStop}%
\bibitem [{\citenamefont {de~Angelis}\ and\ \citenamefont
  {Pimenta}(2018)}]{DeAngelis2018}%
  \BibitemOpen
  \bibfield  {author} {\bibinfo {author} {\bibfnamefont {A.}~\bibnamefont
  {de~Angelis}}\ and\ \bibinfo {author} {\bibfnamefont {M.}~\bibnamefont
  {Pimenta}},\ }\href@noop {} {\emph {\bibinfo {title} {Introduction to
  Particle and Astroparticle Physics: Multimessenger Astronomy and its Particle
  Physics Foundations}}}\ (\bibinfo  {publisher} {Springer},\ \bibinfo
  {address} {Berlin},\ \bibinfo {year} {2018})\BibitemShut {NoStop}%
\bibitem [{\citenamefont {Abraham}\ \emph {et~al.}(2004)\citenamefont {Abraham}
  \emph {et~al.}}]{Abraham2004}%
  \BibitemOpen
  \bibfield  {author} {\bibinfo {author} {\bibfnamefont {J.}~\bibnamefont
  {Abraham}} \emph {et~al.},\ }\href@noop {} {\bibfield  {journal} {\bibinfo
  {journal} {Nucl. Instrum. Methods Phys. Res. A}\ }\textbf
  {\bibinfo {volume} {523}},\ \bibinfo {pages} {50} (\bibinfo {year}
  {2004})}\BibitemShut {NoStop}%
\bibitem [{\citenamefont {Malthus}(1798)}]{Malthus1798}%
  \BibitemOpen
  \bibfield  {author} {\bibinfo {author} {\bibfnamefont {T.~R.}\ \bibnamefont
  {Malthus}},\ }\href@noop {} {\emph {\bibinfo {title} {An Essay on the
  Principle of Population}}}\ (\bibinfo  {publisher} {J. Johnson},\ \bibinfo
  {address} {London},\ \bibinfo {year} {1798})\BibitemShut {NoStop}%
\bibitem [{\citenamefont {Riordan}\ \emph {et~al.}(2002)\citenamefont
  {Riordan}, \citenamefont {Hoddeson},\ and\ \citenamefont {Kolb}}]{SSC1993}%
  \BibitemOpen
  \bibfield  {author} {\bibinfo {author} {\bibfnamefont {M.}~\bibnamefont
  {Riordan}}, \bibinfo {author} {\bibfnamefont {L.}~\bibnamefont {Hoddeson}}, \
  and\ \bibinfo {author} {\bibfnamefont {A.}~\bibnamefont {Kolb}},\ }\href@noop
  {} {\bibfield  {journal} {\bibinfo  {journal} {Hist. Stud. Phys. Biol. Sci.}\
  }\textbf {\bibinfo {volume} {32}},\ \bibinfo {pages} {319} (\bibinfo
  {year} {2002})}\BibitemShut {NoStop}%
\bibitem [{\citenamefont {Goldstone}(1961)}]{Goldstone1961}%
  \BibitemOpen
  \bibfield  {author} {\bibinfo {author} {\bibfnamefont {J.}~\bibnamefont
  {Goldstone}},\ }\href {\doibase 10.1007/BF02812722} {\bibfield  {journal}
  {\bibinfo  {journal} {Nuovo Cimento}\ }\textbf {\bibinfo {volume} {19}},\
  \bibinfo {pages} {154} (\bibinfo {year} {1961})}\BibitemShut {NoStop}%
\bibitem [{\citenamefont {Veneziano}(1968)}]{Veneziano1968}%
  \BibitemOpen
  \bibfield  {author} {\bibinfo {author} {\bibfnamefont {G.}~\bibnamefont
  {Veneziano}},\ }\href {\doibase 10.1007/BF02824451} {\bibfield  {journal}
  {\bibinfo  {journal} {Nuovo Cimento A}\ }\textbf {\bibinfo {volume}
  {57}},\ \bibinfo {pages} {190} (\bibinfo {year} {1968})}\BibitemShut
  {NoStop}%
\bibitem [{\citenamefont {Desrochers}\ \emph {et~al.}(2025)\citenamefont
  {Desrochers}, \citenamefont {Marchand},\ and\ \citenamefont
  {Stamp}}]{Desrochers2025}%
  \BibitemOpen
  \bibfield  {author} {\bibinfo {author} {\bibfnamefont {M.~J.}\ \bibnamefont
  {Desrochers}}, \bibinfo {author} {\bibfnamefont {D.}~\bibnamefont
  {Marchand}}, \ and\ \bibinfo {author} {\bibfnamefont {P.~C.~E.}\ \bibnamefont
  {Stamp}},\ }\href {\doibase 10.1073/pnas.2421273122} {\bibfield  {journal}
  {\bibinfo  {journal} {Proc. Natl. Acad. Sci.}\ }\textbf {\bibinfo {volume}
  {122}} (\bibinfo {year} {2025}),\ 10.1073/pnas.2421273122}\BibitemShut
  {NoStop}%
\bibitem [{\citenamefont {Amati}\ \emph {et~al.}(1989)\citenamefont {Amati},
  \citenamefont {Ciafaloni},\ and\ \citenamefont {Veneziano}}]{Amati1989}%
  \BibitemOpen
  \bibfield  {author} {\bibinfo {author} {\bibfnamefont {D.}~\bibnamefont
  {Amati}}, \bibinfo {author} {\bibfnamefont {M.}~\bibnamefont {Ciafaloni}}, \
  and\ \bibinfo {author} {\bibfnamefont {G.}~\bibnamefont {Veneziano}},\
  }\href@noop {} {\bibfield  {journal} {\bibinfo  {journal} {Phys. Lett.
  B}\ }\textbf {\bibinfo {volume} {216}},\ \bibinfo {pages} {41} (\bibinfo
  {year} {1989})}\BibitemShut {NoStop}%
\bibitem [{\citenamefont {Amati}\ \emph {et~al.}(1991)\citenamefont {Amati},
  \citenamefont {Ciafaloni},\ and\ \citenamefont
  {Veneziano}}]{AmatiVeneziano1991}%
  \BibitemOpen
  \bibfield  {author} {\bibinfo {author} {\bibfnamefont {D.}~\bibnamefont
  {Amati}}, \bibinfo {author} {\bibfnamefont {M.}~\bibnamefont {Ciafaloni}}, \
  and\ \bibinfo {author} {\bibfnamefont {G.}~\bibnamefont {Veneziano}},\
  }\href@noop {} {\bibfield  {journal} {\bibinfo  {journal} {Nucl. Phys.
  B}\ }\textbf {\bibinfo {volume} {360}},\ \bibinfo {pages} {237} (\bibinfo
  {year} {1991})}\BibitemShut {NoStop}%
\bibitem [{\citenamefont {Kempf}\ \emph {et~al.}(1995)\citenamefont {Kempf},
  \citenamefont {Mangano},\ and\ \citenamefont {Mann}}]{Kempf1995}%
  \BibitemOpen
  \bibfield  {author} {\bibinfo {author} {\bibfnamefont {A.}~\bibnamefont
  {Kempf}}, \bibinfo {author} {\bibfnamefont {G.}~\bibnamefont {Mangano}}, \
  and\ \bibinfo {author} {\bibfnamefont {R.~B.}\ \bibnamefont {Mann}},\
  }\href@noop {} {\bibfield  {journal} {\bibinfo  {journal} {Phys. Rev.
  D}\ }\textbf {\bibinfo {volume} {52}},\ \bibinfo {pages} {1108} (\bibinfo
  {year} {1995})}\BibitemShut {NoStop}%
\bibitem [{\citenamefont {Pikovski}\ \emph {et~al.}(2012)\citenamefont
  {Pikovski}, \citenamefont {Vanner}, \citenamefont {Aspelmeyer}, \citenamefont
  {Kim},\ and\ \citenamefont {Brukner}}]{Pikovski2012}%
  \BibitemOpen
  \bibfield  {author} {\bibinfo {author} {\bibfnamefont {I.}~\bibnamefont
  {Pikovski}}, \bibinfo {author} {\bibfnamefont {M.~R.}\ \bibnamefont
  {Vanner}}, \bibinfo {author} {\bibfnamefont {M.}~\bibnamefont {Aspelmeyer}},
  \bibinfo {author} {\bibfnamefont {M.}~\bibnamefont {Kim}}, \ and\ \bibinfo
  {author} {\bibfnamefont {{\v C}.}~\bibnamefont {Brukner}},\ }\href@noop {}
  {\bibfield  {journal} {\bibinfo  {journal} {Nat. Phys.}\ }\textbf
  {\bibinfo {volume} {8}},\ \bibinfo {pages} {393} (\bibinfo {year}
  {2012})}\BibitemShut {NoStop}%
\bibitem [{\citenamefont {Bawaj}\ \emph {et~al.}(2015)\citenamefont {Bawaj}
  \emph {et~al.}}]{Bawaj2015}%
  \BibitemOpen
  \bibfield  {author} {\bibinfo {author} {\bibfnamefont {M.}~\bibnamefont
  {Bawaj}} \emph {et~al.},\ }\href@noop {} {\bibfield  {journal} {\bibinfo
  {journal} {Nat. Commun.}\ }\textbf {\bibinfo {volume} {6}},\ \bibinfo
  {pages} {7503} (\bibinfo {year} {2015})}\BibitemShut {NoStop}%
\bibitem [{\citenamefont {Bushev}\ \emph {et~al.}(2019)\citenamefont {Bushev},
  \citenamefont {Bourhill}, \citenamefont {Goryachev}, \citenamefont
  {Kukharchyk}, \citenamefont {Ivanov}, \citenamefont {Galliou}, \citenamefont
  {Tobar},\ and\ \citenamefont {Danilishin}}]{Bushev2019}%
  \BibitemOpen
  \bibfield  {author} {\bibinfo {author} {\bibfnamefont {P.~A.}\ \bibnamefont
  {Bushev}}, \bibinfo {author} {\bibfnamefont {J.}~\bibnamefont {Bourhill}},
  \bibinfo {author} {\bibfnamefont {M.}~\bibnamefont {Goryachev}}, \bibinfo
  {author} {\bibfnamefont {N.}~\bibnamefont {Kukharchyk}}, \bibinfo {author}
  {\bibfnamefont {E.}~\bibnamefont {Ivanov}}, \bibinfo {author} {\bibfnamefont
  {S.}~\bibnamefont {Galliou}}, \bibinfo {author} {\bibfnamefont {M.~E.}\
  \bibnamefont {Tobar}}, \ and\ \bibinfo {author} {\bibfnamefont
  {S.}~\bibnamefont {Danilishin}},\ }\href@noop {} {\bibfield  {journal}
  {\bibinfo  {journal} {Phys. Rev. D}\ }\textbf {\bibinfo {volume}
  {100}},\ \bibinfo {pages} {066020} (\bibinfo {year} {2019})}\BibitemShut
  {NoStop}%
\bibitem [{\citenamefont {Bekenstein}(2012)}]{Bekenstein2012}%
  \BibitemOpen
  \bibfield  {author} {\bibinfo {author} {\bibfnamefont {J.~D.}\ \bibnamefont
  {Bekenstein}},\ }\href@noop {} {\bibfield  {journal} {\bibinfo  {journal}
  {Phys. Rev. D}\ }\textbf {\bibinfo {volume} {86}},\ \bibinfo {pages}
  {124040} (\bibinfo {year} {2012})}\BibitemShut {NoStop}%
\bibitem [{\citenamefont {Bekenstein}(2014)}]{Bekenstein2014}%
  \BibitemOpen
  \bibfield  {author} {\bibinfo {author} {\bibfnamefont {J.~D.}\ \bibnamefont
  {Bekenstein}},\ }\href@noop {} {\bibfield  {journal} {\bibinfo  {journal}
  {Found. Phys.}\ }\textbf {\bibinfo {volume} {44}},\ \bibinfo {pages}
  {452} (\bibinfo {year} {2014})}\BibitemShut {NoStop}%
\bibitem [{\citenamefont {Hogan}(2012)}]{Hogan2012}%
  \BibitemOpen
  \bibfield  {author} {\bibinfo {author} {\bibfnamefont {C.~J.}\ \bibnamefont
  {Hogan}},\ }\href@noop {} {\bibfield  {journal} {\bibinfo  {journal}
  {Phys. Rev. D}\ }\textbf {\bibinfo {volume} {85}},\ \bibinfo {pages}
  {064007} (\bibinfo {year} {2012})}\BibitemShut {NoStop}%
\bibitem [{\citenamefont {Das}\ and\ \citenamefont
  {Vagenas}(2008)}]{DasVagenas2008}%
  \BibitemOpen
  \bibfield  {author} {\bibinfo {author} {\bibfnamefont {S.}~\bibnamefont
  {Das}}\ and\ \bibinfo {author} {\bibfnamefont {E.~C.}\ \bibnamefont
  {Vagenas}},\ }\href@noop {} {\bibfield  {journal} {\bibinfo  {journal}
  {Phys. Rev. Lett.}\ }\textbf {\bibinfo {volume} {101}},\ \bibinfo {pages}
  {221301} (\bibinfo {year} {2008})}\BibitemShut {NoStop}%
\bibitem [{\citenamefont {Howl}\ \emph {et~al.}(2023)\citenamefont {Howl},
  \citenamefont {Cooper},\ and\ \citenamefont {Hackerm\"uller}}]{Howl2023}%
  \BibitemOpen
  \bibfield  {author} {\bibinfo {author} {\bibfnamefont {R.}~\bibnamefont
  {Howl}}, \bibinfo {author} {\bibfnamefont {N.}~\bibnamefont {Cooper}}, \ and\
  \bibinfo {author} {\bibfnamefont {L.}~\bibnamefont {Hackerm\"uller}},\
  }\href@noop {} {\bibfield  {journal} {\bibinfo  {journal} {arXiv e-prints}\ }
  (\bibinfo {year} {2023})},\ \Eprint
  {http://arxiv.org/abs/2304.00734}{2304.00734 [quant-ph]}\BibitemShut
  {NoStop}%
\bibitem [{\citenamefont {Aziz}\ and\ \citenamefont {Howl}(2025)}]{Aziz2025}%
  \BibitemOpen
  \bibfield  {author} {\bibinfo {author} {\bibfnamefont {J.}~\bibnamefont
  {Aziz}}\ and\ \bibinfo {author} {\bibfnamefont {R.}~\bibnamefont {Howl}},\
  }\href@noop {} {\bibfield  {journal} {\bibinfo  {journal} {Nature}\ }\textbf
  {\bibinfo {volume} {646}},\ \bibinfo {pages} {813} (\bibinfo {year}
  {2025})},\ \Eprint {http://arxiv.org/abs/2510.19714}{2510.19714}\BibitemShut
  {NoStop}%
\bibitem [{\citenamefont {Unruh}(1981)}]{Unruh1981}%
  \BibitemOpen
  \bibfield  {author} {\bibinfo {author} {\bibfnamefont {W.~G.}\ \bibnamefont
  {Unruh}},\ }\href@noop {} {\bibfield  {journal} {\bibinfo  {journal}
  {Phys. Rev. Lett.}\ }\textbf {\bibinfo {volume} {46}},\ \bibinfo {pages}
  {1351} (\bibinfo {year} {1981})}\BibitemShut {NoStop}%
\bibitem [{\citenamefont {Barcel\'o}\ \emph {et~al.}(2011)\citenamefont
  {Barcel\'o}, \citenamefont {Liberati},\ and\ \citenamefont
  {Visser}}]{Barcelo2011}%
  \BibitemOpen
  \bibfield  {author} {\bibinfo {author} {\bibfnamefont {C.}~\bibnamefont
  {Barcel\'o}}, \bibinfo {author} {\bibfnamefont {S.}~\bibnamefont {Liberati}},
  \ and\ \bibinfo {author} {\bibfnamefont {M.}~\bibnamefont {Visser}},\
  }\href@noop {} {\bibfield  {journal} {\bibinfo  {journal} {Living Rev.
  Relativ.}\ }\textbf {\bibinfo {volume} {14}},\ \bibinfo {pages} {3}
  (\bibinfo {year} {2011})},\ \Eprint
  {http://arxiv.org/abs/gr-qc/0505065}{gr-qc/0505065}\BibitemShut {NoStop}%
\bibitem [{\citenamefont {Balbinot}\ \emph {et~al.}(2005)\citenamefont
  {Balbinot}, \citenamefont {Fagnocchi}, \citenamefont {Fabbri},\ and\
  \citenamefont {Procopio}}]{Balbinot2005}%
  \BibitemOpen
  \bibfield  {author} {\bibinfo {author} {\bibfnamefont {R.}~\bibnamefont
  {Balbinot}}, \bibinfo {author} {\bibfnamefont {A.}~\bibnamefont {Fagnocchi}},
  \bibinfo {author} {\bibfnamefont {A.}~\bibnamefont {Fabbri}}, \ and\ \bibinfo
  {author} {\bibfnamefont {G.~P.}\ \bibnamefont {Procopio}},\ }\href@noop {}
  {\bibfield  {journal} {\bibinfo  {journal} {Phys. Rev. Lett.}\
  }\textbf {\bibinfo {volume} {94}},\ \bibinfo {pages} {161302} (\bibinfo
  {year} {2005})},\ \Eprint
  {http://arxiv.org/abs/gr-qc/0405096}{gr-qc/0405096}\BibitemShut {NoStop}%
\bibitem [{\citenamefont {Carusotto}\ \emph {et~al.}(2008)\citenamefont
  {Carusotto}, \citenamefont {Fagnocchi}, \citenamefont {Recati}, \citenamefont
  {Balbinot},\ and\ \citenamefont {Fabbri}}]{Carusotto2008}%
  \BibitemOpen
  \bibfield  {author} {\bibinfo {author} {\bibfnamefont {I.}~\bibnamefont
  {Carusotto}}, \bibinfo {author} {\bibfnamefont {A.}~\bibnamefont
  {Fagnocchi}}, \bibinfo {author} {\bibfnamefont {A.}~\bibnamefont {Recati}},
  \bibinfo {author} {\bibfnamefont {R.}~\bibnamefont {Balbinot}}, \ and\
  \bibinfo {author} {\bibfnamefont {A.}~\bibnamefont {Fabbri}},\ }\href@noop {}
  {\bibfield  {journal} {\bibinfo  {journal} {New J. Phys.}\ }\textbf
  {\bibinfo {volume} {10}},\ \bibinfo {pages} {103001} (\bibinfo {year}
  {2008})},\ \Eprint {http://arxiv.org/abs/0803.0507}{0803.0507}\BibitemShut
  {NoStop}%
\bibitem [{\citenamefont {Balbinot}\ \emph {et~al.}(2008)\citenamefont
  {Balbinot}, \citenamefont {Fabbri}, \citenamefont {Fagnocchi}, \citenamefont
  {Recati},\ and\ \citenamefont {Carusotto}}]{Balbinot2008}%
  \BibitemOpen
  \bibfield  {author} {\bibinfo {author} {\bibfnamefont {R.}~\bibnamefont
  {Balbinot}}, \bibinfo {author} {\bibfnamefont {A.}~\bibnamefont {Fabbri}},
  \bibinfo {author} {\bibfnamefont {A.}~\bibnamefont {Fagnocchi}}, \bibinfo
  {author} {\bibfnamefont {A.}~\bibnamefont {Recati}}, \ and\ \bibinfo {author}
  {\bibfnamefont {I.}~\bibnamefont {Carusotto}},\ }\href@noop {} {\bibfield
  {journal} {\bibinfo  {journal} {Phys. Rev. A}\ }\textbf {\bibinfo {volume}
  {78}},\ \bibinfo {pages} {021603} (\bibinfo {year} {2008})},\ \Eprint
  {http://arxiv.org/abs/0711.4520}{0711.4520}\BibitemShut {NoStop}%
\bibitem [{\citenamefont {Lahav}\ \emph {et~al.}(2010)\citenamefont {Lahav},
  \citenamefont {Itah}, \citenamefont {Blumkin}, \citenamefont {Gordon},\ and\
  \citenamefont {Steinhauer}}]{Lahav2010}%
  \BibitemOpen
  \bibfield  {author} {\bibinfo {author} {\bibfnamefont {O.}~\bibnamefont
  {Lahav}}, \bibinfo {author} {\bibfnamefont {A.}~\bibnamefont {Itah}},
  \bibinfo {author} {\bibfnamefont {A.}~\bibnamefont {Blumkin}}, \bibinfo
  {author} {\bibfnamefont {C.}~\bibnamefont {Gordon}}, \ and\ \bibinfo {author}
  {\bibfnamefont {J.}~\bibnamefont {Steinhauer}},\ }\href@noop {} {\bibfield
  {journal} {\bibinfo  {journal} {Phys. Rev. Lett.}\ }\textbf {\bibinfo
  {volume} {105}},\ \bibinfo {pages} {240401} (\bibinfo {year} {2010})},\
  \Eprint {http://arxiv.org/abs/0906.1337}{0906.1337}\BibitemShut {NoStop}%
\bibitem [{\citenamefont {Steinhauer}(2014)}]{Steinhauer2014}%
  \BibitemOpen
  \bibfield  {author} {\bibinfo {author} {\bibfnamefont {J.}~\bibnamefont
  {Steinhauer}},\ }\href@noop {} {\bibfield  {journal} {\bibinfo  {journal}
  {Nat. Phys.}\ }\textbf {\bibinfo {volume} {10}},\ \bibinfo {pages} {864}
  (\bibinfo {year} {2014})},\ \Eprint
  {http://arxiv.org/abs/1409.6550}{1409.6550}\BibitemShut {NoStop}%
\bibitem [{\citenamefont {Steinhauer}(2016)}]{Steinhauer2016}%
  \BibitemOpen
  \bibfield  {author} {\bibinfo {author} {\bibfnamefont {J.}~\bibnamefont
  {Steinhauer}},\ }\href@noop {} {\bibfield  {journal} {\bibinfo  {journal}
  {Nat. Phys.}\ }\textbf {\bibinfo {volume} {12}},\ \bibinfo {pages} {959}
  (\bibinfo {year} {2016})}\BibitemShut {NoStop}%
\bibitem [{\citenamefont {Carusotto}\ and\ \citenamefont
  {Balbinot}(2016)}]{Carusotto2016}%
  \BibitemOpen
  \bibfield  {author} {\bibinfo {author} {\bibfnamefont {I.}~\bibnamefont
  {Carusotto}}\ and\ \bibinfo {author} {\bibfnamefont {R.}~\bibnamefont
  {Balbinot}},\ }\href@noop {} {\bibfield  {journal} {\bibinfo  {journal}
  {Nat. Phys.}\ }\textbf {\bibinfo {volume} {12}},\ \bibinfo {pages} {897}
  (\bibinfo {year} {2016})}\BibitemShut {NoStop}%
\bibitem [{\citenamefont {Mu\~noz~de Nova}\ \emph {et~al.}(2019)\citenamefont
  {Mu\~noz~de Nova}, \citenamefont {Golubkov}, \citenamefont {Kolobov},\ and\
  \citenamefont {Steinhauer}}]{MunozDeNova2019}%
  \BibitemOpen
  \bibfield  {author} {\bibinfo {author} {\bibfnamefont {J.~R.}\ \bibnamefont
  {Mu\~noz~de Nova}}, \bibinfo {author} {\bibfnamefont {K.}~\bibnamefont
  {Golubkov}}, \bibinfo {author} {\bibfnamefont {V.~I.}\ \bibnamefont
  {Kolobov}}, \ and\ \bibinfo {author} {\bibfnamefont {J.}~\bibnamefont
  {Steinhauer}},\ }\href@noop {} {\bibfield  {journal} {\bibinfo  {journal}
  {Nature}\ }\textbf {\bibinfo {volume} {569}},\ \bibinfo {pages} {688}
  (\bibinfo {year} {2019})}\BibitemShut {NoStop}%
\bibitem [{\citenamefont {Kolobov}\ \emph {et~al.}(2021)\citenamefont
  {Kolobov}, \citenamefont {Golubkov}, \citenamefont {Mu\~noz~de Nova},\ and\
  \citenamefont {Steinhauer}}]{Kolobov2021}%
  \BibitemOpen
  \bibfield  {author} {\bibinfo {author} {\bibfnamefont {V.~I.}\ \bibnamefont
  {Kolobov}}, \bibinfo {author} {\bibfnamefont {K.}~\bibnamefont {Golubkov}},
  \bibinfo {author} {\bibfnamefont {J.~R.}\ \bibnamefont {Mu\~noz~de Nova}}, \
  and\ \bibinfo {author} {\bibfnamefont {J.}~\bibnamefont {Steinhauer}},\
  }\href@noop {} {\bibfield  {journal} {\bibinfo  {journal} {Nat. Phys.}\
  }\textbf {\bibinfo {volume} {17}},\ \bibinfo {pages} {362} (\bibinfo {year}
  {2021})}\BibitemShut {NoStop}%
\bibitem [{\citenamefont {Horstmann}\ \emph {et~al.}(2011)\citenamefont
  {Horstmann}, \citenamefont {Sch\"utzhold}, \citenamefont {Reznik},
  \citenamefont {Fagnocchi},\ and\ \citenamefont {Cirac}}]{Horstmann2011}%
  \BibitemOpen
  \bibfield  {author} {\bibinfo {author} {\bibfnamefont {B.}~\bibnamefont
  {Horstmann}}, \bibinfo {author} {\bibfnamefont {R.}~\bibnamefont
  {Sch\"utzhold}}, \bibinfo {author} {\bibfnamefont {B.}~\bibnamefont
  {Reznik}}, \bibinfo {author} {\bibfnamefont {S.}~\bibnamefont {Fagnocchi}}, \
  and\ \bibinfo {author} {\bibfnamefont {J.~I.}\ \bibnamefont {Cirac}},\
  }\href@noop {} {\bibfield  {journal} {\bibinfo  {journal} {New J. Phys.}\
  }\textbf {\bibinfo {volume} {13}},\ \bibinfo {pages} {045008}
  (\bibinfo {year} {2011})},\ \Eprint
  {http://arxiv.org/abs/1008.3494}{1008.3494}\BibitemShut {NoStop}%
\bibitem [{\citenamefont {Volovik}(1999)}]{Volovik1999}%
  \BibitemOpen
  \bibfield  {author} {\bibinfo {author} {\bibfnamefont {G.~E.}\ \bibnamefont
  {Volovik}},\ }\href@noop {} {\bibfield  {journal} {\bibinfo  {journal}
  {JETP Lett.}\ }\textbf {\bibinfo {volume} {69}},\ \bibinfo {pages} {705}
  (\bibinfo {year} {1999})},\ \Eprint
  {http://arxiv.org/abs/gr-qc/9901077}{gr-qc/9901077}\BibitemShut {NoStop}%
\bibitem [{\citenamefont {Volovik}(2001)}]{Volovik2001}%
  \BibitemOpen
  \bibfield  {author} {\bibinfo {author} {\bibfnamefont {G.~E.}\ \bibnamefont
  {Volovik}},\ }\href@noop {} {\bibfield  {journal} {\bibinfo  {journal}
  {Pis'ma Zh. Eksp. Teor. Fiz.}\ }\textbf {\bibinfo {volume} {73}},\ \bibinfo
  {pages} {721} (\bibinfo {year} {2001})},\ \Eprint
  {http://arxiv.org/abs/gr-qc/0104088}{gr-qc/0104088}\BibitemShut {NoStop}%
\bibitem [{\citenamefont {Volovik}(2021)}]{Volovik2021}%
  \BibitemOpen
  \bibfield  {author} {\bibinfo {author} {\bibfnamefont {G.~E.}\ \bibnamefont
  {Volovik}},\ }\href@noop {} {\bibfield  {journal} {\bibinfo  {journal}
  {Mod. Phys. Lett. A}\ }\textbf {\bibinfo {volume} {36}},\ \bibinfo
  {pages} {2150177} (\bibinfo {year} {2021})},\ \Eprint
  {http://arxiv.org/abs/2107.11193}{2107.11193}\BibitemShut {NoStop}%
\bibitem [{\citenamefont {Belgiorno}\ \emph {et~al.}(2010)\citenamefont
  {Belgiorno}, \citenamefont {Cacciatori}, \citenamefont {Clerici},
  \citenamefont {Gorini}, \citenamefont {Ortenzi} \emph
  {et~al.}}]{Belgiorno2010}%
  \BibitemOpen
  \bibfield  {author} {\bibinfo {author} {\bibfnamefont {F.}~\bibnamefont
  {Belgiorno}}, \bibinfo {author} {\bibfnamefont {S.~L.}\ \bibnamefont
  {Cacciatori}}, \bibinfo {author} {\bibfnamefont {M.}~\bibnamefont {Clerici}},
  \bibinfo {author} {\bibfnamefont {V.}~\bibnamefont {Gorini}}, \bibinfo
  {author} {\bibfnamefont {G.}~\bibnamefont {Ortenzi}},  \emph {et~al.},\
  }\href@noop {} {\bibfield  {journal} {\bibinfo  {journal} {Phys. Rev.
  Lett.}\ }\textbf {\bibinfo {volume} {105}},\ \bibinfo {pages} {203901}
  (\bibinfo {year} {2010})}\BibitemShut {NoStop}%
\bibitem [{\citenamefont {Sch\"utzhold}\ and\ \citenamefont
  {Unruh}(2011)}]{Schutzhold2011}%
  \BibitemOpen
  \bibfield  {author} {\bibinfo {author} {\bibfnamefont {R.}~\bibnamefont
  {Sch\"utzhold}}\ and\ \bibinfo {author} {\bibfnamefont {W.~G.}\ \bibnamefont
  {Unruh}},\ }\href@noop {} {\bibfield  {journal} {\bibinfo  {journal}
  {Phys. Rev. Lett.}\ }\textbf {\bibinfo {volume} {107}},\ \bibinfo {pages}
  {149401} (\bibinfo {year} {2011})},\ \Eprint
  {http://arxiv.org/abs/1012.2686}{1012.2686}\BibitemShut {NoStop}%


\bibitem [{\citenamefont {Belgiorno}\ \emph {et~al.}(2011)\citenamefont
  {Belgiorno}, \citenamefont {Cacciatori}, \citenamefont {Clerici},
  \citenamefont {Gorini}, \citenamefont {Ortenzi} \emph
  {et~al.}}]{Belgiorno2011}%
  \BibitemOpen
  \bibfield  {author} {\bibinfo {author} {\bibfnamefont {F.}~\bibnamefont
  {Belgiorno}}, \bibinfo {author} {\bibfnamefont {S.~L.}\ \bibnamefont
  {Cacciatori}}, \bibinfo {author} {\bibfnamefont {M.}~\bibnamefont {Clerici}},
  \bibinfo {author} {\bibfnamefont {V.}~\bibnamefont {Gorini}}, \bibinfo
  {author} {\bibfnamefont {G.}~\bibnamefont {Ortenzi}},  \emph {et~al.},\
  }\href@noop {} {\bibfield  {journal} {\bibinfo  {journal} {Phys. Rev.
  Lett.}\ }\textbf {\bibinfo {volume} {107}},\ \bibinfo {pages} {149402}
  (\bibinfo {year} {2011})},\ \Eprint
  {http://arxiv.org/abs/1012.5062}{1012.5062}\BibitemShut {NoStop}%
\bibitem [{\citenamefont {Torres}\ \emph {et~al.}(2017)\citenamefont {Torres},
  \citenamefont {Patrick}, \citenamefont {Coutant}, \citenamefont {Richartz},\
  and\ \citenamefont {Tedford}}]{Torres2017}%
  \BibitemOpen
  \bibfield  {author} {\bibinfo {author} {\bibfnamefont {T.}~\bibnamefont
  {Torres}}, \bibinfo {author} {\bibfnamefont {S.}~\bibnamefont {Patrick}},
  \bibinfo {author} {\bibfnamefont {A.}~\bibnamefont {Coutant}}, \bibinfo
  {author} {\bibfnamefont {M.}~\bibnamefont {Richartz}}, \ and\ \bibinfo
  {author} {\bibfnamefont {E.~W.}\ \bibnamefont {Tedford}},\ }\href@noop {}
  {\bibfield  {journal} {\bibinfo  {journal} {Nat. Phys.}\ }\textbf
  {\bibinfo {volume} {13}},\ \bibinfo {pages} {833} (\bibinfo {year}
  {2017})}\BibitemShut {NoStop}%
\bibitem [{\citenamefont {Misner}(1972)}]{Misner1972}%
  \BibitemOpen
  \bibfield  {author} {\bibinfo {author} {\bibfnamefont {C.~W.}\ \bibnamefont
  {Misner}},\ }\href@noop {} {\bibfield  {journal} {\bibinfo  {journal}
  {Phys. Rev. Lett.}\ }\textbf {\bibinfo {volume} {28}},\ \bibinfo {pages}
  {994} (\bibinfo {year} {1972})}\BibitemShut {NoStop}%
\bibitem [{\citenamefont {Starobinskii}(1973)}]{Starobinsky1973}%
  \BibitemOpen
  \bibfield  {author} {\bibinfo {author} {\bibfnamefont {A.~A.}\ \bibnamefont
  {Starobinskii}},\ }\href@noop {} {\bibfield  {journal} {\bibinfo  {journal}
  {Sov. Phys. JETP}\ }\textbf {\bibinfo {volume} {37}},\ \bibinfo {pages}
  {28} (\bibinfo {year} {1973})}\BibitemShut {NoStop}%
\bibitem [{\citenamefont {Unruh}(1974)}]{Unruh1974}%
  \BibitemOpen
  \bibfield  {author} {\bibinfo {author} {\bibfnamefont {W.~G.}\ \bibnamefont
  {Unruh}},\ }\href@noop {} {\bibfield  {journal} {\bibinfo  {journal}
  {Phys. Rev. D}\ }\textbf {\bibinfo {volume} {10}},\ \bibinfo {pages}
  {3194} (\bibinfo {year} {1974})}\BibitemShut {NoStop}%
\bibitem [{\citenamefont {Press}\ and\ \citenamefont
  {Teukolsky}(1972)}]{PressTeukolsky1972}%
  \BibitemOpen
  \bibfield  {author} {\bibinfo {author} {\bibfnamefont {W.~H.}\ \bibnamefont
  {Press}}\ and\ \bibinfo {author} {\bibfnamefont {S.~A.}\ \bibnamefont
  {Teukolsky}},\ }\href@noop {} {\bibfield  {journal} {\bibinfo  {journal}
  {Nature}\ }\textbf {\bibinfo {volume} {238}},\ \bibinfo {pages} {211}
  (\bibinfo {year} {1972})}\BibitemShut {NoStop}%
\bibitem [{\citenamefont {Zel'dovich}(1971)}]{Zeldovich1971}%
  \BibitemOpen
  \bibfield  {author} {\bibinfo {author} {\bibfnamefont {Y.~B.}\ \bibnamefont
  {Zel'dovich}},\ }\href@noop {} {\bibfield  {journal} {\bibinfo  {journal}
  {JETP Lett.}\ }\textbf {\bibinfo {volume} {14}},\ \bibinfo {pages} {180}
  (\bibinfo {year} {1971})}\BibitemShut {NoStop}%
\bibitem [{\citenamefont {Zel'dovich}(1972)}]{Zeldovich1972}%
  \BibitemOpen
  \bibfield  {author} {\bibinfo {author} {\bibfnamefont {Y.~B.}\ \bibnamefont
  {Zel'dovich}},\ }\href@noop {} {\bibfield  {journal} {\bibinfo  {journal}
  {Sov. Phys. JETP}\ }\textbf {\bibinfo {volume} {35}},\ \bibinfo {pages}
  {1085} (\bibinfo {year} {1972})}\BibitemShut {NoStop}%
\bibitem [{\citenamefont {Volovik}(2006)}]{Volovik2006}%
  \BibitemOpen
  \bibfield  {author} {\bibinfo {author} {\bibfnamefont {G.~E.}\ \bibnamefont
  {Volovik}},\ }\href@noop {} {\emph {\bibinfo {title} {The Universe in a
  Helium Droplet}}},\ International Series of Monographs on Physics\ (\bibinfo
  {publisher} {Clarendon Press},\ \bibinfo {address} {Oxford},\ \bibinfo {year}
  {2006})\BibitemShut {NoStop}%
\bibitem [{\citenamefont {Cabrera}(1982)}]{Cabrera1982}%
  \BibitemOpen
  \bibfield  {author} {\bibinfo {author} {\bibfnamefont {B.}~\bibnamefont
  {Cabrera}},\ }\href {\doibase 10.1103/PhysRevLett.48.1378} {\bibfield
  {journal} {\bibinfo  {journal} {Phys. Rev. Lett.}\ }\textbf {\bibinfo
  {volume} {48}},\ \bibinfo {pages} {1378} (\bibinfo {year}
  {1982})}\BibitemShut {NoStop}%
\bibitem [{\citenamefont {Dirac}(1931)}]{Dirac1931}%
  \BibitemOpen
  \bibfield  {author} {\bibinfo {author} {\bibfnamefont {P.~A.~M.}\
  \bibnamefont {Dirac}},\ }\href@noop {} {\bibfield  {journal} {\bibinfo
  {journal} {Proc. Roy. Soc. London A}\ }\textbf {\bibinfo {volume} {133}},\
  \bibinfo {pages} {60} (\bibinfo {year} {1931})}\BibitemShut {NoStop}%
\bibitem [{\citenamefont {Carroll}(2019)}]{Carroll2019}%
  \BibitemOpen
  \bibfield  {author} {\bibinfo {author} {\bibfnamefont {S.~M.}\ \bibnamefont
  {Carroll}},\ }\href {\doibase 10.1017/9781108770385} {\emph {\bibinfo {title}
  {Spacetime and Geometry: An Introduction to General Relativity}}}\ (\bibinfo
  {publisher} {Cambridge University Press},\ \bibinfo {year}
  {2019})\BibitemShut {NoStop}%
\bibitem [{\citenamefont {Maldacena}(1998)}]{Maldacena1998}%
  \BibitemOpen
  \bibfield  {author} {\bibinfo {author} {\bibfnamefont {J.}~\bibnamefont
  {Maldacena}},\ }\href@noop {} {\bibfield  {journal} {\bibinfo  {journal}
  {Adv. Theor. Math. Phys.}\ }\textbf {\bibinfo {volume} {2}},\ \bibinfo
  {pages} {231} (\bibinfo {year} {1998})},\ \Eprint
  {http://arxiv.org/abs/hep-th/9711200}{hep-th/9711200}\BibitemShut {NoStop}%
\bibitem [{\citenamefont {Witten}(1998)}]{Witten1998}%
  \BibitemOpen
  \bibfield  {author} {\bibinfo {author} {\bibfnamefont {E.}~\bibnamefont
  {Witten}},\ }\href@noop {} {\bibfield  {journal} {\bibinfo  {journal}
  {Adv. Theor. Math. Phys.}\ }\textbf {\bibinfo {volume} {2}},\ \bibinfo
  {pages} {253} (\bibinfo {year} {1998})},\ \Eprint
  {http://arxiv.org/abs/hep-th/9802150}{hep-th/9802150}\BibitemShut {NoStop}%
\bibitem [{\citenamefont {Gubser}\ \emph {et~al.}(1998)\citenamefont
  {Gubser}, \citenamefont {Klebanov},\ and\ \citenamefont
  {Polyakov}}]{Gubser1998}%
  \BibitemOpen
  \bibfield  {author} {\bibinfo {author} {\bibfnamefont {S.~S.}\ \bibnamefont
  {Gubser}}, \bibinfo {author} {\bibfnamefont {I.~R.}\ \bibnamefont
  {Klebanov}}, \ and\ \bibinfo {author} {\bibfnamefont {A.~M.}\ \bibnamefont
  {Polyakov}},\ }\href@noop {} {\bibfield  {journal} {\bibinfo  {journal}
  {Phys. Lett. B}\ }\textbf {\bibinfo {volume} {428}},\ \bibinfo {pages}
  {105} (\bibinfo {year} {1998})},\ \Eprint
  {http://arxiv.org/abs/hep-th/9802109}{hep-th/9802109}\BibitemShut {NoStop}%
\bibitem [{\citenamefont {Kovtun}\ \emph {et~al.}(2005)\citenamefont {Kovtun},
  \citenamefont {Son},\ and\ \citenamefont {Starinets}}]{Kovtun2005}%
  \BibitemOpen
  \bibfield  {author} {\bibinfo {author} {\bibfnamefont {P.}~\bibnamefont
  {Kovtun}}, \bibinfo {author} {\bibfnamefont {D.~T.}\ \bibnamefont {Son}}, \
  and\ \bibinfo {author} {\bibfnamefont {A.~O.}\ \bibnamefont {Starinets}},\
  }\href@noop {} {\bibfield  {journal} {\bibinfo  {journal} {Phys. Rev.
  Lett.}\ }\textbf {\bibinfo {volume} {94}},\ \bibinfo {pages} {111601}
  (\bibinfo {year} {2005})},\ \Eprint
  {http://arxiv.org/abs/hep-th/0405231}{hep-th/0405231}\BibitemShut {NoStop}%
\bibitem [{\citenamefont {Luzum}\ and\ \citenamefont
  {Romatschke}(2008)}]{Luzum2008}%
  \BibitemOpen
  \bibfield  {author} {\bibinfo {author} {\bibfnamefont {M.}~\bibnamefont
  {Luzum}}\ and\ \bibinfo {author} {\bibfnamefont {P.}~\bibnamefont
  {Romatschke}},\ }\href@noop {} {\bibfield  {journal} {\bibinfo  {journal}
  {Phys. Rev. C}\ }\textbf {\bibinfo {volume} {78}},\ \bibinfo {pages}
  {034915} (\bibinfo {year} {2008})},\ \Eprint
  {http://arxiv.org/abs/0804.4015}{0804.4015}\BibitemShut {NoStop}%
\bibitem [{\citenamefont {McLerran}(2007)}]{McLerran2007}%
  \BibitemOpen
  \bibfield  {author} {\bibinfo {author} {\bibfnamefont {L.}~\bibnamefont
  {McLerran}},\ }\href@noop {} {\bibfield  {journal} {\bibinfo  {journal}
  {J. Phys. G}\ }\textbf {\bibinfo {volume} {34}},\ \bibinfo {pages}
  {S583} (\bibinfo {year} {2007})},\ \Eprint
  {http://arxiv.org/abs/hep-ph/0702004}{hep-ph/0702004}\BibitemShut {NoStop}%
\bibitem [{\citenamefont {Anderson}(2013)}]{Anderson2013}%
  \BibitemOpen
  \bibfield  {author} {\bibinfo {author} {\bibfnamefont {P.~W.}\ \bibnamefont
  {Anderson}},\ }\href@noop {} {\bibfield  {journal} {\bibinfo  {journal}
  {Phys. Today}\ }\textbf {\bibinfo {volume} {66}},\ \bibinfo {pages} {9}
  (\bibinfo {year} {2013})}\BibitemShut {NoStop}%
\bibitem [{\citenamefont {Nicolini}(2010)}]{Nicolini2010}%
  \BibitemOpen
  \bibfield  {author} {\bibinfo {author} {\bibfnamefont {P.}~\bibnamefont
  {Nicolini}},\ }\href@noop {} {\bibfield  {journal} {\bibinfo  {journal}
  {Phys. Rev. D}\ }\textbf {\bibinfo {volume} {82}},\ \bibinfo {pages}
  {044030} (\bibinfo {year} {2010})},\ \Eprint
  {http://arxiv.org/abs/1005.2996}{1005.2996}\BibitemShut {NoStop}%
\bibitem [{\citenamefont {Verlinde}(2011)}]{Verlinde2011}%
  \BibitemOpen
  \bibfield  {author} {\bibinfo {author} {\bibfnamefont {E.~P.}\ \bibnamefont
  {Verlinde}},\ }\href@noop {} {\bibfield  {journal} {\bibinfo  {journal}
  {JHEP}\ }\textbf {\bibinfo {volume} {04}},\ \bibinfo {pages} {029}
  (\bibinfo {year} {2011})},\ \Eprint
  {http://arxiv.org/abs/1001.0785}{1001.0785}\BibitemShut {NoStop}%
\bibitem [{\citenamefont {Adler}(2010)}]{Adler2010}%
  \BibitemOpen
  \bibfield  {author} {\bibinfo {author} {\bibfnamefont {R.~J.}\ \bibnamefont
  {Adler}},\ }\href@noop {} {\emph {\bibinfo {title} {Six Easy Roads to the
  {Planck} Scale}}}\ (\bibinfo  {publisher} {IOP Publishing},\ \bibinfo
  {address} {Bristol},\ \bibinfo {year} {2010})\BibitemShut {NoStop}%
\bibitem [{\citenamefont {Padmanabhan}(1997)}]{Padmanabhan1997}%
  \BibitemOpen
  \bibfield  {author} {\bibinfo {author} {\bibfnamefont {T.}~\bibnamefont
  {Padmanabhan}},\ }\href@noop {} {\bibfield  {journal} {\bibinfo  {journal}
  {Phys. Rev. Lett.}\ }\textbf {\bibinfo {volume} {78}},\ \bibinfo {pages}
  {1854} (\bibinfo {year} {1997})}\BibitemShut {NoStop}%
\bibitem [{\citenamefont {Maggiore}(1993)}]{Maggiore1993}%
  \BibitemOpen
  \bibfield  {author} {\bibinfo {author} {\bibfnamefont {M.}~\bibnamefont
  {Maggiore}},\ }\href@noop {} {\bibfield  {journal} {\bibinfo  {journal}
  {Phys. Lett. B}\ }\textbf {\bibinfo {volume} {304}},\ \bibinfo {pages}
  {65} (\bibinfo {year} {1993})}\BibitemShut {NoStop}%
\bibitem [{\citenamefont {Modesto}\ and\ \citenamefont
  {Nicolini}(2010)}]{ModestoNicolini2010}%
  \BibitemOpen
  \bibfield  {author} {\bibinfo {author} {\bibfnamefont {L.}~\bibnamefont
  {Modesto}}\ and\ \bibinfo {author} {\bibfnamefont {P.}~\bibnamefont
  {Nicolini}},\ }\href {\doibase 10.1103/PhysRevD.81.104040} {\bibfield
  {journal} {\bibinfo  {journal} {Phys. Rev. D}\ }\textbf {\bibinfo {volume}
  {81}},\ \bibinfo {pages} {104040} (\bibinfo {year} {2010})},\ \Eprint
  {http://arxiv.org/abs/0912.0220}{arXiv:0912.0220 [hep-th]}\BibitemShut
  {NoStop}%
\bibitem [{\citenamefont {Seiberg}\ and\ \citenamefont
  {Witten}(1999)}]{SeibergWitten1999}%
  \BibitemOpen
  \bibfield  {author} {\bibinfo {author} {\bibfnamefont {N.}~\bibnamefont
  {Seiberg}}\ and\ \bibinfo {author} {\bibfnamefont {E.}~\bibnamefont
  {Witten}},\ }\href@noop {} {\bibfield  {journal} {\bibinfo  {journal}
  {JHEP}\ }\textbf {\bibinfo {volume} {1999}},\ \bibinfo {pages} {032}
  (\bibinfo {year} {1999})},\ \Eprint
  {http://arxiv.org/abs/hep-th/9908142}{hep-th/9908142}\BibitemShut {NoStop}%
\bibitem [{\citenamefont {Ambj{\o}rn}\ \emph {et~al.}(2005)\citenamefont
  {Ambj{\o}rn}, \citenamefont {Jurkiewicz},\ and\ \citenamefont
  {Loll}}]{Ambjorn2005}%
  \BibitemOpen
  \bibfield  {author} {\bibinfo {author} {\bibfnamefont {J.}~\bibnamefont
  {Ambj{\o}rn}}, \bibinfo {author} {\bibfnamefont {J.}~\bibnamefont
  {Jurkiewicz}}, \ and\ \bibinfo {author} {\bibfnamefont {R.}~\bibnamefont
  {Loll}},\ }\href@noop {} {\bibfield  {journal} {\bibinfo  {journal}
  {Phys. Rev. Lett.}\ }\textbf {\bibinfo {volume} {95}},\ \bibinfo {pages}
  {171301} (\bibinfo {year} {2005})},\ \Eprint
  {http://arxiv.org/abs/hep-th/0404156}{hep-th/0404156}\BibitemShut {NoStop}%
\bibitem [{\citenamefont {'t~Hooft}(1993)}]{tHooft1993}%
  \BibitemOpen
  \bibfield  {author} {\bibinfo {author} {\bibfnamefont {G.}~\bibnamefont
  {'t~Hooft}},\ }\href@noop {} {\bibfield  {journal} {\bibinfo  {journal}
  {arXiv e-prints}\ } (\bibinfo {year} {1993})},\ \bibinfo {note} {salamfest
  1993},\ \Eprint
  {http://arxiv.org/abs/gr-qc/9310026}{gr-qc/9310026}\BibitemShut {NoStop}%
\bibitem [{\citenamefont {Georgi}(2007{\natexlab{a}})}]{Georgi2007a}%
  \BibitemOpen
  \bibfield  {author} {\bibinfo {author} {\bibfnamefont {H.}~\bibnamefont
  {Georgi}},\ }\href@noop {} {\bibfield  {journal} {\bibinfo  {journal}
  {Phys. Rev. Lett.}\ }\textbf {\bibinfo {volume} {98}},\ \bibinfo {pages}
  {221601} (\bibinfo {year} {2007}{\natexlab{a}})},\ \Eprint
  {http://arxiv.org/abs/hep-ph/0703260}{hep-ph/0703260}\BibitemShut {NoStop}%
\bibitem [{\citenamefont {Georgi}(2007{\natexlab{b}})}]{Georgi2007b}%
  \BibitemOpen
  \bibfield  {author} {\bibinfo {author} {\bibfnamefont {H.}~\bibnamefont
  {Georgi}},\ }\href@noop {} {\bibfield  {journal} {\bibinfo  {journal}
  {Phys. Lett. B}\ }\textbf {\bibinfo {volume} {650}},\ \bibinfo {pages}
  {275} (\bibinfo {year} {2007}{\natexlab{b}})},\ \Eprint
  {http://arxiv.org/abs/0704.2457}{0704.2457}\BibitemShut {NoStop}%
\bibitem [{\citenamefont {van~der Bij}\ and\ \citenamefont
  {Dilcher}(2006)}]{vanderBij2006}%
  \BibitemOpen
  \bibfield  {author} {\bibinfo {author} {\bibfnamefont {J.~J.}\ \bibnamefont
  {van~der Bij}}\ and\ \bibinfo {author} {\bibfnamefont {S.}~\bibnamefont
  {Dilcher}},\ }\href@noop {} {\bibfield  {journal} {\bibinfo  {journal}
  {Phys. Lett. B}\ }\textbf {\bibinfo {volume} {638}},\ \bibinfo {pages}
  {234} (\bibinfo {year} {2006})},\ \Eprint
  {http://arxiv.org/abs/hep-ph/0603231}{hep-ph/0603231}\BibitemShut {NoStop}%
\bibitem [{\citenamefont {Hill}\ and\ \citenamefont {van~der
  Bij}(1987)}]{Hill1987}%
  \BibitemOpen
  \bibfield  {author} {\bibinfo {author} {\bibfnamefont {A.}~\bibnamefont
  {Hill}}\ and\ \bibinfo {author} {\bibfnamefont {J.~J.}\ \bibnamefont {van~der
  Bij}},\ }\href@noop {} {\bibfield  {journal} {\bibinfo  {journal}
  {Phys. Rev. D}\ }\textbf {\bibinfo {volume} {36}},\ \bibinfo {pages}
  {3463} (\bibinfo {year} {1987})}\BibitemShut {NoStop}%
\bibitem [{\citenamefont {Banks}\ and\ \citenamefont
  {Zaks}(1982)}]{BanksZaks1982}%
  \BibitemOpen
  \bibfield  {author} {\bibinfo {author} {\bibfnamefont {T.}~\bibnamefont
  {Banks}}\ and\ \bibinfo {author} {\bibfnamefont {A.}~\bibnamefont {Zaks}},\
  }\href@noop {} {\bibfield  {journal} {\bibinfo  {journal} {Nucl. Phys.
  B}\ }\textbf {\bibinfo {volume} {196}},\ \bibinfo {pages} {189} (\bibinfo
  {year} {1982})}\BibitemShut {NoStop}%
\bibitem [{\citenamefont {Cheung}\ \emph {et~al.}(2007)\citenamefont {Cheung},
  \citenamefont {Keung},\ and\ \citenamefont {Yuan}}]{Cheung2007}%
  \BibitemOpen
  \bibfield  {author} {\bibinfo {author} {\bibfnamefont {K.}~\bibnamefont
  {Cheung}}, \bibinfo {author} {\bibfnamefont {W.-Y.}\ \bibnamefont {Keung}}, \
  and\ \bibinfo {author} {\bibfnamefont {T.-C.}\ \bibnamefont {Yuan}},\
  }\href@noop {} {\bibfield  {journal} {\bibinfo  {journal} {Phys. Rev.
  Lett.}\ }\textbf {\bibinfo {volume} {99}},\ \bibinfo {pages} {051803}
  (\bibinfo {year} {2007})}\BibitemShut {NoStop}%
\bibitem [{\citenamefont {Liao}(2007)}]{Liao2007}%
  \BibitemOpen
  \bibfield  {author} {\bibinfo {author} {\bibfnamefont {Y.}~\bibnamefont
  {Liao}},\ }\href@noop {} {\bibfield  {journal} {\bibinfo  {journal}
  {Phys. Rev. D}\ }\textbf {\bibinfo {volume} {76}},\ \bibinfo {pages}
  {056006} (\bibinfo {year} {2007})}\BibitemShut {NoStop}%
\bibitem [{\citenamefont {Goldberg}\ and\ \citenamefont
  {Nath}(2008)}]{Goldberg2008}%
  \BibitemOpen
  \bibfield  {author} {\bibinfo {author} {\bibfnamefont {H.}~\bibnamefont
  {Goldberg}}\ and\ \bibinfo {author} {\bibfnamefont {P.}~\bibnamefont
  {Nath}},\ }\href@noop {} {\bibfield  {journal} {\bibinfo  {journal}
  {Phys. Rev. Lett.}\ }\textbf {\bibinfo {volume} {100}},\ \bibinfo {pages}
  {031803} (\bibinfo {year} {2008})}\BibitemShut {NoStop}%
\bibitem [{\citenamefont {Mureika}(2008)}]{Mureika2008}%
  \BibitemOpen
  \bibfield  {author} {\bibinfo {author} {\bibfnamefont {J.~R.}\ \bibnamefont
  {Mureika}},\ }\href@noop {} {\bibfield  {journal} {\bibinfo  {journal}
  {Phys. Lett. B}\ }\textbf {\bibinfo {volume} {660}},\ \bibinfo {pages}
  {561} (\bibinfo {year} {2008})}\BibitemShut {NoStop}%
\bibitem [{\citenamefont {Mureika}(2009)}]{Mureika2009}%
  \BibitemOpen
  \bibfield  {author} {\bibinfo {author} {\bibfnamefont {J.~R.}\ \bibnamefont
  {Mureika}},\ }\href@noop {} {\bibfield  {journal} {\bibinfo  {journal}
  {Phys. Rev. D}\ }\textbf {\bibinfo {volume} {79}},\ \bibinfo {pages}
  {056003} (\bibinfo {year} {2009})}\BibitemShut {NoStop}%
\bibitem [{\citenamefont {Gaete}\ \emph {et~al.}(2010)\citenamefont {Gaete},
  \citenamefont {Helay{\"e}l-Neto},\ and\ \citenamefont
  {Spallucci}}]{Gaete2010}%
  \BibitemOpen
  \bibfield  {author} {\bibinfo {author} {\bibfnamefont {P.}~\bibnamefont
  {Gaete}}, \bibinfo {author} {\bibfnamefont {J.~A.}\ \bibnamefont
  {Helay{\"e}l-Neto}}, \ and\ \bibinfo {author} {\bibfnamefont
  {E.}~\bibnamefont {Spallucci}},\ }\href@noop {} {\bibfield  {journal}
  {\bibinfo  {journal} {Phys. Lett. B}\ }\textbf {\bibinfo {volume}
  {693}},\ \bibinfo {pages} {155} (\bibinfo {year} {2010})}\BibitemShut
  {NoStop}%
\bibitem [{\citenamefont {Mureika}\ and\ \citenamefont
  {Spallucci}(2010)}]{MureikaSpallucci2010}%
  \BibitemOpen
  \bibfield  {author} {\bibinfo {author} {\bibfnamefont {J.~R.}\ \bibnamefont
  {Mureika}}\ and\ \bibinfo {author} {\bibfnamefont {E.}~\bibnamefont
  {Spallucci}},\ }\href@noop {} {\bibfield  {journal} {\bibinfo  {journal}
  {Phys. Lett. B}\ }\textbf {\bibinfo {volume} {693}},\ \bibinfo {pages}
  {129} (\bibinfo {year} {2010})}\BibitemShut {NoStop}%
\bibitem [{\citenamefont {Wondrak}\ \emph {et~al.}(2016)\citenamefont
  {Wondrak}, \citenamefont {Nicolini},\ and\ \citenamefont
  {Bleicher}}]{Wondrak2016}%
  \BibitemOpen
  \bibfield  {author} {\bibinfo {author} {\bibfnamefont {M.~F.}\ \bibnamefont
  {Wondrak}}, \bibinfo {author} {\bibfnamefont {P.}~\bibnamefont {Nicolini}}, \
  and\ \bibinfo {author} {\bibfnamefont {M.}~\bibnamefont {Bleicher}},\
  }\href@noop {} {\bibfield  {journal} {\bibinfo  {journal} {Phys. Lett.
  B}\ }\textbf {\bibinfo {volume} {759}},\ \bibinfo {pages} {589} (\bibinfo
  {year} {2016})}\BibitemShut {NoStop}%
\bibitem [{\citenamefont {Davoudiasl}(2007)}]{Davoudiasl2007}%
  \BibitemOpen
  \bibfield  {author} {\bibinfo {author} {\bibfnamefont {H.}~\bibnamefont
  {Davoudiasl}},\ }\href@noop {} {\bibfield  {journal} {\bibinfo  {journal}
  {Phys. Rev. Lett.}\ }\textbf {\bibinfo {volume} {99}},\ \bibinfo {pages}
  {141301} (\bibinfo {year} {2007})}\BibitemShut {NoStop}%
\bibitem [{\citenamefont {Freitas}\ and\ \citenamefont
  {Wyler}(2007)}]{Freitas2007}%
  \BibitemOpen
  \bibfield  {author} {\bibinfo {author} {\bibfnamefont {A.}~\bibnamefont
  {Freitas}}\ and\ \bibinfo {author} {\bibfnamefont {D.}~\bibnamefont
  {Wyler}},\ }\href@noop {} {\bibfield  {journal} {\bibinfo  {journal}
  {JHEP}\ }\textbf {\bibinfo {volume} {12}},\ \bibinfo {pages} {033}
  (\bibinfo {year} {2007})}\BibitemShut {NoStop}%
\bibitem [{\citenamefont {Nicolini}\ and\ \citenamefont
  {Spallucci}(2011)}]{NicoliniSpallucci2011}%
  \BibitemOpen
  \bibfield  {author} {\bibinfo {author} {\bibfnamefont {P.}~\bibnamefont
  {Nicolini}}\ and\ \bibinfo {author} {\bibfnamefont {E.}~\bibnamefont
  {Spallucci}},\ }\href {\doibase 10.1016/j.physletb.2010.10.041} {\bibfield
  {journal} {\bibinfo  {journal} {Phys. Lett. B}\ }\textbf {\bibinfo {volume}
  {695}},\ \bibinfo {pages} {290} (\bibinfo {year} {2011})},\ \Eprint
  {http://arxiv.org/abs/1005.1509}{arXiv:1005.1509 [hep-th]}\BibitemShut
  {NoStop}%
\bibitem [{\citenamefont {Frassino}\ \emph {et~al.}(2017)\citenamefont
  {Frassino}, \citenamefont {Nicolini},\ and\ \citenamefont
  {Panella}}]{Frassino2017}%
  \BibitemOpen
  \bibfield  {author} {\bibinfo {author} {\bibfnamefont {A.~M.}\ \bibnamefont
  {Frassino}}, \bibinfo {author} {\bibfnamefont {P.}~\bibnamefont {Nicolini}},
  \ and\ \bibinfo {author} {\bibfnamefont {O.}~\bibnamefont {Panella}},\
  }\href@noop {} {\bibfield  {journal} {\bibinfo  {journal} {Phys. Lett.
  B}\ }\textbf {\bibinfo {volume} {772}},\ \bibinfo {pages} {675} (\bibinfo
  {year} {2017})}\BibitemShut {NoStop}%
\bibitem [{\citenamefont {LeBlanc}\ and\ \citenamefont
  {Grushin}(2015)}]{LeBlanc2015}%
  \BibitemOpen
  \bibfield  {author} {\bibinfo {author} {\bibfnamefont {J.~P.~F.}\
  \bibnamefont {LeBlanc}}\ and\ \bibinfo {author} {\bibfnamefont {A.~G.}\
  \bibnamefont {Grushin}},\ }\href@noop {} {\bibfield  {journal} {\bibinfo
  {journal} {New J. Phys.}\ }\textbf {\bibinfo {volume} {17}},\ \bibinfo
  {pages} {033039} (\bibinfo {year} {2015})}\BibitemShut {NoStop}%
\bibitem [{\citenamefont {Karch}\ \emph {et~al.}(2016)\citenamefont {Karch},
  \citenamefont {Limtragool},\ and\ \citenamefont {Phillips}}]{Karch2016}%
  \BibitemOpen
  \bibfield  {author} {\bibinfo {author} {\bibfnamefont {A.}~\bibnamefont
  {Karch}}, \bibinfo {author} {\bibfnamefont {K.}~\bibnamefont {Limtragool}}, \
  and\ \bibinfo {author} {\bibfnamefont {P.~W.}\ \bibnamefont {Phillips}},\
  }\href@noop {} {\bibfield  {journal} {\bibinfo  {journal} {JHEP}\
  }\textbf {\bibinfo {volume} {2016}},\ \bibinfo {pages} {175} (\bibinfo
  {year} {2016})},\ \Eprint
  {http://arxiv.org/abs/1511.02868}{1511.02868}\BibitemShut {NoStop}%
\bibitem [{\citenamefont {Abbott}\ and\ \citenamefont
  {Wise}(1981)}]{AbbottWise1981}%
  \BibitemOpen
  \bibfield  {author} {\bibinfo {author} {\bibfnamefont {L.~F.}\ \bibnamefont
  {Abbott}}\ and\ \bibinfo {author} {\bibfnamefont {M.~B.}\ \bibnamefont
  {Wise}},\ }\href@noop {} {\bibfield  {journal} {\bibinfo  {journal} {Amer.
  J. Phys.}\ }\textbf {\bibinfo {volume} {49}},\ \bibinfo {pages} {37}
  (\bibinfo {year} {1981})}\BibitemShut {NoStop}%
\bibitem [{\citenamefont {Nicolini}\ and\ \citenamefont
  {Niedner}(2011)}]{NicoliniNiedner2011}%
  \BibitemOpen
  \bibfield  {author} {\bibinfo {author} {\bibfnamefont {P.}~\bibnamefont
  {Nicolini}}\ and\ \bibinfo {author} {\bibfnamefont {B.}~\bibnamefont
  {Niedner}},\ }\href {\doibase 10.1103/PhysRevD.83.024017} {\bibfield
  {journal} {\bibinfo  {journal} {Phys. Rev. D}\ }\textbf {\bibinfo {volume}
  {83}},\ \bibinfo {pages} {024017} (\bibinfo {year} {2011})},\ \Eprint
  {http://arxiv.org/abs/1009.3267}{arXiv:1009.3267 [gr-qc]}\BibitemShut
  {NoStop}%
\bibitem [{\citenamefont {Eguchi}(1980)}]{Eguchi1980}%
  \BibitemOpen
  \bibfield  {author} {\bibinfo {author} {\bibfnamefont {T.}~\bibnamefont
  {Eguchi}},\ }\href@noop {} {\bibfield  {journal} {\bibinfo  {journal}
  {Phys. Rev. Lett.}\ }\textbf {\bibinfo {volume} {44}},\ \bibinfo {pages}
  {126} (\bibinfo {year} {1980})}\BibitemShut {NoStop}%
\bibitem [{\citenamefont {Ansoldi}\ \emph {et~al.}(1997)\citenamefont
  {Ansoldi}, \citenamefont {Aurilia},\ and\ \citenamefont
  {Spallucci}}]{Ansoldi1997}%
  \BibitemOpen
  \bibfield  {author} {\bibinfo {author} {\bibfnamefont {S.}~\bibnamefont
  {Ansoldi}}, \bibinfo {author} {\bibfnamefont {A.}~\bibnamefont {Aurilia}}, \
  and\ \bibinfo {author} {\bibfnamefont {E.}~\bibnamefont {Spallucci}},\ }\href
  {\doibase 10.1103/PhysRevD.56.2352} {\bibfield  {journal} {\bibinfo
  {journal} {Phys. Rev. D}\ }\textbf {\bibinfo {volume} {56}},\ \bibinfo
  {pages} {2352} (\bibinfo {year} {1997})},\ \Eprint
  {http://arxiv.org/abs/hep-th/9705010}{hep-th/9705010}\BibitemShut {NoStop}%
\bibitem [{\citenamefont {Ansoldi}\ \emph {et~al.}(1999)\citenamefont
  {Ansoldi}, \citenamefont {Aurilia},\ and\ \citenamefont
  {Spallucci}}]{Ansoldi1999}%
  \BibitemOpen
  \bibfield  {author} {\bibinfo {author} {\bibfnamefont {S.}~\bibnamefont
  {Ansoldi}}, \bibinfo {author} {\bibfnamefont {A.}~\bibnamefont {Aurilia}}, \
  and\ \bibinfo {author} {\bibfnamefont {E.}~\bibnamefont {Spallucci}},\ }\href
  {\doibase 10.1016/S0960-0779(98)00115-5} {\bibfield  {journal} {\bibinfo
  {journal} {Chaos Soliton. Fract.}\ }\textbf {\bibinfo {volume} {10}},\
  \bibinfo {pages} {197} (\bibinfo {year} {1999})},\ \Eprint
  {http://arxiv.org/abs/hep-th/9803229}{hep-th/9803229}\BibitemShut {NoStop}%
\bibitem [{\citenamefont {Zeeman}(1976)}]{ZeemanCatastrophe}%
  \BibitemOpen
  \bibfield  {author} {\bibinfo {author} {\bibfnamefont {E.~C.}\ \bibnamefont
  {Zeeman}},\ }\href@noop {} {\enquote {\bibinfo {title} {Catastrophe
  theory},}\ } (\bibinfo {year} {1976}),\ \bibinfo {note} {Scientific American
  234, 65--83 (April 1976)}\BibitemShut {NoStop}%
\bibitem [{\citenamefont {Sakharov}(1968)}]{Sakharov1968}%
  \BibitemOpen
  \bibfield  {author} {\bibinfo {author} {\bibfnamefont {A.~D.}\ \bibnamefont
  {Sakharov}},\ }\href@noop {} {\bibfield  {journal} {\bibinfo  {journal}
  {Sov. Phys. Dokl.}\ }\textbf {\bibinfo {volume} {12}},\ \bibinfo {pages}
  {1040} (\bibinfo {year} {1968})}\BibitemShut {NoStop}%
\bibitem [{\citenamefont {Birrell}\ and\ \citenamefont
  {Davies}(1982)}]{BirrellDavies1982}%
  \BibitemOpen
  \bibfield  {author} {\bibinfo {author} {\bibfnamefont {N.~D.}\ \bibnamefont
  {Birrell}}\ and\ \bibinfo {author} {\bibfnamefont {P.~C.~W.}\ \bibnamefont
  {Davies}},\ }\href@noop {} {\emph {\bibinfo {title} {Quantum Fields in Curved
  Space}}}\ (\bibinfo  {publisher} {Cambridge University Press},\ \bibinfo
  {address} {Cambridge},\ \bibinfo {year} {1982})\BibitemShut {NoStop}%
\bibitem [{\citenamefont {Ord}(1983)}]{Ord1983}%
  \BibitemOpen
  \bibfield  {author} {\bibinfo {author} {\bibfnamefont {G.~N.}\ \bibnamefont
  {Ord}},\ }\href@noop {} {\bibfield  {journal} {\bibinfo  {journal} {J.
  Phys. A}\ }\textbf {\bibinfo {volume} {16}},\ \bibinfo {pages} {1869}
  (\bibinfo {year} {1983})}\BibitemShut {NoStop}%
\bibitem [{\citenamefont {Cannata}\ and\ \citenamefont
  {Ferrari}(1988)}]{Cannata1988}%
  \BibitemOpen
  \bibfield  {author} {\bibinfo {author} {\bibfnamefont {F.}~\bibnamefont
  {Cannata}}\ and\ \bibinfo {author} {\bibfnamefont {L.}~\bibnamefont
  {Ferrari}},\ }\href@noop {} {\bibfield  {journal} {\bibinfo  {journal}
  {Amer. J. Phys.}\ }\textbf {\bibinfo {volume} {56}},\ \bibinfo {pages}
  {721} (\bibinfo {year} {1988})}\BibitemShut {NoStop}%
\bibitem [{\citenamefont {Nottale}(1989)}]{Nottale1989}%
  \BibitemOpen
  \bibfield  {author} {\bibinfo {author} {\bibfnamefont {L.}~\bibnamefont
  {Nottale}},\ }\href@noop {} {\bibfield  {journal} {\bibinfo  {journal}
  {Int. J. Mod. Phys. A}\ }\textbf {\bibinfo {volume} {4}},\ \bibinfo
  {pages} {5047} (\bibinfo {year} {1989})}\BibitemShut {NoStop}%
\bibitem [{\citenamefont {Nottale}(1992)}]{Nottale1992}%
  \BibitemOpen
  \bibfield  {author} {\bibinfo {author} {\bibfnamefont {L.}~\bibnamefont
  {Nottale}},\ }\href@noop {} {\emph {\bibinfo {title} {Fractal Spacetime and
  Microphysics}}}\ (\bibinfo  {publisher} {World Scientific},\ \bibinfo
  {address} {Singapore},\ \bibinfo {year} {1992})\BibitemShut {NoStop}%
\bibitem [{\citenamefont {Casadio}\ and\ \citenamefont
  {Scardigli}(2014)}]{Casadio2014}%
  \BibitemOpen
  \bibfield  {author} {\bibinfo {author} {\bibfnamefont {R.}~\bibnamefont
  {Casadio}}\ and\ \bibinfo {author} {\bibfnamefont {F.}~\bibnamefont
  {Scardigli}},\ }\href@noop {} {\bibfield  {journal} {\bibinfo  {journal}
  {Eur. Phys. J. C}\ }\textbf {\bibinfo {volume} {74}},\ \bibinfo {pages}
  {2685} (\bibinfo {year} {2014})},\ \Eprint
  {http://arxiv.org/abs/1306.5298}{1306.5298}\BibitemShut {NoStop}%
\bibitem [{\citenamefont {Casadio}\ \emph {et~al.}(2014)\citenamefont
  {Casadio}, \citenamefont {Micu},\ and\ \citenamefont
  {Scardigli}}]{Casadio2014b}%
  \BibitemOpen
  \bibfield  {author} {\bibinfo {author} {\bibfnamefont {R.}~\bibnamefont
  {Casadio}}, \bibinfo {author} {\bibfnamefont {O.}~\bibnamefont {Micu}}, \
  and\ \bibinfo {author} {\bibfnamefont {F.}~\bibnamefont {Scardigli}},\
  }\href@noop {} {\bibfield  {journal} {\bibinfo  {journal} {Phys. Lett.
  B}\ }\textbf {\bibinfo {volume} {732}},\ \bibinfo {pages} {105} (\bibinfo
  {year} {2014})},\ \Eprint
  {http://arxiv.org/abs/1311.5698}{1311.5698}\BibitemShut {NoStop}%
\bibitem [{\citenamefont {Casadio}\ \emph {et~al.}(2015)\citenamefont
  {Casadio}, \citenamefont {Micu},\ and\ \citenamefont
  {Nicolini}}]{Casadio2015}%
  \BibitemOpen
  \bibfield  {author} {\bibinfo {author} {\bibfnamefont {R.}~\bibnamefont
  {Casadio}}, \bibinfo {author} {\bibfnamefont {O.}~\bibnamefont {Micu}}, \
  and\ \bibinfo {author} {\bibfnamefont {P.}~\bibnamefont {Nicolini}},\ }in\
  \href@noop {} {\emph {\bibinfo {booktitle} {Quantum Aspects of Black
  Holes}}},\ \bibinfo {series} {Fundamental Theories of Physics}, Vol.\
  \bibinfo {volume} {178},\ \bibinfo {editor} {edited by\ \bibinfo {editor}
  {\bibfnamefont {X.}~\bibnamefont {Calmet}}}\ (\bibinfo  {publisher}
  {Springer},\ \bibinfo {year} {2015})\ pp.\ \bibinfo {pages} {293--322},\
  \Eprint {http://arxiv.org/abs/1405.1692}{1405.1692}\BibitemShut {NoStop}%
\bibitem [{\citenamefont {Casadio}\ \emph {et~al.}(2016)\citenamefont
  {Casadio}, \citenamefont {Giugno},\ and\ \citenamefont {Micu}}]{Casadio2016}%
  \BibitemOpen
  \bibfield  {author} {\bibinfo {author} {\bibfnamefont {R.}~\bibnamefont
  {Casadio}}, \bibinfo {author} {\bibfnamefont {A.}~\bibnamefont {Giugno}}, \
  and\ \bibinfo {author} {\bibfnamefont {O.}~\bibnamefont {Micu}},\ }\href@noop
  {} {\bibfield  {journal} {\bibinfo  {journal} {Int. J. Mod. Phys.
  D}\ }\textbf {\bibinfo {volume} {25}},\ \bibinfo {pages} {1630006} (\bibinfo
  {year} {2016})},\ \Eprint
  {http://arxiv.org/abs/1512.04071}{1512.04071}\BibitemShut {NoStop}%
\bibitem [{\citenamefont {Spallucci}\ and\ \citenamefont
  {Smailagic}(2016)}]{Spallucci2016}%
  \BibitemOpen
  \bibfield  {author} {\bibinfo {author} {\bibfnamefont {E.}~\bibnamefont
  {Spallucci}}\ and\ \bibinfo {author} {\bibfnamefont {A.}~\bibnamefont
  {Smailagic}},\ }\href@noop {} {\bibfield  {journal} {\bibinfo  {journal}
  {arXiv e-prints}\ } (\bibinfo {year} {2016})},\ \Eprint
  {http://arxiv.org/abs/1601.06004}{1601.06004 [hep-th]}\BibitemShut {NoStop}%
\bibitem [{\citenamefont {Spallucci}\ and\ \citenamefont
  {Smailagic}(2017)}]{Spallucci2017}%
  \BibitemOpen
  \bibfield  {author} {\bibinfo {author} {\bibfnamefont {E.}~\bibnamefont
  {Spallucci}}\ and\ \bibinfo {author} {\bibfnamefont {A.}~\bibnamefont
  {Smailagic}},\ }in\ \href@noop {} {\emph {\bibinfo {booktitle} {Quantum
  Gravity: Theory and Research}}},\ \bibinfo {editor} {edited by\ \bibinfo
  {editor} {\bibfnamefont {B.}~\bibnamefont {Mitchell}}}\ (\bibinfo
  {publisher} {Nova Science Publishers},\ \bibinfo {year} {2017})\ pp.\
  \bibinfo {pages} {1--32},\ \Eprint
  {http://arxiv.org/abs/1605.05911}{1605.05911}\BibitemShut {NoStop}%
\bibitem [{\citenamefont {Dvali}\ and\ \citenamefont
  {Gomez}(2013)}]{DvaliGomez2013}%
  \BibitemOpen
  \bibfield  {author} {\bibinfo {author} {\bibfnamefont {G.}~\bibnamefont
  {Dvali}}\ and\ \bibinfo {author} {\bibfnamefont {C.}~\bibnamefont {Gomez}},\
  }\href@noop {} {\bibfield  {journal} {\bibinfo  {journal} {Fortschr. Phys.}\
  }\textbf {\bibinfo {volume} {61}},\ \bibinfo {pages} {742} (\bibinfo
  {year} {2013})},\ \Eprint
  {http://arxiv.org/abs/1112.3359}{1112.3359}\BibitemShut {NoStop}%
\bibitem [{\citenamefont {Dvali}\ and\ \citenamefont
  {Gomez}(2014)}]{DvaliGomez2014}%
  \BibitemOpen
  \bibfield  {author} {\bibinfo {author} {\bibfnamefont {G.}~\bibnamefont
  {Dvali}}\ and\ \bibinfo {author} {\bibfnamefont {C.}~\bibnamefont {Gomez}},\
  }\href@noop {} {\bibfield  {journal} {\bibinfo  {journal} {Eur. Phys. J.
  C}\ }\textbf {\bibinfo {volume} {74}},\ \bibinfo {pages} {2752} (\bibinfo
  {year} {2014})},\ \Eprint
  {http://arxiv.org/abs/1207.4059}{1207.4059}\BibitemShut {NoStop}%
\bibitem [{\citenamefont {Aurilia}\ and\ \citenamefont
  {Spallucci}(2013)}]{AuriliaSpallucci2013}%
  \BibitemOpen
  \bibfield  {author} {\bibinfo {author} {\bibfnamefont {A.}~\bibnamefont
  {Aurilia}}\ and\ \bibinfo {author} {\bibfnamefont {E.}~\bibnamefont
  {Spallucci}},\ }\href@noop {} {\bibfield  {journal} {\bibinfo  {journal}
  {arXiv e-prints}\ } (\bibinfo {year} {2013})},\ \Eprint
  {http://arxiv.org/abs/1309.7186}{1309.7186 [gr-qc]}\BibitemShut {NoStop}%
\bibitem [{\citenamefont {Carr}(2016)}]{Carr2016}%
  \BibitemOpen
  \bibfield  {author} {\bibinfo {author} {\bibfnamefont {B.~J.}\ \bibnamefont
  {Carr}},\ }\href {\doibase 10.1007/978-3-319-20046-0_19} {\bibfield
  {journal} {\bibinfo  {journal} {Springer Proc. Phys.}\ }\textbf
  {\bibinfo {volume} {170}},\ \bibinfo {pages} {159} (\bibinfo {year}
  {2016})},\ \Eprint {http://arxiv.org/abs/1402.1427}{1402.1427}\BibitemShut
  {NoStop}%
\bibitem [{\citenamefont {Carr}\ \emph {et~al.}(2015)\citenamefont {Carr},
  \citenamefont {Mureika},\ and\ \citenamefont {Nicolini}}]{Carr2015}%
  \BibitemOpen
  \bibfield  {author} {\bibinfo {author} {\bibfnamefont {B.~J.}\ \bibnamefont
  {Carr}}, \bibinfo {author} {\bibfnamefont {J.}~\bibnamefont {Mureika}}, \
  and\ \bibinfo {author} {\bibfnamefont {P.}~\bibnamefont {Nicolini}},\
  }\href@noop {} {\bibfield  {journal} {\bibinfo  {journal} {arXiv e-prints}\
  } (\bibinfo {year} {2015})},\ \Eprint
  {http://arxiv.org/abs/1504.07637}{1504.07637 [gr-qc]}\BibitemShut {NoStop}%
\bibitem [{\citenamefont {Dvali}\ \emph {et~al.}(2011)\citenamefont {Dvali},
  \citenamefont {Folkerts},\ and\ \citenamefont {Germani}}]{Dvali2011}%
  \BibitemOpen
  \bibfield  {author} {\bibinfo {author} {\bibfnamefont {G.}~\bibnamefont
  {Dvali}}, \bibinfo {author} {\bibfnamefont {S.}~\bibnamefont {Folkerts}}, \
  and\ \bibinfo {author} {\bibfnamefont {C.}~\bibnamefont {Germani}},\
  }\href@noop {} {\bibfield  {journal} {\bibinfo  {journal} {Phys. Rev.
  D}\ }\textbf {\bibinfo {volume} {84}},\ \bibinfo {pages} {024039} (\bibinfo
  {year} {2011})},\ \Eprint
  {http://arxiv.org/abs/1006.0984}{1006.0984}\BibitemShut {NoStop}%
\bibitem [{\citenamefont {Barrau}\ \emph {et~al.}(2014)\citenamefont {Barrau},
  \citenamefont {Rovelli},\ and\ \citenamefont {Vidotto}}]{Barrau2014}%
  \BibitemOpen
  \bibfield  {author} {\bibinfo {author} {\bibfnamefont {A.}~\bibnamefont
  {Barrau}}, \bibinfo {author} {\bibfnamefont {C.}~\bibnamefont {Rovelli}}, \
  and\ \bibinfo {author} {\bibfnamefont {F.}~\bibnamefont {Vidotto}},\
  }\href@noop {} {\bibfield  {journal} {\bibinfo  {journal} {Phys. Rev.
  D}\ }\textbf {\bibinfo {volume} {90}},\ \bibinfo {pages} {127503} (\bibinfo
  {year} {2014})},\ \Eprint
  {http://arxiv.org/abs/1404.7153}{1404.7153}\BibitemShut {NoStop}%
\bibitem [{\citenamefont {Jacobson}(1995)}]{Jacobson1995}%
  \BibitemOpen
  \bibfield  {author} {\bibinfo {author} {\bibfnamefont {T.}~\bibnamefont
  {Jacobson}},\ }\href {\doibase 10.1103/PhysRevLett.75.1260} {\bibfield
  {journal} {\bibinfo  {journal} {Phys. Rev. Lett.}\ }\textbf {\bibinfo
  {volume} {75}},\ \bibinfo {pages} {1260} (\bibinfo {year} {1995})},\ \Eprint
  {http://arxiv.org/abs/gr-qc/9504004}{gr-qc/9504004}\BibitemShut {NoStop}%
\bibitem [{\citenamefont {Padmanabhan}(2002)}]{Padmanabhan2002}%
  \BibitemOpen
  \bibfield  {author} {\bibinfo {author} {\bibfnamefont {T.}~\bibnamefont
  {Padmanabhan}},\ }\href {\doibase 10.1088/0264-9381/19/21/306} {\bibfield
  {journal} {\bibinfo  {journal} {Class. Quant. Grav.}\ }\textbf
  {\bibinfo {volume} {19}},\ \bibinfo {pages} {5387} (\bibinfo {year}
  {2002})},\ \Eprint
  {http://arxiv.org/abs/gr-qc/0204019}{gr-qc/0204019}\BibitemShut {NoStop}%
\bibitem [{\citenamefont {Padmanabhan}(2010)}]{Padmanabhan2010}%
  \BibitemOpen
  \bibfield  {author} {\bibinfo {author} {\bibfnamefont {T.}~\bibnamefont
  {Padmanabhan}},\ }\href {\doibase 10.1088/0034-4885/73/4/046901} {\bibfield
  {journal} {\bibinfo  {journal} {Rep. Prog. Phys.}\ }\textbf {\bibinfo
  {volume} {73}},\ \bibinfo {pages} {046901} (\bibinfo {year}
  {2010})},\ \Eprint {http://arxiv.org/abs/0911.5004}{0911.5004}\BibitemShut
  {NoStop}%
\bibitem [{\citenamefont {Bianconi}(2025)}]{Bianconi2025}%
  \BibitemOpen
  \bibfield  {author} {\bibinfo {author} {\bibfnamefont {G.}~\bibnamefont
  {Bianconi}},\ }\href {\doibase 10.1103/PhysRevD.111.066001} {\bibfield
  {journal} {\bibinfo  {journal} {Phys. Rev. D}\ }\textbf {\bibinfo {volume}
  {111}},\ \bibinfo {pages} {066001} (\bibinfo {year} {2025})},\ \Eprint
  {http://arxiv.org/abs/2408.14391}{2408.14391 [gr-qc]}\BibitemShut {NoStop}%
\bibitem [{\citenamefont {Dvali}\ and\ \citenamefont
  {Gomez}(2012)}]{DvaliGomez2012}%
  \BibitemOpen
  \bibfield  {author} {\bibinfo {author} {\bibfnamefont {G.}~\bibnamefont
  {Dvali}}\ and\ \bibinfo {author} {\bibfnamefont {C.}~\bibnamefont {Gomez}},\
  }\href@noop {} {\bibfield  {journal} {\bibinfo  {journal} {arXiv e-prints}\
  } (\bibinfo {year} {2012})},\ \Eprint
  {http://arxiv.org/abs/1212.0765}{1212.0765 [hep-th]}\BibitemShut {NoStop}%


\bibitem [{\citenamefont {Giddings}(2017)}]{Giddings2017}%
  \BibitemOpen
  \bibfield  {author} {\bibinfo {author} {\bibfnamefont {S.~B.}\ \bibnamefont
  {Giddings}},\ }\href {\doibase 10.1038/s41550-017-0067} {\bibfield  {journal}
  {\bibinfo  {journal} {Nat. Astron.}\ }\textbf {\bibinfo {volume} {1}},\
  \bibinfo {pages} {0067} (\bibinfo {year} {2017})},\ \Eprint
  {http://arxiv.org/abs/1703.03387}{1703.03387 [gr-qc]}\BibitemShut
  {NoStop}%
\bibitem [{\citenamefont {Nicolini}\ and\ \citenamefont
  {Spallucci}(2014)}]{NicoliniSpallucci2014}%
  \BibitemOpen
  \bibfield  {author} {\bibinfo {author} {\bibfnamefont {P.}~\bibnamefont
  {Nicolini}}\ and\ \bibinfo {author} {\bibfnamefont {E.}~\bibnamefont
  {Spallucci}},\ }\href {\doibase 10.1155/2014/805684} {\bibfield  {journal}
  {\bibinfo  {journal} {Adv. High Energy Phys.}\ }\textbf {\bibinfo {volume}
  {2014}},\ \bibinfo {pages} {805684} (\bibinfo {year} {2014})},\ \Eprint
  {http://arxiv.org/abs/1210.0015}{1210.0015 [hep-th]}\BibitemShut {NoStop}%
\bibitem [{\citenamefont {Nicolini}(2018)}]{Nicolini2018}%
  \BibitemOpen
  \bibfield  {author} {\bibinfo {author} {\bibfnamefont {P.}~\bibnamefont
  {Nicolini}},\ }\href {\doibase 10.1016/j.physletb.2018.01.013} {\bibfield
  {journal} {\bibinfo  {journal} {Phys. Lett. B}\ }\textbf {\bibinfo {volume}
  {778}},\ \bibinfo {pages} {88} (\bibinfo {year} {2018})},\ \Eprint
  {http://arxiv.org/abs/1712.05062}{1712.05062 [gr-qc]}\BibitemShut {NoStop}%
\bibitem [{\citenamefont {Kleinert}(1987)}]{Kleinert1987}%
  \BibitemOpen
  \bibfield  {author} {\bibinfo {author} {\bibfnamefont {H.}~\bibnamefont
  {Kleinert}},\ }\href {\doibase 10.1002/andp.19874990206} {\bibfield  {journal}
  {\bibinfo  {journal} {Ann. Phys. (Leipzig)}\ }\textbf {\bibinfo {volume}
  {44}},\ \bibinfo {pages} {117} (\bibinfo {year} {1987})}\BibitemShut
  {NoStop}%
\bibitem [{\citenamefont {Aurilia}\ \emph {et~al.}(2002)\citenamefont
  {Aurilia}, \citenamefont {Ansoldi},\ and\ \citenamefont
  {Spallucci}}]{AuriliaAnsoldiSpallucci2002}%
  \BibitemOpen
  \bibfield  {author} {\bibinfo {author} {\bibfnamefont {A.}~\bibnamefont
  {Aurilia}}, \bibinfo {author} {\bibfnamefont {S.}~\bibnamefont {Ansoldi}}, \
  and\ \bibinfo {author} {\bibfnamefont {E.}~\bibnamefont {Spallucci}},\ }\href
  {\doibase 10.1088/0264-9381/19/12/307} {\bibfield  {journal} {\bibinfo
  {journal} {Class. Quant. Grav.}\ }\textbf {\bibinfo {volume} {19}},\ \bibinfo
  {pages} {3207} (\bibinfo {year} {2002})},\ \Eprint
  {http://arxiv.org/abs/hep-th/0205028}{hep-th/0205028}\BibitemShut {NoStop}%
\bibitem [{\citenamefont {Nicolai}(2014)}]{Nicolai2014}%
  \BibitemOpen
  \bibfield  {author} {\bibinfo {author} {\bibfnamefont {H.}~\bibnamefont
  {Nicolai}},\ }\href {\doibase 10.1007/978-3-319-06349-2\_18} {\bibfield
  {journal} {\bibinfo  {journal} {Fundamental Theories of Physics}\ }\textbf
  {\bibinfo {volume} {177}},\ \bibinfo {pages} {369} (\bibinfo {year}
  {2014})},\ \Eprint {http://arxiv.org/abs/1301.5481}{1301.5481
  [gr-qc]}\BibitemShut {NoStop}%
\bibitem [{\citenamefont {Ackermann}\ \emph {et~al.}(2009)\citenamefont
  {Ackermann} \emph {et~al.}}]{Ackermann2009}%
  \BibitemOpen
  \bibfield  {author} {\bibinfo {author} {\bibfnamefont {M.}~\bibnamefont
  {Ackermann}} \emph {et~al.},\ }\href {\doibase 10.1038/nature08574}
  {\bibfield  {journal} {\bibinfo  {journal} {Nature}\ }\textbf {\bibinfo
  {volume} {462}},\ \bibinfo {pages} {331} (\bibinfo {year} {2009})},\ \Eprint
  {http://arxiv.org/abs/0908.1832}{0908.1832 [astro-ph.HE]}\BibitemShut
  {NoStop}%
\bibitem [{\citenamefont {Abdo}\ \emph {et~al.}(2009)\citenamefont {Abdo} \emph
  {et~al.}}]{Abdo2009}%
  \BibitemOpen
  \bibfield  {author} {\bibinfo {author} {\bibfnamefont {A.~A.}\ \bibnamefont
  {Abdo}} \emph {et~al.},\ }\href {\doibase 10.1126/science.1169101} {\bibfield
  {journal} {\bibinfo  {journal} {Science}\ }\textbf {\bibinfo {volume}
  {323}},\ \bibinfo {pages} {1688} (\bibinfo {year} {2009})}\BibitemShut
  {NoStop}%
\bibitem [{\citenamefont {Amelino-Camelia}\ \emph {et~al.}(1998)\citenamefont
  {Amelino-Camelia}, \citenamefont {Ellis}, \citenamefont {Mavromatos},
  \citenamefont {Nanopoulos},\ and\ \citenamefont
  {Sarkar}}]{AmelinoCamelia1998}%
  \BibitemOpen
  \bibfield  {author} {\bibinfo {author} {\bibfnamefont {G.}~\bibnamefont
  {Amelino-Camelia}}, \bibinfo {author} {\bibfnamefont {J.}~\bibnamefont
  {Ellis}}, \bibinfo {author} {\bibfnamefont {N.~E.}\ \bibnamefont
  {Mavromatos}}, \bibinfo {author} {\bibfnamefont {D.~V.}\ \bibnamefont
  {Nanopoulos}}, \ and\ \bibinfo {author} {\bibfnamefont {S.}~\bibnamefont
  {Sarkar}},\ }\href {\doibase 10.1038/31647} {\bibfield  {journal} {\bibinfo
  {journal} {Nature}\ }\textbf {\bibinfo {volume} {393}},\ \bibinfo {pages}
  {763} (\bibinfo {year} {1998})},\ \Eprint
  {http://arxiv.org/abs/astro-ph/9712103}{astro-ph/9712103
  [astro-ph]}\BibitemShut {NoStop}%
\end{thebibliography}
\end{document}